\documentclass[11pt,superscriptaddress,amsmath,amssymb,nofootinbib,longbibliography]{revtex4-2}
\usepackage{graphicx}
\usepackage{dcolumn}
\usepackage{bm}
\usepackage{amssymb}
\usepackage{amsmath}
\usepackage{color}
\usepackage[normalem]{ulem}
\usepackage{dsfont}

\newcommand{\bwt}{\begin{widetext}}
\newcommand{\ewt}{\end{widetext}}
\newcommand{\newc}{\newcommand}
\newc{\hc}{\dagger}
\newc{\pd}{\partial}
\newc{\beq}{\begin{equation}}
\newc{\eeq}{\end{equation}}
\newc{\beqa}{\begin{eqnarray}}
\newc{\eeqa}{\end{eqnarray}}
\newc{\bi}{\begin{itemize}}
\newc{\ei}{\end{itemize}}
\newc{\ra}{\rightarrow}
\newc{\la}{\leftarrow}
\newc{\lra}{\longrightarrow}
\newc{\lla}{\longleftarrow}
\newc{\Lra}{\Longrightarrow}
\newc{\Lla}{\Longleftarrow}
\newc{\half}{\frac{1}{2}}
\newc{\fth}{\frac{1}{4}}
\newc{\hchecked}{\backslash\!\!\!\!\checkmark }
\newc{\del}{\delta}
\newc{\Del}{\Delta}
\newc{\gm}{\gamma}
\newc{\Gm}{\Gamma}
\newc{\lam}{\lambda}
\newc{\kap}{\kappa}
\newc{\tri}{\triangle}
\newc{\eps}{\epsilon}
\newc{\epsp}{\epsilon^\prime}
\newc{\ot}{\frac{1}{3}}
\newc{\tth}{\frac{2}{3}}
\newc{\ft}{\frac{4}{3}}
\newc{\wt}{\widetilde}
\newc{\ovl}{\overline}
\newc{\tchi}{\tilde{\chi}}
\newc{\ds}{\displaystyle}
\newc{\pmt}{\pm\!\pm}
\newc{\PL}{\hat{L}}
\newc{\PR}{\hat{R}}

\newc{\msm}{\mathrm{SM}}
\newc{\msh}{\mathrm{sh}}
\newc{\mtev}{\mathrm{TeV}}
\newc{\mgev}{\mathrm{GeV}}
\newc{\mmev}{\mathrm{MeV}}
\newc{\mkev}{\mathrm{keV}}
\newc{\mev}{\mathrm{eV}}
\newc{\Tr}{\mathrm{Tr}}
\newc{\nonr}{\nonumber}
\newc{\clbl}{\color{blue}}
\newc{\clg}{\color{green}}
\newc{\clr}{\color{red}}
\mathchardef\mhyphen="2D
\newc{\SL}{\not\!\!}
\usepackage{bm}
\usepackage{amsfonts}
\usepackage{xcolor}
\usepackage{amscd}
\usepackage{amsmath}

\begin{document}

\title{One colorful resolution to the neutrino mass generation, three lepton flavor universality anomalies, and the  Cabibbo angle anomaly}
\date{\today}

\author{We-Fu Chang}
\email{wfchang@phys.nthu.edu.tw}
\affiliation{Department of Physics, National Tsing Hua University, Hsinchu, Taiwan 30013, R.O.C. }

\begin{abstract}

We propose a simple model to simultaneously explain four observed flavor anomalies  while  generating the neutrino mass at the one-loop level. Specifically, we address the measured anomalous magnetic dipole moments of the muon, $\Delta a_\mu$ , and electron, $\Delta a_e$,
the observed anomaly of  $b\ra s l^+ l^-$ in the $B$-meson decays, and the Cabibbo-angle anomaly.
The model consists of four colorful new degrees of freedom: three scalar leptoquarks with the Standard Model quantum numbers $(3,3,-1/3),(3,2,1/6)$, and $(3,1,2/3)$, and one pair of down-quark-like vector fermion in $(3,1,-1/3)$. The baryon number is assumed to be conserved for simplicity.

Phenomenologically viable solutions with the minimal number of real parameters can be found to accommodate all the above-mentioned anomalies and produce the approximate, close to $1\sigma$, neutrino oscillation pattern at the same time.
From general consideration, the model robustly predicts: (1) neutrino mass is of the normal hierarchy type, and (2) ${\cal M}^\nu_{ee}\lesssim 3\times 10^{-4}\mev$.

The possible UV origin to explain the flavor pattern of the found viable parameter space is briefly discussed.
The parameter space can be well reproduced within a simple split fermion toy model.

\end{abstract}
\maketitle

\section{Introduction}
Despite of the amazing success of the Standard Model (SM) of particle physics,
new physics (NP) beyond the SM   is ineluctable because  compelling evidence showing that at least two active neutrinos are massive\cite{Zyla:2020zbs}.
In addition to the nonzero neutrino masses,  several recent experimental measurements prominently deviated from the SM predictions are also suggestive to NP.
In particular,
\bi

\item
 The measured anomalous magnetic moment of muon $(g-2)_\mu$\cite{Zyla:2020zbs, Bennett:2006fi} differs from the most recent SM prediction\cite{Blum:2018mom, Davier:2019can} by an amount of $\sim 3.7\sigma$:
 \beq
 \tri a_\mu =a_\mu^{exp} -a_\mu^{SM} \simeq (27.4\pm7.3)\times 10^{-10}\,,
 \eeq
where the uncertainty is the quadratic combination of the experimental and theoretical ones.
The most recent  $0.46 ppm$ measurement conducted at Fermilab\cite{Abi:2021gix} gives
\beq
a_\mu^{exp}= 116 592 040(54)\times 10^{-11}\,,
\eeq
which agrees with the previous measurements.
And the new experimental average of $a_\mu^{exp}= 116 592 061(41) \times 10^{-11}$ 
drives the deviation to $4.2\sigma$ with
 \beq
 \tri a_\mu \simeq (25.1\pm 5.9)\times 10^{-10}\,.
 \eeq

Another new measurement at the J-PARC\cite{Saito:2012zz}
is also expected to improve the experimental uncertainty in the near future.

\item
With the determination of the fine-structure constant by using the cesium atom\cite{Parker:2018vye}, the measured electron $(g-2)_e$  \cite{Hanneke:2010au} shows a $\sim 2.4 \sigma$ discrepancy from the SM prediction\cite{Aoyama:2017uqe}:
\beq
\tri a_e^{Cs} =a_e^{exp} -a_e^{SM} \simeq (-8.7\pm 3.6)\times 10^{-13}\,.
\label{eq:deltaAe_minus}
\eeq
Note that $\tri a_e^{Cs}$ and $\tri a_\mu$ have opposite signs. New models\cite{Liu:2018xkx, Crivellin:2018qmi, Endo:2019bcj, Bauer:2019gfk, Badziak:2019gaf, Abdullah:2019ofw, Hiller:2019mou, Cornella:2019uxs, Haba:2020gkr, Bigaran:2020jil, Jana:2020pxx, Calibbi:2020emz, Yang:2020bmh, Chen:2020jvl, Hati:2020fzp, Dutta:2020scq, Chen:2020tfr,Chun:2020uzw,Li:2020dbg,Dorsner:2020aaz,Keung:2021rps} have been constructed to accommodate both $\tri a_e^{Cs}$ and $\tri a_\mu$.
 Moreover, \cite{Arbelaez:2020rbq,Jana:2020joi,Escribano:2021css} also attempt to incorporate the neutrino mass generation with the observed $\tri a_{e,\mu}$.

  However, the  most recent $\alpha$ determination by using the rubidium atom\cite{Morel:2020dww} yields
a different outcome
\beq
\tri a_e^{Rb}  \simeq (+4.8\pm 3.0)\times 10^{-13}\,,
\label{eq:deltaAe_plus}
\eeq
whose sign is different from the one of $\tri a_e^{Cs}$.
These two highly precise values of $\alpha$ differ by a tantalizing $\sim 5\sigma$.
More independent investigations or measurements are required to resolve this tension.
At this moment, we should consider both cases in this work. 

\item
From the global fits\cite{Capdevila:2017bsm, Altmannshofer:2017yso, DAmico:2017mtc, Hiller:2017bzc, Ciuchini:2017mik, Geng:2017svp, Hurth:2017hxg, Alok:2017sui, Alguero:2019ptt, Aebischer:2019mlg, Ciuchini:2019usw, Datta:2019zca} to various $b\ra s l^+ l^-$ data\cite{Aaij:2019wad, Aaij:2017vbb, Abdesselam:2019wac, Abdesselam:2019lab, Aad:2014fwa,Aaij:2015esa, Wei:2009zv, Aaltonen:2011ja, Khachatryan:2015isa, Abdesselam:2016llu, Aaij:2015oid, Wehle:2016yoi, Sirunyan:2017dhj, Aaboud:2018krd, Lees:2015ymt}, the discrepancy is more than $5\sigma$ from the SM predictions. The new result\cite{Aaij:2021vac} further strengthens the lepton flavor universality violation\cite{Altmannshofer:2021qrr, Geng:2021nhg}.
 This anomaly alone convincingly indicates NP, and stimulates many investigations to address this deviation.
For example,  a new gauge sector was introduced in \cite{Capdevila:2020rrl, Altmannshofer:2019xda, Gauld:2013qja},
 leptoquark has been employed in \cite{Bauer:2015knc,ColuccioLeskow:2016dox, Angelescu:2018tyl, Crivellin:2019dwb, Fuentes-Martin:2020bnh, Saad:2020ucl, Balaji:2019kwe,Babu:2020hun, Dev:2020qet}, assisted by the 1-loop contributions from exotic particles in \cite{Gripaios:2015gra, Arnan:2016cpy, Li:2018rax, Hu:2019ahp, Arnan:2019uhr},
 and more references can be found in \cite{Capdevila:2017bsm, Li:2018lxi, Bifani:2018zmi}.

\item
The so-called Cabibbo angle anomaly refers to the unexpected shortfall in the first row Cabibbo-Kobayashi-Maskawa(CKM) unitarity\cite{Zyla:2020zbs},
\beq
|V_{ud}|^2+|V_{us}|^2+|V_{ub}|^2 = 0.9985\pm 0.0005\,.
\label{eq:PDGCKM}
\eeq
The above value is smaller than one, and the inconsistence is now at the level of $\simeq 2-4 \sigma$ level\cite{Grossman:2019bzp, Seng:2020wjq}.
There are tensions among different determinations of the Cabibbo angle from tau decays\cite{Amhis:2019ckw}, kaon decays\cite{Aoki:2019cca}, and super-allowed $\beta$ decay (by using CKM unitarity and the theoretical input\cite{Seng:2018yzq, Czarnecki:2019mwq}).
The potential NP involving vector quarks or the origins of lepton flavor universality violation  have been discussed in \cite{Belfatto:2019swo, Cheung:2020vqm, Crivellin:2020oup, Coutinho:2019aiy, Crivellin:2020lzu, Crivellin:2020ebi, Kirk:2020wdk, Alok:2020jod,Belfatto:2021jhf}.

\ei

Whether these $2-5\, \sigma$ anomalies will persist is not predictable;  the future improvement on the theoretical predictions\footnote{  For example, the recent lattice study\cite{Borsanyi:2020mff}  suggests the SM prediction used to derive $\tri a_\mu$ needs to be revised, and there is no
significant tension with the recent FNAL experimental determination. } and the  experimental measurements will be the ultimate  arbiters.
However, at this moment, it is interesting to speculate whether all the above-mentioned anomalies and the neutrino mass can be explained simultaneously.
In this paper, we point out one of such resolutions.
With the addition of three scalar leptoquarks,   $T(3,3,-1/3)$, $D(3,2,1/6)$, and $S(3,1,2/3)$, and a pair of vector fermion, $b'_{L,R}(3,1,-1/3)$, the plethoric new parameters ( mostly the Yukawa couplings) allow one to reconcile all data contemporarily.
Parts of the particle content of this model had been employed in the past  to accommodate some of the anomalies.
However,  to our best knowledge, this model as whole is new to comprehensively interpret all the observed deviations from the SM.

The paper is laid out as follows: in Sec.\ref{sec:model} we spell out the model and the relevant ingredients. Following that in Sec.\ref{sec:howitworks} we explain how each anomaly and the neutrino mass generation works in our model. Next we discuss the various phenomenological constraint and provide some model parameter samples in Sec.\ref{sec:pheno} . In Sec.\ref{sec:discussion} we discuss some phenomenological consequences, and the UV origin of the flavor pattern of the parameter space.  Then comes our conclusion in Sec.\ref{sec:conclusion}. Some details of our notation and the low energy effective Hamiltonian are collected in the Appendix.

\section{The Model}
\label{sec:model}
 In this model, three scalar leptoquarks,    $T(3,3,-1/3)$, $D(3,2,1/6)$, and $S(3,1,2/3)$\footnote{ In the literature\cite{Buchmuller:1986zs}, the corresponding notations for $D(3,2,1/6)$ and $S(3,1,2/3)$
  are  $\wt{R}_2$ and $(\bar{S}_1)^*$, respectively. The one closely related to our    $T(3,3,-1/3)$ is $S_3^*$ }. , and a pair of down-quark-like vector fermion, $b'_{L,R}(3,1,-1/3)$, are augmented on top of the SM. Our notation for the SM particle content and the exotics are listed in Tab.\ref{table:SMparticle} and Tab.\ref{table:newparticle}, respectively.

\begin{table}
\begin{center}
\begin{tabular}{|c||ccccc|c|}\hline
          & \multicolumn{5}{c|}{ SM Fermion}  &  \multicolumn{1}{c|}{ SM Scalar} \\\hline
 Fields   &  $Q_L= \begin{pmatrix} u_L \\ d_L\end{pmatrix}
 $  &$u_R$  &$d_R$   & $L_L=\begin{pmatrix} \nu_L \\ e_L\end{pmatrix}$
   &$e_R$   & $ H=\begin{pmatrix} H^+ \\ H^0\end{pmatrix}$
      \\\hline
$SU(3)_c$ &  $3$  &$3$  &$3$ & $1$ & $1$  & $1$  \\
$SU(2)_L$ &  $2$  & $1$ & $1$ & $2$ & $1$  & $2$ \\
$U(1)_Y$  & $\frac{1}{6}$  & $\tth$ &$-\ot$ & $-\frac{1}{2}$   &$-1$  & $\frac{1}{2}$ \\\hline
\end{tabular}
\caption{The SM field content and quantum number assignment under the SM gauge symmetries $SU(3)_c\otimes SU(2)_L \otimes U(1)_Y $, where $L,R$ stand for the chirality of the fermion. For simplicity, all the generation indices associated with the fermions are suppressed. }
\label{table:SMparticle}
\end{center}
\end{table}


\begin{table}
\begin{center}
\begin{tabular}{|c||c|ccc|}\hline
       &   \multicolumn{1}{c|}{ New Fermion}  &  \multicolumn{3}{c|}{ New Scalar}\\\hline
 Fields & $b'_{L,R}$
  & $T=\begin{pmatrix} T^\tth \\ T^{-\ot} \\ T^{-\ft}\end{pmatrix}$
   & $D=\begin{pmatrix} D^\tth \\ D^{-\ot}\end{pmatrix}$ & $S^{\tth}$
    \\\hline
$SU(3)_c$ & $3$ & $3$ & $3$ & $3$ \\
$SU(2)_L$ & $1$ & $3$ & $2$  & $1$ \\
$U(1)_Y$  & $-\ot$ & $-\ot$ & $\frac{1}{6}$ & $\tth$\\
lepton number & $0$  & $1$ & $-1$ & $-1$ \\
baryon number & $\ot$ & $\ot$ & $\ot$ & $\ot$\\
\hline
\end{tabular}
\caption{New field content and quantum number assignment under the SM gauge symmetries $SU(3)_c\otimes SU(2)_L \otimes U(1)_Y $,  and the global lepton/baryon numbers. }
\label{table:newparticle}
\end{center}
\end{table}

Like most models beyond the SM, the complete Lagrangian is lengthy, and not illuminating. In this work, we only focus on the new gauge invariant interactions relevant to addressing the flavor anomalies.
For simplicity, we also assume the model Lagrangian respects the global baryon number symmetry, and  both $T$ and $S$ carry one third of baryon-number to avoid their possible di-quark couplings.
Moreover, we do not consider the possible CP violating signals in this model.

For the scalar couplings, we have\footnote{To simplify the notation, we use ``$\{,\}$'' and ``$[,]$'' to denote the $SU(2)_L$ triplet and singlet constructed from two given $SU(2)_L$ doublets, respectively. Also, ``$\odot$'' means forming an $SU(2)_L$ singlet from two given triplets; see Appendix for the details.}
\beqa
{\cal L} &\supset& \mu_3 \left\{H ,\tilde{D} \right\}\odot T+ \mu_1 \left[ H ,D \right] S^{-\tth} +H.c.\\
&=&  \mu_3\left[ H^+ D^{\ot}T^{-\ft} -\frac{1}{\sqrt{2}}\left(H^0D^{\ot}-H^+D^{-\tth}\right)T^{-\ot}-H^0D^{-\tth}T^{\tth}\right]\nonr\\
&& - \mu_1  \frac{1}{\sqrt{2}} \left(H^0 D^{\tth}- H^+D^{-\ot}\right)S^{-\tth}+H.c.
\eeqa
The couplings $\mu_1$ and $\mu_3$ are unknown dimensionful parameters.
Note that the $\mu_3$ coupling softly breaks the global lepton number by two units, which is crucial for the neutrino mass generation.
On the other hand,  the lepton-number conserving $\mu_1$ triple scalar interaction is essential for explaining $\tri a_e$ and $\tri a_\mu$ ( to be discussed in the following sections).
As it will be clear later, to fit all the data, $\mu_3$ turns out to be very small, $\sim {\cal O}(0.5\mkev)$, and $\mu_{1}\sim {\cal O}(\mtev)$.

After electroweak spontaneous symmetry breaking (SSB), $\langle H^0 \rangle = v_0/\sqrt{2}$ and the Goldstone $H^\pm$ are  eaten by the $W^\pm$ bosons. Below the electroweak scale, it becomes:
\beq
-\frac{\mu_3 v_0}{2} D^{\ot} T^{-\ot} -\frac{\mu_3 v_0}{\sqrt{2}} D^{-\tth} T^{\tth} - \frac{\mu_1 v_0}{2} D^{\tth} S^{-\tth}  +H.c.
\eeq
Comparing to their tree-level masses, $\widetilde{M}_{T,D,S} \simeq M_{LQ}$\footnote{ Our notation is ${\cal L} \supset -\tilde{M}^2_T T^\dag T -\tilde{M}^2_D D^\dag D -\tilde{M}^2_S S^\dag S $. In order to preserve the $SU(3)_c$ symmetry, $T,D,S$ cannot develop nonzero vacuum expectation values. }, we expect the mixings are  small and suppressed by the factor of ${\cal O}( \mu_{LQ} v_0/M_{LQ}^2)$. However, these mixings break the isospin multiplet mass degeneracy of $T$ and $D$. After the mass diagonalization, we have two charge-$\ot$, three charge-$\tth$, and one charge-$\ft$ physical scalar leptoquarks.

In addition to the SM Yukawa interactions in the form of $\bar{Q}d_R H$, $\bar{Q}u_R \tilde{H}$, and $\bar{L}e_R H$, this model has the following  new Yukawa couplings ( in the interaction basis):
\beqa
{\cal L} &\supset& -\widetilde{\lambda}_T T^\dag\cdot \left\{\bar{L}^c , Q \right\} - \widetilde{\lambda}_D \bar{d}_R\left[ L ,D \right] - \widetilde{\lambda}'_D \bar{b'}_R\left[ L ,D \right]
  - \widetilde{\lambda}_S  \bar{e}_R b'_L S^{-\tth} -\widetilde{Y}_d' \bar{Q}b'_R H +H.c.\\
&=&  -\widetilde{\lambda}_T \left[ \bar{\nu}^c u_L T^{-\tth}+ \left(\bar{\nu}^c d_L +\bar{e}^c u_L\right)\frac{ T^{\ot} }{\sqrt{2}}+ \bar{e}^c d_L T^{\ft}\right] - \widetilde{Y}_d' (\bar{u}_L H^++\bar{d}_L H^0) b'_R \nonr\\
&&- \widetilde{\lambda}_D \frac{ \bar{d}_R}{\sqrt{2}} \left(\nu_L D^{-\ot}-e_L D^{\tth}\right)
- \widetilde{\lambda}'_D  \frac{ \bar{b'}_R}{\sqrt{2}} \left(\nu_L D^{-\ot}-e_L D^{\tth}\right)
- \widetilde{\lambda}_S  \bar{e}_R b'_L S^{-\tth} +H.c.\,,
\label{eq:LQYukawa_coupling}
\eeqa
where all the generation indices are suppressed to keep the notation simple and it should be understood that all the Yukawa couplings are matrices.
Moreover, the model allows two kinds of tree-level Dirac mass term:
\beq
{\cal L} \supset M_1 \bar{b}'_R b'_L  + M_2 \bar{d}_R b'_L + H.c.
\eeq

With the introduction of $b'$, the mass matrix  for down-quark-like fermions after the electroweak SSB becomes:
\beq
{\cal L} \supset - (\bar{d_R},\bar{b'_R}) {\cal M}^d \left(\begin{array}{c} d_L\\ b'_L\end{array} \right)+H.c.\,,\,\,
{\cal M}^d =  \left(
              \begin{array}{cc}
               \frac{\widetilde{Y}_d v_0}{\sqrt{2}}  & M_2 \\
                \frac{\widetilde{Y}'_d v_0}{\sqrt{2}}   &  M_1 \\
              \end{array}
            \right)\,,
\eeq
where $\widetilde{Y}_d$ is the SM down-quark three-by-three Yukawa coupling matrix in the interaction basis. Note that ${\cal M}^d$ is now a four-by-four matrix.
This matrix can be diagonalized by the bi-unitary transformation, ${\cal M}^d_{diag}=U^d_R {\cal M}^d (U^d_L)^\dag=\mbox{diag}(m_d,m_s,m_b,M_{b'})$, and
\beqa
&&U^d_R  {\cal M}^d ({\cal M}^d)^\dag (U^d_R)^\dag =
U^d_L  ({\cal M}^d)^\dag{\cal M}^d  (U^d_L )^\dag=
({\cal M}_{diag}^d)^2\,,\\
&& 
(   d_1,d_2,d_3, d_4)_{L/R}=  ( d, s,b,b')_{L/R} (U^d_{L/R})^*\,,
\eeqa
where $(   d_1, d_2, d_3, d_4)$ and $(d,s,b,b')$ stand for the interaction and mass eigenstates, respectively.
The new notation, $d_4$, is designated for the interaction basis of the singlet $b'_{L,R}$, and $b'$ is recycled to represent the heaviest mass eigenstate of down-type quark. One will see that the mass and interaction eigenstates of $b'$ are very close to each other from the later phenomenology study.

Similarly, the SM up-type quarks and the charged leptons can be brought to their mass eigenstates by $U^u_{L/R}$ and $U^e_{L/R}$, respectively\footnote{Note that  the SM neutrinos are still  massless at the tree-level.}.
Since $\widetilde{\lambda}$'s are unknown in the first place, one can focus on the couplings in the charged fermion  mass basis,
denoted as $\lambda^{T,D,S}$, which are more phenomenologically useful.
However, note that the mass diagonalization  matrices are in general different for the left-handed (LH) up- and LH down-quark sectors.
If we pick the flavor indices of $\lambda^T$ to label the charged lepton and down quark mass states,
the up-type quark in the triplet leptoquark coupling will receive an extra factor to compensate the difference between $U^d_L$ and $U^u_L$.
Explicitly,
\beqa
{\cal L} \supset  &-& \sum_{l=e,\mu,\tau}\sum_{p=d,s,b,b'}(\lambda^T)_{l p} \sum_{r=u,c,t}\tilde{A}^\dag_{pr} \left[ \bar{\nu}_l^c T^{-\tth}+ \bar{e}_l^c \frac{ T^{\ot} }{\sqrt{2}}\right]u_{L,r}\nonr\\
&-& \sum_{l=e,\mu,\tau}\sum_{p=d,s,b,b'}(\lambda^T)_{l p} \left[  \bar{e}_l^c  T^{\ft}+\bar{\nu}_l^c \frac{ T^{\ot} }{\sqrt{2}} \right]d_{L,p} + H.c.
\eeqa
with the four-by-three matrix
\beq
\tilde{A}^\dag_{pr}= \sum_{j=1}^4 (U^d_L)_{p j} (U^u_L)^\dag_{j r}\,.
\eeq
As will be discussed in below,  $\tilde{A}$ is the extended CKM rotation matrix, $\tilde{V}$, and $\tilde{A}\ra (V_{CKM})$ if $b'_L$ decouples. Now, all $\lambda^T$, $\lambda^D$, and $\lambda^S$  are three-by-four matrices.

In the interaction basis,  only the LH doublets participate in the charged-current(CC) interaction.
Thus, the SM $W^\pm$ interaction for the quark sector is
\beq
{\cal L} \supset \frac{g_2}{\sqrt{2}} \sum_{i=1}^3 \left(\bar{u}_i \gamma^\alpha \PL d_i \right)W^+_\alpha +H.c.
\eeq
However,  the singlet $b'_{L}$ mixes with other LH down-type-quarks and change the SM CC interaction.
In the mass basis, it becomes
\beq
{\cal L} \supset \frac{g_2}{\sqrt{2}} (\bar{u},\bar{c},\bar{t})\gamma^\alpha \PL
\widetilde{V}\left( \begin{array}{c} d \\s\\b\\b' \end{array} \right)  W^+_\alpha +H.c.
\eeq
where
\beq
\widetilde{V}=\left(
                \begin{array}{cccc}
                  \wt{V}_{ud} &   \wt{V}_{us} & \wt{V}_{ub} & \wt{V}_{u b'} \\
                  \wt{V}_{cd} &   \wt{V}_{cs} & \wt{V}_{cb} & \wt{V}_{c b'} \\
                  \wt{V}_{td} &   \wt{V}_{ts} & \wt{V}_{tb} & \wt{V}_{t b'} \\
                \end{array}
              \right)\,,\; \mbox{ and }\,\,
\widetilde{V}_{pq}\equiv \sum_{i=1}^3  (U^u_L)_{pi}(U^d_L)^\dag_{iq} \,.
\eeq
Therefore, the SM three-by-three unitary CKM matrix changes into a three-by-four matrix in our model.
When the $b'_L$ decouples, the coupling matrix $\widetilde{V}$ reduces to the SM $V_{CKM}$.

Instead of dealing with a three-by-four matrix, it is helpful to consider an auxiliary unitary four-by-four matrix
\beq
\wt{V}_4 \equiv \left( \begin{array}{cc} U^u_L & 0  \\ 0 & 1  \end{array}   \right)
\cdot (U^d_L)^\dag =
\left(
                \begin{array}{cccc}
                  \wt{V}_{ud} &   \wt{V}_{us} & \wt{V}_{ub} & \wt{V}_{u b'} \\
                  \wt{V}_{cd} &   \wt{V}_{cs} & \wt{V}_{cb} & \wt{V}_{c b'} \\
                  \wt{V}_{td} &   \wt{V}_{ts} & \wt{V}_{tb} & \wt{V}_{t b'} \\
                  (U^d_L)_{d4}^* &  (U^d_L)_{s4}^* &(U^d_L)_{b4}^* &(U^d_L)_{b'4}^*\\
                \end{array}
              \right)\,.
\label{eq:V4}
\eeq

To quantify the NP effect, one can parameterize the four-by-four unitary matrix $U^d_L$  by a unitary  three-by-three sub-matrix,
$U^d_{L3}$, and three rotations as:
\beq
(U^d_L)^\dag =\left(
         \begin{array}{cc}
           (U^d_{L3})^\dag &0 \\
           0& 1 \\
         \end{array}
       \right)\cdot R_4\,,\;\;\mbox{where}\;\;
 R_4 =
       \left(
       \begin{array}{cccc}
         c_1 & 0 & 0 & s_1\\
          -s_1 s_2 & c_2 & 0 & c_1 s_2 \\
           -s_1 c_2 s_3 & -s_2 s_3 & c_3 & c_1 c_2 s_3\\
            -s_1 c_2 c_3 & -s_2 c_3 & -s_3 & c_1 c_2 c_3\\
       \end{array}
       \right)\,,
       \label{eq:R4}
\eeq
where $s_i(c_i)$ stands for $\sin \theta_i (\cos \theta_i) $, and $\theta_i$ is the mixing angle between $d_{iL}$ and $b'_L$.
In this work, we assume there is no new CP violation phase beyond the SM CKM phase for simplicity.
Now, Eq.(\ref{eq:V4}) can be parameterized as
\beq
\wt{V}_4 = \left(
         \begin{array}{cc}
           V_{CKM} &0 \\
           0& 1 \\
         \end{array}
       \right)\cdot R_4\,.
\eeq

Again, we use $d,s,b,b'$ to denote the mass eigenstates with $m_d\simeq  4.7\mmev$, $m_s\simeq 96\mmev$, $m_b\simeq 4.18\mgev$, and $M_{b'}$, the mass of $b'$, unknown.
The null result of direct searching for the singlet $b'$ at ATLAS sets a limit that $M_{b'} > 1.22 TeV$ \cite{Aaboud:2018pii} (by assuming only three 2-body decays: $b'\ra Wt,bZ,bH$ ), and similar limits have obtained by CMS\cite{Sirunyan:2018qau,Sirunyan:2019sza}. We take $M_{b'}=1.5\mtev$ as a reference in this paper.
Moreover, all the direct searches for the scalar leptoquarks at the colliders strongly depend on the assumption of their decay modes.
Depending on the working assumptions, the exclusion  limits range from $\sim 0.5\mtev$ to $\sim 1.6\mtev$\cite{Zyla:2020zbs}.
Instead of making simple assumptions, it will be more motivated to associate the leptoquark branching ratios to neutrino mass generation\cite{Chang:2016zll} or the $b$-anomalies \cite{Diaz:2017lit,Schmaltz:2018nls}.
In this paper, we take $m_{T,D,S}\sim M_{LQ} = 1\mtev$  as the reference point. And the constraint we obtained can be easily scaled for a different $M_{LQ}$ or $M_{b'}$.

Since all the new color degrees of freedom are heavier than $\gtrsim TeV$, it is straightforward to integrate them out and perform the Fierz transformation to get the low energy effective Hamiltonian, see Appendix \ref{sec:H_eff}.

\section{Explaining the anomalies}
\label{sec:howitworks}

\begin{table}
\begin{center}
\begin{tabular}{|c|cccc|c|}
  \hline
  Anomaly $\backslash$ Field & $T(3,3,  -\ot)$ & $D(3,2,\frac{1}{6})$ & $S(3,1,\tth)$  & $b'(3,1,-\ot)$  & Remark\\
  \hline
  Neutrino mass & $\checkmark$ & $\checkmark$ & - &  $\checkmark$ & 1-loop\\
  Cabibbo angle anomaly & - & - & - & $\checkmark$ & extended CKM \\
  $\tri a_e$ & $\times$ &$\checkmark$ &$\checkmark$ & $\checkmark$ & 1-loop \\
    $\tri a_\mu$ & $\hchecked$ &$\checkmark$ &$\checkmark$ & $\checkmark$ & 1-loop \\
    $b\ra sl^+l^-$ & $\checkmark$ & - & -  & $\checkmark$ & box diagram \\
 \hline
\end{tabular}
\caption{The anomalies and the fields to accommodate them in this model. The meaning of the legends: $\checkmark$: essential, $\hchecked$: helpful but not important or required, $\times$: negative effect, $-$: irrelevant.  }
\end{center}
\end{table}

\subsection{Neutrino mass}
\begin{figure}[htb]
\centering
\includegraphics[width=0.6\textwidth]{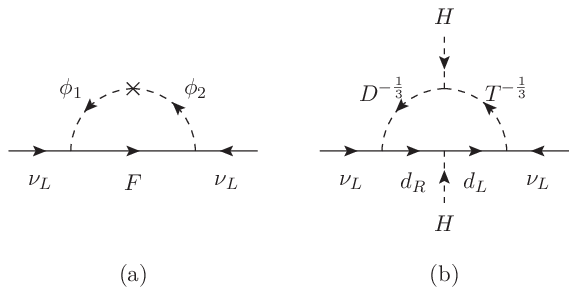}
\caption{ The Feynman diagrams for
 the neutrino mass generation.   (a) General case, where $\phi_{1,2}$ are in their interaction basis, and $F$ is in its mass eigenstate, and (b) for this model, where the fields are in the interaction basis. Here all the flavor indices are omitted.}
\label{fig:nu_mass}
\end{figure}
Instead of using the bi-lepton $SU(2)$ singlet and a charged scalar without lepton number as first proposed in Ref.\cite{Zee:1980ai}, we employ two leptoquarks which carry different lepton numbers to break the lepton number and generate the neutrino mass radiactively.
We start with a general discussion on the 1-loop neutrino mass generation.
If there are two scalars $\phi_{1,2}$ which interact with fermion $F_k$  and the neutrino via a general Yukawa coupling parameterized as
\beq
{\cal L}\supset  \lambda_{ij} \bar{F}_j \nu_{L i}\phi_1  + \kappa_{ij} \bar{F}_j \nu^c_i \phi_2 + H.c.\,,
\eeq
where the fermion $F$ can carry arbitrary lepton number and baryon number $(L_F, B_F)$.
If the two scalars do not mix, $\phi_1/ \phi_2$ can be assigned with the lepton-number and baryon number $ (L_F-1, L_B)/(L_F+1, L_B)$ and the Lagrangian enjoys both the global lepton-number $U(1)_L$ and the global baryon-number $U(1)_B$ symmetries\footnote{ For the discussion of the pure leptonic gauge symmetry $U(1)_L$, see for example \cite{Schwaller:2013hqa, Chao:2010mp, Chang:2018nid, Chang:2018wsw, Chang:2018vdd}.}.
Without losing the generality, $F_k$ is assumed to be in its mass eigenstate with a mass $m_k$.
If $\phi_{1,2}$ can mix with each other, the lepton number is broken by two units, and the Weinberg operator\cite{Weinberg:1979sa} can be generated radiatively.
Let's denote $\phi_{h(l)}$ as the heavier(lighter) mass state with mass $m_h(m_l)$, and parameterize their mixing as
$\phi_1 = c_\alpha \phi_l +s_\alpha \phi_h$ and   $\phi_2 = -s_\alpha \phi_l +c_\alpha \phi_h$, where $s_\alpha(c_\alpha)$ is the shorthand notation for $\sin\alpha(\cos\alpha)$ and $\alpha$ is the mixing angle.
The resultant neutrino mass from Fig.\ref{fig:nu_mass}(a) can be calculated as
\beq
M^\nu_{ij}= \sum_k {N^F_c m_k \over 16\pi^2} s_\alpha c_\alpha (\kappa_{ik}\lambda_{jk}+\kappa_{jk}\lambda_{ik})
\left[ {m_h^2 \over m_h^2 -m_k^2}\ln \frac{m_h^2}{m_k^2} -  {m_l^2 \over m_l^2 -m_k^2}\ln \frac{m_l^2}{m_k^2} \right]\,,
\eeq
which is exact and free of divergence. Note that for the diagonal element, the combination in the bracket should be replaced by
$2 \mbox{Re}(\kappa_{ik}\lambda_{ik})$.
 When the mixing is small, this result can also be approximately calculated in the interaction  basis of $\phi_1$ and $\phi_2$.

In our model, the mass eigenstate $F$ can be the SM down-type quark or the exotic $b'$, and $D^{\ot}/T^\ot$ plays the role of $\phi_1/\phi_2$,  as depicted in Fig.\ref{fig:nu_mass}(b). Assume the $D\mhyphen T$ mixing is small, then
\beq
M^\nu_{ij} \simeq  \sum_{k=d,s,b,b'} {3 m_k \over 32\pi^2} (\lambda^T_{ik}\lambda^D_{jk}+\lambda^T_{jk}\lambda^D_{ik})
{\mu_3 v_0 \over M_D^2-M_T^2}\ln \frac{M_T^2}{M_D^2}
\label{eq:nu_mass}
\eeq
for $i\neq j$, and $2 \mbox{Re}(\lambda^T_{ik}\lambda^D_{ik})$ should be used in the bracket for the  diagonal elements.
To have sub-eV neutrino masses, we need roughly
\beq
\mu_3 m_b\lambda^D \lambda^T\,,\, \mu_3  M_{b'}\lambda^D \lambda^T  \simeq {\cal O}(10^{-5})\times\left( \frac{M_{LQ}}{\mbox{TeV}}\right)^2 (\mgev)^2
\eeq
if $b^{(')}$-quark contribution dominates. More comprehensive numerical consideration with other phenomenology will be given in section \ref{sec:pheno}.

\subsection{$(g-2)$ of charged leptons }
\label{sec:g-2}
\begin{figure}[htb]
\centering
\includegraphics[width=0.57\textwidth]{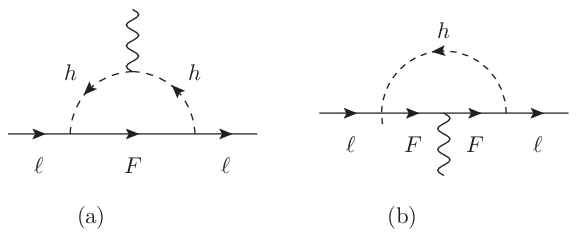}
\caption{ The Feynman diagrams in the mass basis for
 the anomalous magnetic dipole moment of charged lepton in general cases.  }
\label{fig:g-2G}
\end{figure}

We also start with a general discussion on  the 1-loop contribution to $(g-2)_\ell$ by adding a fermion, $F$, and a charged scalar, $h$.
The $F\mhyphen \ell \mhyphen h$  Yukawa interaction can be parameterized as
\beq
{\cal L}\supset \bar{F} (y^l_R \hat{R} + y^l_L \hat{L}) \ell h + H.c.
\label{eq:gen_eF_Yukawa}
\eeq
where both $F$ and $\ell$ are in their mass eigenstates.
Here, we  have suppressed the flavor indices but it should be understood that both $y_R$ and $y_L$ are in general flavor dependent.
Then, the resulting 1-loop anomalous magnetic moment depicted in Fig.\ref{fig:g-2G}(a,b) can be calculated as
\beqa
\tri a^h_l &=& \frac{- N^F_c(1+Q_F) m_l^2}{8\pi^2}\int^1_0 dx\, x(1-x){ x \frac{ |y^l_L|^2+|y^l_R|^2 }{2} + \frac{m_F}{m_l}\Re[(y^l_R)^* y^l_L]
\over x^2 m_l^2 + x ( m_h^2-m_l^2) +(1-x) m_F^2}\,,\\
\tri a^F_l &=& \frac{- N^F_c Q_F m_l^2}{8\pi^2}\int^1_0 dx\, x^2 { (1-x) \frac{ |y^l_L|^2+|y^l_R|^2 }{2}  + \frac{m_F}{m_l}\Re[(y^l_R)^* y^l_L]
\over x^2 m_l^2+ x ( m_F^2-m_l^2)+(1-x)m_h^2 }\,,
\eeqa
where $Q_F$ is the electric charge of $F$, and $\tri a_l^F (\tri a_l^h)$ is the contribution with the external photon attached to the fermion (scalar) inside the loop.
We keep $\tri a_l^F (\tri a_l^h)$ in the integral form since the analytic expression of resulting integration is not illuminating at all.
The physics is also clear from the above expression that one needs $m_F\gg m_l$ also both $y^l_R$ and $y^l_L$ nonzero to make $\tri a_e$ and $\tri a_\mu$ of opposite sign.
For $m_F\gg m_l$, we have
\beqa
\tri a_l^h &\simeq& \Re[(y^l_R)^* y^l_L]\left( \frac{m_l}{ m_F}\right) \frac{- N^F_c(1+Q_F)}{8\pi^2}\int^1_0 dx\,{   x(1-x)
\over  x   +(1-x)\frac{m_h^2}{m_F^2} }\,,\\
\tri a_l^F &\simeq& \Re[(y^l_R)^* y_L] \left( \frac{m_l}{ m_F}\right) \frac{-N^F_c Q_F }{8\pi^2}\int^1_0 dx\,  { x^2
\over x + (1- x) \frac{m_h^2}{m_F^2}   }\,.
\eeqa
Namely,
\beq
\tri a_l = \tri a_l^F+ \tri a_l^h \simeq  -\frac{N^F_c \Re[(y^l_R)^* y_L] }{8\pi^2}  \left( \frac{m_l}{m_F} \right) {\cal J}_{Q_F}\left(\frac{m_h^2}{m_F^2}\right)\,,
\eeq
where
\beqa
{\cal J}_Q(\alpha )&=&\int^1_0 dx\,{   x(1-x) +x\, Q \over  x   +(1-x)\alpha  }\nonr\\
&=&{ 2 Q(1-\alpha)(1-\alpha+\alpha\ln\alpha)+ (1-\alpha^2+2\alpha\ln\alpha) \over 2(1-\alpha)^3 }\,.
\label{eq:a2J}
\eeqa
From Eq.(\ref{eq:a2J}), it is clear that ${\cal J}_Q(0)=(1+2Q)/2$, ${\cal J}_Q(1)=(1+3Q)/6$,
and ${\cal J}_Q(\alpha )\ra ( Q\ln\alpha-1/2)/\alpha$ for $\alpha \gg 1$.

Similar calculation leads to a $l\ra l' \gamma$ transition amplitude:
\beq
i{\cal M} \simeq i e \frac{m_l N_c^F}{16\pi^2 m_F}  {\cal J}_{Q_F}(\beta_h )\times
\overline{u_{l'}}(p-k)\left[\frac{i \sigma^{\alpha\beta} k_\beta \epsilon^*_\alpha}{m_l}\left(A^{l l'}_M+A^{l l'}_E\gamma^5\right)\right ] u_l(p)\,,
\eeq
where $\beta_h=(m_h/m_F)^2 $, $\epsilon$ is the polarization of the photon, and
\beq
A^{l l'}_M = \frac{1}{2}\left[(y^{l'}_R)^* y^l_L + (y^{l'}_L)^* y^l_R \right]\,,\;
A^{l l'}_E = \frac{1}{2}\left[(y^{l'}_R)^* y^l_L - (y^{l'}_L)^* y^l_R \right]\,.
\eeq
For $l=\mu$ and $l'=e$, the above  transition amplitude results in the $\mu\ra e \gamma$ branching ratio\cite{Chang:2005ag}
\beq
Br(\mu\ra e \gamma)= {3 \alpha  (N^F_c)^2 \over 8 \pi G_F^2 m_F^2 m_\mu^2 }
\left(  \left|A^{\mu e}_M\right|^2 +\left|A^{\mu e}_E\right|^2 \right)\,,
\label{eq:brmeg}
\eeq
and it must complies with the current experimental limit, $Br(\mu\ra e \gamma)< 4.2\times 10^{-13}$\cite{TheMEG:2016wtm}, or $|A^{\mu e}_{E,M}| \lesssim {\cal O}(10^{-8})$.
Moreover, if the dipole transition is dominate, then
\beq
{Br(\mu \ra 3 e) \over Br(   \mu \ra e \gamma)}= \frac{2 \alpha}{3\pi}\left[\ln \frac{m_\mu}{m_e}-\frac{11}{8}\right] \simeq 6.12\times 10^{-3}\,,
\eeq
thus can be ignored.

\begin{figure}[htb]
\centering
\includegraphics[width=0.57\textwidth]{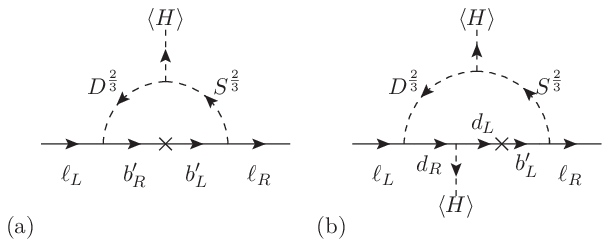}
\caption{ The Feynman diagrams, in the interaction basis, for
 the anomalous magnetic dipole moment of charged lepton in this model.
 Here all the flavor indices are omitted.   The external photon (not shown ) can be attached to any charged particle in the loop.  }
\label{fig:g-2}
\end{figure}

In our model,  the vector fermion $b'_{L,R}(3,1,-1/3)$,  Fig.\ref{fig:g-2}(a), and/or the SM b-quark, Fig.\ref{fig:g-2}(b), can play the role of $F$ both carrying an electric charge $-\ot$.
The  function ${\cal J}_{-\ot}(\alpha )$ takes a value in the range from $-0.022$ to $0.087$ for $\alpha\in[0.1,10.0]$.
In the interaction basis,  $T(3,3,-1/3)$ does not couple to $b'$, $D(3,2,1/6)$ only couples to left-handed charged lepton, and $S(3,1,2/3)$ only couples to the right-handed charged lepton. Due to the $D\mhyphen S$ and $D\mhyphen T$ mixings, the three charge-$2/3$ physical mass states acquire both the LH and RH Yukawa couplings as shown in Eq.(\ref{eq:gen_eF_Yukawa}). However, the physical state dominated by the $T$ component gets double suppression form $D\mhyphen T$ and $d\mhyphen b'$ mixings, thus not important here.
Assuming small $D\mhyphen S$ mixing in our model, the anomalous magnetic moment of charged lepton becomes
\beqa
\tri a_l &\simeq & ( \lambda^D_{l b'} \lambda^S_{l b'} ) {3 \mu_1 v_0\over 16\sqrt{2}\pi^2}\frac{m_l}{M^3_{b'}}
\times {\cal K}\left(\frac{M_D^2}{M^2_{b'}}, \frac{M_S^2}{M^2_{b'}}\right)\nonr\\
&+ &  ( \lambda^D_{l b} \lambda^S_{l b} ) {3 \mu_1 v_0\over 16\sqrt{2}\pi^2}\frac{m_l}{m^3_{b}}
\times {\cal K}\left(\frac{M_D^2}{m^2_{b}}, \frac{M_S^2}{m^2_{b}}\right) \,,
\eeqa
where
\beq
{\cal K}(a,b)\equiv {{\cal J}_{-\ot}(a)- {\cal J}_{-\ot}(b) \over b-a }\,.
\eeq
When $a \simeq b$, the function ${\cal K}$ takes a limit
\beq
{\cal K}(a,b) \stackrel{b\ra a}{\Rightarrow} - \left. \frac{d }{d\alpha}{\cal J}_{-\ot}(\alpha )\right|_{\alpha=a}
= -{11-4 a-7 a^2+2[2+a(6+a)]\ln a \over 6(1-a)^4}\,.
\eeq
For $a \simeq b \simeq 1$, it can be approximated by ${\cal K}(a,b) \simeq 1/36-(a+b-2)/45$, and
${\cal K}(a,b) \simeq -\ln a/ (3 a^2)$ for $ a\simeq b \gg 1$.

If factoring out the $M_F=M_{b'}$, the dipole transition coefficients in Eq.(\ref{eq:brmeg}) are given by
\beq
A^{\mu e}_{M/E} \simeq -{\mu_1 v_0 \over 4\sqrt{2} M_{b'}^2}\left\{
\left[ (  \lambda^S_{e b'})^*\lambda^D_{\mu b'} \pm (\lambda^D_{e b'})^*\lambda^S_{\mu b'}  \right]{\cal K}(\beta_D,\beta_S)+ \frac{M_{b'}^3}{m_b^3}\left[  (\lambda^S_{e b})^*\lambda^D_{\mu b} \pm (\lambda^D_{e b})^*\lambda^S_{\mu b} \right]{\cal K}(b_D,b_S) \right\}\,,
\eeq
where $b_{D,S}\equiv  ( M_{D,S}/m_b )^2$, and $\beta_{D,S}= (M_{D,S}/M_{b'})^2$.
The current upper bound of  $Br(\mu\ra e \gamma)$ amounts to a stringent limit that the relevant $|\lambda^S \lambda^D| \lesssim 10^{-5}$.
Instead of making the product of  Yukawa couplings small, the $\mu\ra e \gamma$ transition from $D\mhyphen S$ mixing, Fig.\ref{fig:g-2}, can be simply arranged to vanish if muon/electron only couples to $b'/b$ or the other way around.

Modulating by the leptoquark masses,
numerically we have either
\beqa
\mbox{Sol-1} &:&\nonr\\
\tri a_e & \simeq &   2.28\times 10^{-5}\times [\lambda^D_{e b} \lambda^S_{e b}  ]\times \left( \frac{\mu_1}{\mbox{GeV}}\right)  \times  {\cal K}(b_D,b_S)\,,\nonr\\
\tri a_\mu & \simeq &  1.03\times 10^{-10} \times [\lambda^D_{\mu b'} \lambda^S_{\mu b'}  ]\times \left( \frac{\mu_1}{\mbox{GeV}}\right)  \left( \frac{1.5 \mtev}{M_{b'}}\right)^3 \times  {\cal K}(\beta_D,\beta_S)\,,
\label{eq:g2sol1}
\eeqa
or
\beqa
\mbox{Sol-2} &:&\nonr\\
\tri a_e & \simeq &  5.00 \times 10^{-13} \times [\lambda^D_{e b'} \lambda^S_{e b'}  ]\times \left( \frac{\mu_1}{\mbox{GeV}}\right) \left( \frac{1.5 \mtev}{M_{b'}}\right)^3 \times  {\cal K}(\beta_D,\beta_S)\,,\nonr\\
\tri a_\mu & \simeq &   4.71\times 10^{-3}  \times [\lambda^D_{\mu b} \lambda^S_{\mu b}  ]\times
 \left( \frac{\mu_1}{\mbox{GeV}}\right)  \times  {\cal K}(b_D,b_S)\,.
\label{eq:g2sol2}
\eeqa

For $M_{b'}=1.5 \mtev$  and $M_{LQ} \simeq 1 \mtev$, then either $\left\{ \mu_1 \lambda^D_{e b} \lambda^S_{e b}\,,\;  \mu_1 \lambda^D_{\mu b'} \lambda^S_{\mu b'} \right \} \simeq \{ 49.4  [-27.3] , 279.9  \} \mgev $ for (Sol-1), or   $\left\{ \mu_1 \lambda^D_{e b'} \lambda^S_{e b'}\,,\;  \mu_1 \lambda^D_{\mu b} \lambda^S_{\mu b} \right \} \simeq \{ -20.1  [11.1] , -689.5  \} \mbox{GeV}$ for (Sol-2)  can accommodate
the observed central values of $\tri a_e^{Cs}  [\tri a_e^{Rb}] $ and $\tri a_\mu$ simultaneously with vanishing $Br(\mu\ra e \gamma)$. However, as will be discussed later, only Sol-2 is viable to simultaneously accommodate the neutrino data.

\subsection{$b\ra s l^+ l^-$}

\begin{figure}[htb]
\centering
\includegraphics[width=0.65\textwidth]{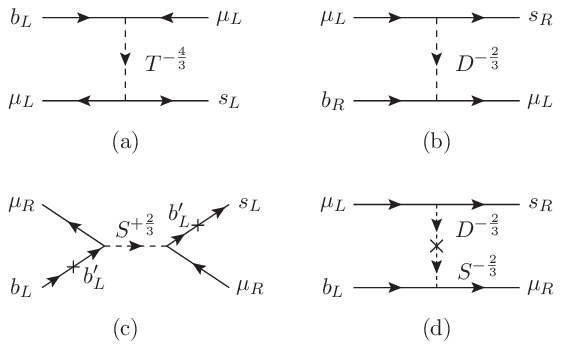}
\caption{ The potential tree-level Feynman diagrams for $b\ra s\mu\bar{\mu}$ transition. }
\label{fig:RK_tree}
\end{figure}

The $b\ra s \mu\bar{\mu}$ transition can be generated by tree-level diagrams mediated by $T^\ft$,  $D^\tth$, $S^\tth$, and the one from  $D\mhyphen S$ mixing, see Fig.\ref{fig:RK_tree}. In Fig.\ref{fig:RK_tree}(c), the crosses represent the mixing between the $b'_L$ and the physical $b$ and $s$ quarks, because $S$ only couples to $b'_L$ in the interaction basis.
From Eq.(\ref{eq:H_eff}), we see that this model can yield $b\ra s \mu\bar{\mu} $ operators in the vector, scalar, and tensor forms.
However, we failed to find a viable parameter space to explain the $b\ra s l^+ l^-$ anomaly and simultaneously comply with other  experimental constraints\footnote{On the other hand, we cannot rule out the possibility of finding such a solution with fine-tuning. }, see Sec.\ref{sec:2q2lHeff}.

\begin{figure}[htb]
\centering
\includegraphics[width=0.65\textwidth]{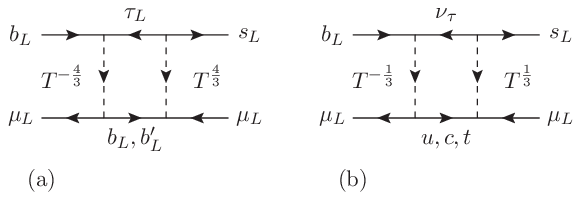}
\caption{ The Feynman diagram for $b\ra s\mu\bar{\mu}$ transition. }
\label{fig:RK}
\end{figure}

Instead, to bypass the stringent experimental bounds and the fine-tuning of the parameters,  we go for the 1-loop box diagram contribution, as shown in Fig.\ref{fig:RK},  which requires only four nonzero triplet Yukawa couplings $\lambda^T_{\tau s}, \lambda^T_{\tau b}, \lambda^T_{\mu b}, \lambda^T_{\mu b'}$.

In the usual convention, the transition is described by a low energy effective Hamiltonian
\beq
{\cal H}^{b\ra s \mu\mu}_{eff} =-\frac{G_F}{\sqrt{2}} \wt{V}_{tb}\wt{V}_{ts}^* \frac{\alpha}{\pi}\sum_i {\cal C}_i {\cal O}_i + H.c.
\eeq
with
\beqa
{\cal O}_9 =\left(\bar{s}\gamma^\alpha \hat{L} b \right) \left(\bar{\mu}\gamma_\alpha \mu\right)\,,\,\,
{\cal O}_{10} =\left(\bar{s}\gamma^\alpha \hat{L} b \right) \left(\bar{\mu}\gamma_\alpha \gamma^5\mu\right)\,,\\
{\cal O}'_9 =\left(\bar{s}\gamma^\alpha \hat{R} b \right) \left(\bar{\mu}\gamma_\alpha \mu\right)\,,\,\,
{\cal O}'_{10} =\left(\bar{s}\gamma^\alpha \hat{R} b \right) \left(\bar{\mu}\gamma_\alpha \gamma^5\mu\right)\,.
\eeqa

Ignoring the tau  mass in the loop, Fig.\ref{fig:RK}(a), the effective Hamiltonian generated by the box-diagram can be easily calculated as
\beq
{\cal H}^{b\ra s \mu\mu}_{eff (a) } \simeq
-  { \lambda^T_{\tau b} (\lambda^T_{\tau s})^* \over 64\pi^2} \left( \frac{|\lambda^T_{\mu b'}|^2}{M_{b'}^2} {\cal G}(\beta'_T)
+\frac{|\lambda^T_{\mu b}|^2}{m_b^2} {\cal G}(\beta_T)  \right)
\left(\bar{s}\gamma^\alpha \PL b \right) \left(\bar{\mu}\gamma_\alpha \PL \mu\right) +H.c.\,,
\eeq
where $\beta'_T\equiv (M_T/M_{b'})^2$, $\beta_T\equiv (M_T/m_{b})^2$, and
\beq
{\cal G}(x)= \left[ \frac{1}{1-x}+ \frac{\ln x}{(1-x)^2}\right]\,.
\eeq
The function has a limit ${\cal G}(x=1)= -1/2$, and ${\cal G}\ra -1/x$ when $x\gg 1$.

The second contribution from the box diagram with $T^{\pm \ot}$ and up-type quark running in the loop yields
\beq
{\cal H}^{b\ra s \mu\mu}_{eff (b)} \simeq
  { \lambda^T_{\tau b} (\lambda^T_{\tau s})^* \over 64\pi^2}
\frac{1}{4 M_T^2}
\left[ |\lambda^T_{\mu b'}|^2 ( s_1^2+s_2^2+s_3^2) + |\lambda^T_{\mu b}|^2 \right]
\left(\bar{s}\gamma^\alpha \PL b \right) \left(\bar{\mu}\gamma_\alpha \PL \mu\right) +H.c.\,.
\eeq
In arriving the above expression, we have made use of the unitarity of $\wt{V}$, namely,
\beq
|\wt{V}_{ub'}|^2+ |\wt{V}_{cb'}|^2+ |\wt{V}_{tb'}|^2 =1 -c_1^2 c_2^2 c_3^2 \simeq s_1^2+s_2^2+s_3^2 \ll 1\,,
\eeq
and
\beq
|\wt{V}_{ub}|^2+ |\wt{V}_{cb}|^2+ |\wt{V}_{tb}|^2 \simeq 1\,.
\eeq
It is clear that the contribution from  Fig.\ref{fig:RK}(b) is dominated by $\lambda^T_{\mu b}$.

And the relevant Wilson coefficients are determined to be
\beqa
{\cal C}_9 &=& -{\cal C}_{10} \simeq \frac{\sqrt{2} }{ 128 \pi\alpha} {  \lambda^T_{\tau b} (\lambda^T_{\tau s})^*  \over V_{tb}V_{ts}^*  G_F M_T^2 }
\left[  |\lambda^T_{\mu b'}|^2 \beta'_T {\cal G}(\beta'_T)-\frac{5}{4} |\lambda^T_{\mu b}|^2 \right]\,,\nonr\\
{\cal C}'_9 &=& -{\cal C}'_{10}=0\,.
\label{eq:C9_10}
\eeqa
For a typical value of $\beta'_T=(1.0 \mbox{TeV}/1.5 \mbox{TeV})^2$, $\beta'_T {\cal G}(\beta'_T)=-0.3677$.

We use the following values,
\beq
({\cal C}_9)^\mu = -({\cal C}_{10})^\mu \simeq   -0.41\pm 0.07\,,
\eeq
for muon, and
\beq
({\cal C}_9)^e \simeq  ({\cal C}_{10})^e \simeq ({\cal C}'_9)^e \simeq ({\cal C}'_{10})^e \simeq 0
\eeq
for the electron counter part,
from the global fit to the $b\ra s l^+ l^-$ data \cite{Altmannshofer:2021qrr}
\footnote{ Similar result is also yielded by \cite{Geng:2021nhg}.
There are other suggestion by the recent studys of\cite{Altmannshofer:2021qrr,Geng:2021nhg}. However, to only produce $C_9^{\mu\mu}$ in our model requires large both $b\mhyphen b'$ and $b'\mhyphen s$ mixings and the tree-level processes, which we discard. }.

If we take $\wt{V}_{tb} \wt{V}_{ts}^* =-0.03975$, then it amounts to
\beq
 \lambda^T_{\tau b} (\lambda^T_{\tau s})^*
\left[ - |\lambda^T_{\mu b'}|^2 \beta'_T {\cal G}(\beta'_T) +\frac{5}{4} |\lambda^T_{\mu b}|^2 \right]
 \simeq  -(0.394 \pm 0.067) \left(\frac{M_T}{\mbox{TeV}}\right)^2 \,.
 \label{eq:bsmm_req}
\eeq
Since the combination in the squared bracket is positive, the product $ \lambda^T_{\tau b} (\lambda^T_{\tau s})^* $ has to be negative.
The constraints from $B_s\mhyphen \overline{B}_s$ mixing and $b\ra s \gamma$  will be carefully discussed in Sec.\ref{sec:pheno}.

\subsection{Cabibbo-angle anomaly}

From the unitarity of $\wt{V}_4$, it is clear that
\beq
|\widetilde{V}_{ud}|^2+|\widetilde{V}_{us}|^2+|\widetilde{V}_{ub}|^2=1 -|\wt{V}_{u b'}|^2\leq 1\,,
\label{eq:V4deficit}
\eeq
and the Cabibbo-angle anomaly(CAA) is naturally embedded in this model.
Moreover, the most commonly discussed unitarity triangle becomes
\beq
\widetilde{V}_{ud}\widetilde{V}_{ub}^*+\widetilde{V}_{cd}\widetilde{V}_{cb}^*+
\widetilde{V}_{td}\widetilde{V}_{tb}^* =  -(U^d_L)_{d4}^* (U^d_L)_{b4}\,.
\eeq

Similarly, this model also predicts that
\beqa
&&|\widetilde{V}_{cd}|^2+|\widetilde{V}_{cs}|^2+|\widetilde{V}_{cb}|^2=1 -|\wt{V}_{c b'}|^2\,,\\
&&|\widetilde{V}_{td}|^2+|\widetilde{V}_{ts}|^2+|\widetilde{V}_{tb}|^2=1 -|\wt{V}_{t b'}|^2\,,\\
&&|\widetilde{V}_{u d}|^2+|\widetilde{V}_{c d}|^2+|\widetilde{V}_{t d}|^2=1-|(U^d_L)_{d4}|^2\,,  \\
&&|\widetilde{V}_{u s}|^2+|\widetilde{V}_{c s}|^2+|\widetilde{V}_{t s}|^2=1-|(U^d_L)_{s4}|^2\,, \\
&&|\widetilde{V}_{u b}|^2+|\widetilde{V}_{c b}|^2+|\widetilde{V}_{t b}|^2=1-|(U^d_L)_{b4}|^2\,,
\eeqa
and all the other SM CKM unitary triangles are no more closed in general.

The matrix elements are easy to read. For example, we have
\beqa
&& \wt{V}_{us} = c_2 V_{us}-s_2 s_3 V_{ub}\,,\;\wt{V}_{u b'} = s_1 V_{ud} +c_1 s_2 V_{us} +c_1 c_2 s_3 V_{ub}\,,\\
&& \left\{ (U^d_L)_{d4},(U^d_L)_{s4},(U^d_L)_{b4},(U^d_L)_{b'4} \right\}= \left\{    -s_1 c_2 c_3,-s_2 c_3, -s_3, c_1 c_2 c_3\right\}\,.
\eeqa
The mixing $\theta_i$ is expected to be small, so a smaller universal
\beq
\left|\wt{V}_{us}\right| \simeq \left|V_{us} \right| \left(1-\frac{\theta_2^2}{2}\right)
\eeq
to leading order is expected as well.
 By using the Wolfenstein parameterization and the central values from global fit\cite{Zyla:2020zbs}, we have
\beq
 \wt{V}_{u b'} \simeq   0.9740 s_1 + 0.2265 c_1 s_2  +0.0036 c_1 c_2 s_3  e^{1.196 i} \,.
\eeq
Therefore, to accommodate the deficit of 1st row CKM unitarity (Eq.(\ref{eq:PDGCKM}) and Eq.(\ref{eq:V4deficit}))
we have
\beq
\left|  s_1 + 0.233 s_2 \right| \simeq 0.039(7)\,.
\label{eq:CKMA_req}
\eeq

\section{Constraints and parameter space}
\label{sec:pheno}
As discussed in the previous section, this model is capable to address neutrino mass generation, $\tri a_{e,\mu}$, $b\ra s\mu\mu$, and the CAA. For readers' convenience, all the requirements are collected and displayed in Table \ref{tab:req_sum}.

\begin{table}
\begin{center}
\begin{tabular}{|c|c|c|}
 \hline
  Anomaly  & Requirement  & Remark\\
  \hline
  $m_\nu$ &
   $\mu_3 m_{b^{(')}}\lambda^D \lambda^T \simeq {\cal O}(10^{-5})(\mgev)^2 $ & Eq.(\ref{eq:nu_mass})\\
  $\tri a_e^{Cs [Rb]}, \tri a_\mu$ ( Sol-1 ) &
  $\left\{ \mu_1 \lambda^D_{e b} \lambda^S_{e b}\,,\;  \mu_1 \lambda^D_{\mu b'} \lambda^S_{\mu b'} \right \} \simeq \{ (49\pm 20)  [-27\pm 17] , (280\pm 66)  \} \mbox{GeV}$
    & Eq.(\ref{eq:g2sol1})  \\
    $\tri a_e^{Cs  [Rb]}, \tri a_\mu$ ( Sol-2 ) &
     $\left\{ \mu_1 \lambda^D_{e b'} \lambda^S_{e b'}\,,\;  \mu_1 \lambda^D_{\mu b} \lambda^S_{\mu b} \right \} \simeq \{ -(20.1\pm8.3)   [+11.1\pm 6.9], -(689 \pm 162)  \} \mbox{GeV}$
  &  Eq.(\ref{eq:g2sol2}) \\
    $b\ra sl^+l^-$ &
    $\lambda^T_{\tau b} (\lambda^T_{\tau s})^* ( |\lambda^T_{\mu b'}|^2 +3.39|\lambda^T_{\mu b}|^2 ) \simeq  -(1.07 \pm 0.18)$
    & Eq.(\ref{eq:C9_10}) \\
  Cabibbo angle anomaly & $\left| s_1 + 0.233 s_2 \right| \simeq 0.039(7)$
   & Eq.(\ref{eq:V4deficit}) \\
 \hline
\end{tabular}
\caption{The requirement for explaining each mechanism/anomaly.  For illustration, we take the following values: $ \wt{V}_{tb} \wt{V}_{ts}^*=-0.03975$,  $M_{LQ}=1.0$ TeV, and $M_{b'}=1.5$ TeV.  }
\end{center}
\label{tab:req_sum}
\end{table}

In this section, we should carefully scrutinize all the existing experimental limits and try to identify the viable model parameter at the end.

\subsection{Low energy $2q2l$ effective Hamiltonian}
\label{sec:2q2lHeff}

\begin{table}
\begin{center}
\begin{tabular}{|lllr|}
\hline
${(\bar{q}_k \gamma^\mu \PL q_l)(\bar{e}_i \gamma_\mu \PL e_j)\over 4 M_T^2}$ & Wilson Coef. & Constraint& Model \\ \hline
$bb\mu\mu$ & $2 |\lambda^T_{\mu b}|^2 $ & $211.1$\cite{Carpentier:2010ue}&  $1.06$ \\ 
$sb\tau\tau$ & $2 \lambda^T_{\tau b} (\lambda^T_{\tau s})^* $ & - &   $-0.14$ \\ 
$sb\mu\mu$ & $0\footnote{There is no such effective operator at tree-level.}$ & - & $0$ \\ 
$sb\mu\tau$& $0$ & - & $0$ \\ 
$sb\tau\mu$ & $2 \lambda^T_{\mu b} (\lambda^T_{\tau s})^* $ & $0.199\footnote{ We update this value by using the new data ${\cal B}(B^+\ra K^+\mu^+ \tau^-)< 4.5\times 10^{-5}$\cite{Zyla:2020zbs}.}$\cite{Carpentier:2010ue}&   $0.11$  \\ \hline
$u u \tau\mu$ & $ \left( \wt{V}_{ub}(\lambda^T_{\tau b})^*+ \wt{V}_{ub'}(\lambda^T_{\tau b'})^*+\wt{V}_{us}(\lambda^T_{\tau s})^*\right)
\times \left(\wt{V}^*_{ub}\lambda^T_{\mu b}+ \wt{V}^*_{ub'}\lambda^T_{\mu b'} \right] $ & $0.13$\cite{Carpentier:2010ue}&   $0.0043$    \\ 
$u u \mu\mu$ & $ \left| \wt{V}_{ub}(\lambda^T_{\mu b})^*+ \wt{V}_{ub'}(\lambda^T_{\mu b'})^* \right|^2 $&$1.03$\cite{Carpentier:2010ue}&   $0.017$   \\ \hline
$u c \mu\mu$ & $ \left( \wt{V}_{ub}(\lambda^T_{\mu b})^*+ \wt{V}_{ub'}(\lambda^T_{\mu b'})^*\right)
\times \left(\wt{V}^*_{cb}\lambda^T_{\mu b}+ \wt{V}^*_{cb'}\lambda^T_{\mu b'} \right) $ &  $0.11\footnote{ We update this value by using the new data ${\cal B}(D^+\ra \pi^+\mu^+ \mu^-)< 7.3\times 10^{-8}$\cite{Zyla:2020zbs}.}
$\cite{Carpentier:2010ue}&$0^*$  \\ \hline
$cc \mu\mu$ & $ \left| \wt{V}_{cb}(\lambda^T_{\mu b})^*+ \wt{V}_{cb'}(\lambda^T_{\mu b'})^* \right|^2 $ & $52.8$\cite{Carpentier:2010ue}&$0^*$  \\ 
$cc \tau\mu$ & $ \left( \wt{V}_{cb}(\lambda^T_{\tau b})^*+ \wt{V}_{cb'}(\lambda^T_{\tau b'})^*+\wt{V}_{cs}(\lambda^T_{\tau s})^*\right)
\times \left(\wt{V}^*_{cb}\lambda^T_{\mu b}+ \wt{V}^*_{cb'}\lambda^T_{\mu b'} \right) $ &  $211.1$\cite{Carpentier:2010ue}&$0^*$\\ \hline
$tc\mu\mu$ & $ \left( \wt{V}_{tb}(\lambda^T_{\mu b})^*+ \wt{V}_{tb'}(\lambda^T_{\mu b'})^*\right)
\times \left(\wt{V}^*_{cb}\lambda^T_{\mu b}+ \wt{V}^*_{cb'}\lambda^T_{\mu b'} \right) $ & -&$0^*$ \\ 
$tc\tau\tau$ & $ \left( \wt{V}_{tb}(\lambda^T_{\tau b})^*+ \wt{V}_{tb'}(\lambda^T_{\tau b'})^*+ \wt{V}_{ts}(\lambda^T_{\tau s})^*\right)
\times \left(\wt{V}^*_{cb}\lambda^T_{\tau b}+ \wt{V}^*_{cb'}\lambda^T_{\tau b'} + \wt{V}^*_{cs}\lambda^T_{\tau s} \right) $ &  - &   $-0.030$ \\ 
$tc\tau\mu$ & $ \left( \wt{V}_{tb}(\lambda^T_{\tau b})^*+ \wt{V}_{tb'}(\lambda^T_{\tau b'})^*+ \wt{V}_{ts}(\lambda^T_{\tau s})^*\right)
\times \left(\wt{V}^*_{cb}\lambda^T_{\mu b}+ \wt{V}^*_{cb'}\lambda^T_{\mu b'} \right) $ &  $11.35\footnote{We obtain the limit by using the top quark decay width, $\Gamma_t=1.42 \mgev$, and ${\cal B}(t\ra q l l')<1.86\times 10^{-5}$\cite{ATLAS:2018avw}.}$&$0^*$ \\ 
$tc\mu\tau$ & $ \left( \wt{V}_{tb}(\lambda^T_{\mu b})^*+ \wt{V}_{tb'}(\lambda^T_{\mu b'})^* \right)
\times \left(\wt{V}^*_{cb}\lambda^T_{\tau b}+ \wt{V}^*_{cb'}\lambda^T_{\tau b'} + \wt{V}^*_{cs}\lambda^T_{\tau s} \right) $ & $11.35$&  $0.02$ \\ \hline
$tu\mu\mu$ & $ \left( \wt{V}_{tb}(\lambda^T_{\mu b})^*+ \wt{V}_{tb'}(\lambda^T_{\mu b'})^*\right)
\times \left(\wt{V}^*_{ub}\lambda^T_{\mu b}+ \wt{V}^*_{ub'}\lambda^T_{\mu b'} \right) $ &  - &   $0.09$  \\ 
$tu\tau\tau$ & $ \left( \wt{V}_{tb}(\lambda^T_{\tau b})^*+ \wt{V}_{tb'}(\lambda^T_{\tau b'})^*+ \wt{V}_{ts}(\lambda^T_{\tau s})^*\right)
\times \left(\wt{V}^*_{ub}\lambda^T_{\tau b}+ \wt{V}^*_{ub'}\lambda^T_{\tau b'} + \wt{V}^*_{us}\lambda^T_{\tau s} \right) $ & -&   $-0.03$   \\ 
$tu\tau\mu$ & $ \left( \wt{V}_{tb}(\lambda^T_{\tau b})^*+ \wt{V}_{tb'}(\lambda^T_{\tau b'})^*+ \wt{V}_{ts}(\lambda^T_{\tau s})^*\right)
\times \left(\wt{V}^*_{ub}\lambda^T_{\mu b}+ \wt{V}^*_{ub'}\lambda^T_{\mu b'} \right) $ &  $11.35$&  $-0.12$  \\ 
$tu\mu\tau$ & $ \left( \wt{V}_{tb}(\lambda^T_{\mu b})^*+ \wt{V}_{tb'}(\lambda^T_{\mu b'})^* \right)
\times \left(\wt{V}^*_{ub}\lambda^T_{\tau b}+ \wt{V}^*_{ub'}\lambda^T_{\tau b'} + \wt{V}^*_{us}\lambda^T_{\tau s} \right) $ &  $11.35$&  $0.02$  \\ \hline
\end{tabular}
\caption{The tree-level NC operators and their Wilson coefficients.
 We take $M_T=1\mtev$ for illustration, and the values in last two columns scale as  $ (M_T/1 \mtev)^2$.
By using the parameter set example of Eq.(\ref{eq:num_result}),  the model predictions, with the signs kept,  are displayed in the last column. In the table, $0^*$ stems from choosing $\wt{V}^*_{cb}\lambda^T_{\mu b}+ \wt{V}^*_{cb'}\lambda^T_{\mu b'} =0$ to retain the $\mu\mhyphen e$ universality in $b\ra c l\nu$ transition as discussed in the text.
   }
   \label{tab:NCTL_list}
\end{center}
\end{table}

\begin{table}
\begin{center}
\begin{tabular}{|lllr|}
\hline
${(\bar{d}_k \gamma^\mu \PL u_l)(\bar{\nu}_i \gamma_\mu \PL e_j)\over 4 M_T^2}$ & Wilson Coef. & Constraint & Model \\ \hline
$su\nu_\mu \mu$ & $0 $ &  & $0$ \\ 
$su\nu_\tau \mu$ & $(\lambda^T_{\tau s})^* \left(\wt{V}^*_{ub}\lambda^T_{\mu b}+ \wt{V}^*_{ub'}\lambda^T_{\mu b'} \right) $ &  $3.96 $&   $ 0.010$  \\ 
$su\nu_\tau \tau$ & $(\lambda^T_{\tau s})^* \left(\wt{V}^*_{ub}\lambda^T_{\tau b}+ \wt{V}^*_{ub'}\lambda^T_{\tau b'}+ \wt{V}^*_{us}\lambda^T_{\tau s}  \right) $ &  $0.79$&   $0.003$ \\ 
$su\nu_\mu \tau$ & $0$ &  &  $0$ \\ \hline
$sc\nu_\mu \mu$ & $0 $ &  &  $0$\\ 
$sc\nu_\tau \mu$ & $(\lambda^T_{\tau s})^* \left(\wt{V}^*_{cb}\lambda^T_{\mu b}+ \wt{V}^*_{cb'}\lambda^T_{\mu b'} \right) $ &  $31.7$& $0^*$ \\ 
$sc\nu_\tau \tau$ & $(\lambda^T_{\tau s})^* \left(\wt{V}^*_{cb}\lambda^T_{\tau b}+ \wt{V}^*_{cb'}\lambda^T_{\tau b'} + \wt{V}^*_{cs}\lambda^T_{\tau s} \right) $ &  $15.8$&   $0.002$  \\ 
$sc\nu_\mu \tau$ & $0$ &  &  $0$ \\ \hline
$b u \nu_\mu\mu$ &  $(\lambda^T_{\mu b})^* \left(\wt{V}^*_{ub}\lambda^T_{\mu b}+ \wt{V}^*_{ub'}\lambda^T_{\mu b'} \right) $ &  $0.51$&   $0.09$  \\ 
$b u \nu_\tau\mu$ &  $(\lambda^T_{\tau b})^* \left(\wt{V}^*_{ub}\lambda^T_{\mu b}+ \wt{V}^*_{ub'}\lambda^T_{\mu b'} \right) $ &  $0.51$&   $-0.12$  \\ 
$b u \nu_\tau\tau$ &  $(\lambda^T_{\tau b})^* \left(\wt{V}^*_{ub}\lambda^T_{\tau b}+ \wt{V}^*_{ub'}\lambda^T_{\tau b'}+\wt{V}^*_{us}\lambda^T_{\tau s} \right) $ &  $0.51$&   $-0.03$  \\ 
$b u \nu_\mu \tau$ &  $(\lambda^T_{\mu b})^* \left(\wt{V}^*_{ub}\lambda^T_{\tau b}+ \wt{V}^*_{ub'}\lambda^T_{\tau b'}+ \wt{V}^*_{us}\lambda^T_{\tau s} \right) $ &  $0.51$&   $0.02$ \\ \hline
$b c \nu_\mu\mu$ &  $(\lambda^T_{\mu b})^* \left(\wt{V}^*_{cb}\lambda^T_{\mu b}+ \wt{V}^*_{cb'}\lambda^T_{\mu b'} \right) $ &  $5.41$&$0^*$ \\ 
$b c \nu_\tau\mu$ &  $(\lambda^T_{\tau b})^* \left(\wt{V}^*_{cb}\lambda^T_{\mu b}+ \wt{V}^*_{cb'}\lambda^T_{\mu b'} \right) $ &  $5.41$&$0^*$ \\ 
$b c \nu_\tau\tau$ &  $(\lambda^T_{\tau b})^* \left(\wt{V}^*_{cb}\lambda^T_{\tau b}+ \wt{V}^*_{cb'}\lambda^T_{\tau b'}+\wt{V}^*_{cs}\lambda^T_{\tau s} \right) $ &  $5.41$& $  -0.03\footnote{This is the effective operator to address the $R(D^{(*)})$ anomaly.}$\\ 
$b c \nu_\mu \tau$ &  $(\lambda^T_{\mu b})^* \left(\wt{V}^*_{cb}\lambda^T_{\tau b}+ \wt{V}^*_{cb'}\lambda^T_{\tau b'}+ \wt{V}^*_{cs}\lambda^T_{\tau s} \right) $ &  $5.41$&   $0.02$  \\ \hline
\end{tabular}
\caption{The  tree-level CC operators and their Wilson coefficients.
All the constraints are taken and derived from\cite{Carpentier:2010ue}. By using the parameter set example of Eq.(\ref{eq:num_result}), the model predictions, with the signs kept, are displayed in the last column.  We take $M_T=1\mtev$ for illustration, and the values in last two columns scale as  $ (M_T/1 \mtev)^2$.
Note that the coefficients for $su(c)\nu_\mu \tau$ and $su(c)\nu_\mu \mu$ are zero at tree-level. In the table, $0^*$ stems from choosing $\wt{V}^*_{cb}\lambda^T_{\mu b}+ \wt{V}^*_{cb'}\lambda^T_{\mu b'} =0$ to retain the $\mu\mhyphen e$ universality in $b\ra c l\nu$ transition as discussed in the text. }
\label{tab:CCTL_list}
\end{center}
\end{table}

In this model, the minimal set ($\mbox{MinS}_T$) of $\lambda^T$ parameters for addressing all the anomalies and neutrino mass consists of five elements:
\beq
\mbox{MinS}_T =\{ \lambda^T_{\tau b}\,,\; \lambda^T_{\tau s}\,,\; \lambda^T_{\tau b'}\,,\; \lambda^T_{\mu b'}\,,\; \lambda^T_{\mu b}\}\,.
\eeq
The following are the consequences of adding other Triplet Yukawa couplings outside the $\mbox{MinS}_T$ :
(1) At tree-level, $\lambda^T_{ed}$ leads to $B^+\ra \pi^+ e\mu$, $B^0\ra \bar{e}\tau$, $\tau\ra e K$, and $\mu\mhyphen e$ conversion.
Then $\lambda^T_{ed}\lesssim 10^{-2}$ must be satisfied if all other $\lambda^T$'s are around ${\cal O}(1)$.
(2)  At tree-level, $\lambda^T_{es}$ leads to $B^+\ra K^+ e\mu$  and $\mu\mhyphen e$ conversion.
Then $\lambda^T_{ed}\lesssim 10^{-2}$ is also required if all other $\lambda^T$'s are around ${\cal O}(1)$.
(3) At tree-level $\lambda^T_{eb}$ also leads to $\mu\mhyphen e$ conversion, but the constraint is weak due to the $|\wt{V}_{ub}|^2$ suppression. On the other hand, at 1-loop level, it generates the unfavored $b\ra s e e$ transition.
Also, note that $\lambda^T_{eb}\neq 0$ is not helpful for generating $M^\nu_{ee}$, which is crucial for the neutrinoless double beta decay. (4) Together with $\lambda^T_{\tau s}$, required for $b\ra s \mu\mu$, any nonzero $\lambda^T_{\ell_i d} (i=e, \mu, \tau)$ gives rise to $K^+\ra \pi^+ \nu \nu$ at the tree-level, and thus strongly constrained.
Moreover,  $\lambda^T_{\tau s}$ and $\lambda^T_{\tau d}$ generate the $K\mhyphen\bar{K}$ mixing via the 1-loop box diagram, and thus
stringently limited.
(5) Together with $\mbox{MinS}_T$, the presence of any of $\lambda^T_{e d_i}, (i=d,s,b,b')$ leads to $l\ra l' \gamma$ transition at the one-loop level.
(6) On the other hand, the introduction of $\lambda^T_{\tau d}$ generates $s\ra d\mu\mu$ transition via the box-diagram
which is severely constraint by the $K_L\ra \mu\mu $ data. So it has to be small too.
(7) In general,  adding $\lambda^T_{\mu s}$ will cause conflict with the precision Kaon data.

From the above discussion, adding any $\lambda^T \not\in \mbox{MinS}_T$ requires fine tuning the parameters.
For simplicity, we set any triplet Yukawa couplings outside the $\mbox{MinS}_T$ to zero.
However, we still need to scrutinize all the phenomenological constraint on the minimal set of parameters.
All the potential detectable effective operators from tree-level contribution of $\mbox{MinS}_T$ are listed in Table \ref{tab:NCTL_list} and Table \ref{tab:CCTL_list}. And one has to make sure all the constraints have to be met.

In addition to the limits considered in Ref\cite{Carpentier:2010ue},  one needs to take into account the
constraint from the lepton universality tests in B decays\cite{Bifani:2018zmi}.
In particular, the $\mu\mhyphen e$ universality in the $b\ra c l_i\nu (i=e,\mu)$ transition has been tested to $\simeq 1\%$ level\cite{Jung:2018lfu}. The $\mbox{MinS}_T$ of $\lambda^T$ introduces two operators,
$(\bar{b}\gamma^\alpha\PL c)(\bar{\nu}_\mu \gamma_\alpha \PL \mu)$ and $(\bar{b}\gamma^\alpha\PL c)(\bar{\nu}_\tau \gamma_\alpha \PL \mu)$, where the first one interferes with the SM CC interaction while the second one does not.  On the other hand, there are no electron counter parts.
Therefore, it is required that the modification to the $b\ra c \mu \nu_j$ transition rate due to the two new operators is less than $\sim 2\%$.
Their Wilson coefficients,  the third and the fourth entities from the end in Table \ref{tab:CCTL_list}, are both proportional to
$\left(\wt{V}^*_{cb}\lambda^T_{\mu b}+ \wt{V}^*_{cb'}\lambda^T_{\mu b'} \right)$. For simplicity, we artificially set this combination to zero to make sure the perfect $\mu\mhyphen e$ universality in $b\ra c l \nu$ at tree level, such that the ratio of $\lambda^T_{\mu b}/\lambda^T_{\mu b'}$ is fixed as well. However, if more parameter space is wanted, this strict relationship can be relaxed as long as the amount of $\mu\mhyphen e$ universality violation is below the experimental precision.

Finally, due to the QCD corrections, the semi-leptonic  effective vector operator for addressing the $b\ra s\mu\mu$ anomaly gets about $\sim +10\%$ enhancement at low energy\cite{Aebischer:2018acj}.
However, all the tree-level $2q2l$ vectors operators listed in  Table \ref{tab:NCTL_list} and Table \ref{tab:CCTL_list}, as the constraint, also get roughly the same enhancement factor. Therefore, we do not consider this RGE running factor at this moment.

Next, we move on to consider the tree-level effects from the doublet leptoquark.
The non-zero $\lambda^D_{\tau b}$ and $\lambda^D_{\mu b}$, required for addressing $\tri a_{e,\mu}$ and neutrino data, lead to the following relevant low energy effective Hamiltonian,
\beq
{\cal H}^{D}_{eff} \supset \frac{[ \bar{b}\gamma^\alpha\PR b ]}{4M_D^2}\left[ |\lambda^D_{\tau b}|^2 (\bar{\tau}\gamma_\alpha \PL \tau)
  + |\lambda^D_{\mu b}|^2 (\bar{\mu}\gamma_\alpha\PL \mu)+ \lambda^D_{\mu b}(\lambda^D_{\tau b})^* (\bar{\tau}\gamma_\alpha \PL\mu)
+ \lambda^D_{\tau b}(\lambda^D_{\mu b})^* (\bar{\mu}\gamma_\alpha\PL \tau)
 \right]\,,
\eeq
and its neutrino counter part as well, see Eq.(\ref{eq:H_eff}). However, the constraint on these operators are rather weak\cite{Carpentier:2010ue} and can be ignored.


\subsection{SM $Z^0$ couplings  }
\label{sec:Zcoupling}
Because $b'_{L,R}$ are charged under $U(1)_Y$ hypercharge, they interact with the $Z^0$ boson.
In the interaction basis\footnote{Here we temporarily switch back to earlier notation that $b'_{L,R}$ represent the interaction basis. }, the SM $Z^0$ interaction for the down quark sector is
\beq
{\cal L} \supset \frac{g_2}{c_W} \left[  g_L^{SM} \sum_{i=1}^3 \bar{d}_{Li} \gamma^\alpha d_{L i}
+g_R^{SM}\sum_{i=1}^3 \bar{d}_{Ri}  \gamma^\alpha d_{R i} +
g_R^{SM}(\bar{b}'_L \gamma^\alpha b'_L + \bar{b}'_R \gamma^\alpha b'_R) \right] Z_\alpha\,,
\eeq
where $g_R^{SM}=\frac{s_W^2}{3}\simeq 0.077$, $g_L^{SM}=(-\frac{1}{2}+\frac{s_W^2}{3})\simeq  -0.423$,  $s_W= \sin\theta_W$, and $\theta_W$ is the Weinberg angle.  If we denote $b'$ as $d_4$, then the above expression can be neatly written as
\beq
\frac{g_2}{c_W} \left[  g_L^{SM} \sum_{i=1}^4 \bar{d}_{Li} \gamma^\alpha d_{L i}
+g_R^{SM}\sum_{i=1}^4 \bar{d}_{Ri}  \gamma^\alpha d_{R i}
+\frac{1}{2}(\bar{b}'_L \gamma^\alpha b'_L)  \right] Z_\alpha\,.
\eeq
When rotating into the mass basis, due to the unitarity of the four-by-four  $U^d_{L,R}$, it becomes
\beq
\frac{g_2}{c_W} \left[   \sum_{\alpha=s,d,b,b'} \bar{d}_{\alpha} \gamma^\alpha (g_L^{SM} \hat{L}+ g_R^{SM} \hat{R} ) d_{\alpha}\right] Z_\mu + \frac{g_2}{2 c_W} \sum_{\alpha,\beta=s,d,b,b'}
\kappa_{\alpha\beta}\left[  (\bar{d}_\alpha \gamma^\alpha \hat{L} d_\beta)  \right] Z_\alpha\,,
\eeq
where $\PL=(1-\gamma^5)/2$,  $\PR=(1+\gamma^5)/2$, and
\beq
\kappa_{\alpha\beta}\equiv (U^d_L)_{\alpha 4}[(U^d_L)_{\beta 4}]^* \,.
\eeq

Using the CP-conserving parametrization introduced in Eq.(\ref{eq:R4}), we have
\beq
\kappa_{sd} = \kappa_{ds}= s_1 s_2 c_2 c_3^2\,,\;
\kappa_{sb} = \kappa_{bs}= s_2 s_3 c_3\,,\;
\kappa_{bd} = \kappa_{db}= s_1 s_3 c_2 c_3\,.
\eeq
It is clear that, with the presence of $b'_L$, the tree-level Flavor-Changing-Neutral-Current (FCNC) in the down sector is inevitable
unless at most one of $\theta_{1,2,3}$ being sizable. For simplicity, we assume one nonvanishing $(U^d_L)_{4d_F}$, where $F$ could be one of $d,s,b$, and all the others are zero.

Let's focus on that specific non-zero flavor diagonal $Z\mhyphen d_F\mhyphen\bar{d_F}$ coupling. The mixing with $b'$ leads to
\beq
g^{SM}_{d_F,R} \Rightarrow  g^{SM}_{d_i,R}\,,\,\,
g^{SM}_{d_F,L} \Rightarrow  g^{SM}_{d_i L} +\frac{1}{2} \left| (U^d_L)_{4d_F}\right|^2\,,
\eeq
The introduction of $b'_{L,R}$  leads to a robust prediction that $(g_{d_F L})^2<(g^{SM}_{d_F L})^2 $ and $(g_{d_F R})^2=(g^{SM}_{d_F R})^2$ for that down-type quark at the tree-level.
Namely, in this model, $A_F$ and $A_F^{FB}$ ( both $\propto [(g_{d_F L})^2 - (g_{d_F R})^2]$ ), and $\Gamma_{d_F}$ ($\propto [(g_{d_F L})^2 + (g_{d_F R})^2]$) are smaller than the SM prediction. This remind us the long standing puzzle of the bottom-quark forward-backward asymmetry, $A^b_{FB}$, which is $2.3 \sigma$ below the SM value\cite{Zyla:2020zbs}. However, if we pick $\theta_3$ to be nonzero, then the CAA cannot be addressed, see Eq.(\ref{eq:CKMA_req}).
Moreover, from our numerical study, only $\theta_1\neq 0$ is viable to satisfy all experimental limits. Thus we set $\theta_2=\theta_3=0$.
From Eq.(\ref{eq:CKMA_req} ), we have
 \beq
 | s_1 | \simeq 0.039(7)\,,
\label{eq:theta2}
\eeq
and
\beq
\wt{V}_{ub'}=s_1 \wt{V}_{ud}  \simeq 0.03798\,,\,
\wt{V}_{cb'}=s_1 \wt{V}_{cd}  \simeq -0.00883\,,\,
\wt{V}_{tb'}=s_1 \wt{V}_{td} \simeq 0.00033\,,
\eeq
 if we take $\theta_1$ to be positive.
This predicts $g_{dL}= g_{dL}^{SM}+s_1^2/2$ at tree-level, but with negligible effect.

On the other hand, one may wonder whether  the introduction of
 $\lambda^T_{\tau b}, \lambda^T_{\mu b}$ and $\lambda^D_{\tau b}$ can lead to sizable non-oblique radiactive
$Z\mhyphen b\mhyphen \bar{b}$ vertex corrections and address both the $A^b_{FB}$ anomaly and $R_b$ with the later one agrees with the SM prediction.
To address the $A^{FB}_b$ anomaly and $R_b$  simultaneously, one needs to increase $g_{bR}^2$ and decrease $g_{bL}^2$ at the same time.
We perform the 1-loop calculation in the $\overline{MS}$ scheme and the on-shell renormalization, and obtain the UV-finite result:
\beqa
\delta g^b_L &\simeq & {|\lambda^T_{\tau b}|^2+|\lambda^T_{\mu b}|^2\over 64\pi^2}\left[ \left(-1+\frac{5}{3}s_W^2\right)\frac{1}{9 \beta_Z}
-s_W^2{ 2\ln \beta_Z +\ot+i\pi/2 \over 3 \beta_Z} \right]\,,\nonr\\
\delta g^b_R &\simeq & {|\lambda^D_{\tau b}|^2 \over 64\pi^2}\left[ \left(-\frac{1}{3}s_W^2\right)\frac{1}{9 \beta_Z}
+s_W^2{ 2\ln \beta_Z +\ot+i\pi/2 \over 3 \beta_Z}\right]\,,
\eeqa
where $\beta_Z=(M_{LQ}/m_Z)^2$. Note the diagrams with $Z$ attached to the lepton in the loop have imaginary parts, and this is due to that the lepton pair can go on-shell. Unfortunately, these loop corrections are too small, $|\delta g^b_{L,R}|\sim {\cal O}(10^{-5})\times |\lambda^{T,D}|^2$, to be detectable. From the above, we conclude that, barring the tree-level FCNC $Z$ coupling, both $A_{FB}^b$ and $R_b$ receive no significant modification in this model.
Of course, future  Z-pole electroweak precision measurements\cite{Abada:2019zxq, Baer:2013cma, CEPCStudyGroup:2018ghi}   will remain the ultimate judge.
If the $A_{FB}^b$ deviation endures, one must go beyond this model.
We note by passing that more complicated model constructions are possible to address the $A^b_{FB}$ anomaly.
For example, this anomaly can be addressed by adding an anomaly-free set of chiral exotic quarks and leptons\cite{Chang:1999zc,Chang:1998pt}, or the vector-like quarks\cite{Choudhury:2001hs, Cheung:2020vqm, Crivellin:2020oup} to the SM.

\subsection{$B_s-\overline{B_s}$ mixing }
\label{sec:BSBSmixing}

One important constraint on the parameters related to $b\ra s \mu\mu$ transition comes from the $B_s\mhyphen \overline{B}_s$ mixing.
In our model, the box diagrams with leptoquark $T$ and lepton running in the loop give a sole effective Hamiltonian
\beq
{\cal H}^{B\bar{B}}_{eff}= {\cal C}_{B\bar{B}} \left(\bar{s}\gamma^\alpha \PL b \right) \left(\bar{s}\gamma_\alpha \PL b \right)\, +H.c.
\eeq
The Wilson coefficient can be easily calculated to be
\beq
{\cal C}_{B\bar{B}}\simeq  { | \lambda^T_{\tau b}|^2 |\lambda^T_{\tau s}|^2  \over 128\pi^2 M_T^2} \left(1+\frac{1}{4}\right)\,,
\eeq
where the one-forth in the parenthesis is the contribution from $T^{\pm \ot}$.
Note that this ${\cal C}_{B\bar{B}}$ and the SM one are of the same sign, and it increases $\tri M_s$, the mass difference between $B_s$ and $\bar{B}_s$. But, the central value of the precisely measured $\tri M_s=17.757(21) \mbox{ps}^{-1}$\cite{Amhis:2019ckw} is smaller than the SM one. On the other hand, the SM prediction has relatively large, $\sim 10\%$\cite{Bona:2006sa, Altmannshofer:2020axr}, uncertainties arising from the hadronic matrix elements.
If putting aside the hadronic uncertainty, this tension could be alleviated in this model by the extended CKM, $V^*_{ts}V_{tb}\Rightarrow \wt{V}^*_{ts}\wt{V}_{tb}=(V^*_{ts}c_2-V^*_{tb}s_2s_3)V_{tb}c_3$, which reduces the SM prediction.
However, it does not work because we set $\theta_3=\theta_2=0$ as discussed in Sec.\ref{sec:Zcoupling}.   Instead, we use the 2$\sigma$  range to constraint the model parameters.
 Following Refs.\cite{Arnan:2019uhr,Huang:2020ris}, the NP contribution can be constrained to be
\beq
\left| 1+\frac{0.8 {\cal C}_{B\bar{B}}(\mu_{LQ}) }{{\cal C}^{SM}_{B\bar{B}}(\mu_b)}\right|-1 = -0.09\pm 0.08\,,\; \mbox{at 1 $\sigma$ C.L.}\,,
\label{eq:delMS}
\eeq
where the factor $0.8$ is the RGE running effect from $\mu_{LQ}\simeq 1\mtev$ to $\mu_b$, and ${\cal C}^{SM}_{B\bar{B}}(\mu_b)\simeq 7.2\times 10^{-11}\mgev^{-2}$ is the SM value at the scale $\mu_b$.
From the above, we obtain
\beq
| \lambda^T_{\tau b} (\lambda^T_{\tau s})^*| <0.0798 \left( M_T \over \mbox{TeV}\right)\,,
\eeq
so that Eq.(\ref{eq:delMS}) can be inside the $2\sigma$ confidence interval.

Together with Eq.(\ref{eq:bsmm_req}) and assuming that $\left(\wt{V}^*_{cb}\lambda^T_{\mu b}+ \wt{V}^*_{cb'}\lambda^T_{\mu b'} \right)=0$, the requirement of the tree-level $\mu\mhyphen e$ universality in $b\ra c l\nu$, one sees that
\beq
|\lambda^T_{\mu b'}|> 3.401 \left[{ C_{10}^\mu(=- C_9^\mu)\over 0.41} \right]^{\frac{1}{2}} \left( {M_T \over \mtev}\right)\,.
\eeq
From this inequality, $M_T$ must be around or smaller than $\mtev$ for this model parameter to stay in the perturbative region.
However, this statement  strongly relies on the SM prediction of $\tri M_s$ and the values of $C^\mu_{9,10}$.

\subsection{  $B\ra K^{(*)}\nu\bar{\nu}$  }
In this model, the  $B\ra K^{(*)}\nu\bar{\nu}$  transition can be mediated by $T$ at tree-level and described by an effective Hamiltonian
\beq
{\cal H}_{eff}^{NP} \supset - \frac{\alpha G_F}{\sqrt2 \pi}  \wt{V}_{tb} \wt{V}_{ts}^* C^\nu_{ij} \left[\bar{s} \gamma^\alpha \PL b\right] \left[\bar{\nu}_i \gamma_\alpha \left(1-\gamma^5\right)\nu_j\right] + H.c\,,
\eeq
where
\beq
C^\nu_{\tau \mu}= { \pi\over \alpha G_F \wt{V}_{tb} \wt{V}_{ts}^* }{\lambda^T_{\mu b} (\lambda^T_{\tau s})^*  \over 4 \sqrt2 M_T^2}\,,\;
C^\nu_{\tau \tau}= { \pi\over \alpha G_F \wt{V}_{tb} \wt{V}_{ts}^* }{\lambda^T_{\tau b} (\lambda^T_{\tau s})^*  \over 4 \sqrt2 M_T^2}\,,
\eeq
and all other Wilson coefficients are zero.
Following \cite{Buras:2014fpa}, the normalized branching ratio for $B\ra K^{(*)}\nu \bar{\nu}$ is given by
\beq
R^\nu_{K^{(*)}} 
= { |C^\nu_{SM} + C^\nu_{\tau\tau}|^2 + 2|C^\nu_{SM}|^2 + | C^\nu_{\tau \mu}|^2 \over 3|C^\nu_{SM}|^2 }\,,
\eeq
where the SM contribution $C^\nu_{SM}\simeq -6.35$ and it is dominated by the Z-penguin.
Using $\wt{V}_{tb} \wt{V}_{ts}^*=-0.03975$ and the current 90\%C.L. limits $R_K^\nu <3.9$ and $R_{K^*}^\nu <2.7$ given by \cite{Belle:2017oht}, we obtain a constraint
\beq
\left|\lambda^T_{\tau b} (\lambda^T_{\tau s})^*+0.03868\right|^2+ \left|\lambda^T_{\mu b} (\lambda^T_{\tau s})^* \right|^2 <  9.127\times 10^{-3}\times \left({M_T \over \mtev}\right)^2\,.
\eeq
This inequality alone implies  $|\lambda^T_{\mu b} (\lambda^T_{\tau s})^*|<0.0955$, which is slightly stronger
than the constraint derived from ${\cal B}(B^+\ra K^+\mu^+ \tau^-)<4.5\times 10^{-5}$, see Table \ref{tab:NCTL_list}.

\subsection{$\tau \ra \mu(e) \gamma$}
\label{sec:tau_mu_p}
Since we also set $\lambda^T_{e d_i}=0 (d_i=d,s,b,b')$, there is no $\mu\ra e \gamma$ transition at 1-loop level by default. Therefore, we only focus on the constraint from $\tau \ra \mu \gamma$ and $\tau \ra e \gamma$.
\begin{figure}[htb]
\centering
\includegraphics[width=0.65\textwidth]{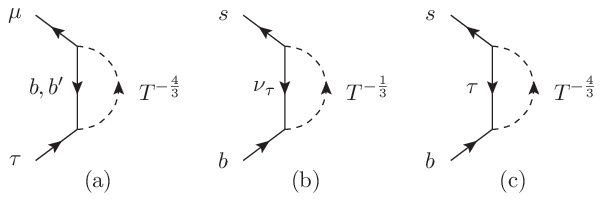}
\caption{ The Feynman diagram for (a) $\tau \ra \mu \gamma$, and (b,c) $b\ra s \gamma(g)$ transition. The external photon (gluon), which is not shown in the illustration, can attach to any charged(color) line in the loop. }
\label{fig:bsg}
\end{figure}

The rare $\tau \ra \mu \gamma$ transition can be induced when both  $\lambda^T_{\tau b'}$ and $\lambda^T_{\mu b'}$ are nonzero.
The 1-loop diagram are shown in Fig.\ref{fig:bsg}(a).

The dipole $\tau \ra \mu \gamma$ amplitude can be parameterized as
\beq
i {\cal M}^\mu = i\, \left[ \bar{\mu}\left( i \sigma^{\mu\nu}k_\nu \right) ( d^{\tau \mu}_R \PR +d^{\tau \mu}_L \PL  )\tau \right]\,,
\eeq
where $k$ is the photon momentum transfer.
If ignoring the muon mass, the partial decay width is given as\cite{Chang:2005ag}
\beq
\Gamma(\tau \ra \mu \gamma) \simeq \frac{m_\tau^3}{16 \pi} ( |d^{\tau \mu}_R|^2 + |d^{\tau \mu}_L|^2)\,.
\eeq

Since the leptoquark $T(D)$ only couples to the LH(RH) charged leptons, it contributes solely to $d^{\tau \mu}_{R(L)}$.

If ignoring the charged lepton masses in the loop,  the  dipole transition coefficient can be easily calculated to be
\beqa
d^{\tau \mu}_R = { e N_c  m_\tau \over 16 \pi^2 M_T^2 }\{ && \lambda^T_{\tau b'} (\lambda^T_{\mu b'})^* \beta'_T
[  - Q_T R_S(\beta'_T) +Q_{(b')^c} R_F(\beta'_T)]\nonr\\
&+&  \lambda^T_{\tau b} (\lambda^T_{\mu b})^* \beta_T
[  - Q_T R_S(\beta_T) +Q_{(b)^c} R_F(\beta_T)]\; \}\,.
\eeqa
In the above, $\beta'_T \equiv (M_T/M_{b'})^2$, $\beta_T \equiv (M_T/m_b)^2$, $Q_T$($Q_{(b^{(')})^c}$) is the electric charge of the scalar(fermion) in the loop,
and the loop functions,
\beqa
R_S(x) &=& {2+3x-6x^2+x^3+6 x \ln x \over 12(1-x)^4}\,,\;\mbox{and}\nonr\\
R_F(x)&=& \frac{R_S(1/x)}{x}={1-6x+3x^2+2x^3- 6 x^2 \ln x \over 12(1-x)^4}\,,
\label{eq:mueg_loop}
\eeqa
correspond to the contributions where the external photon attached to the scalar and fermion line in the loop, respectively.
Both functions have the same limit $1/24$ when $x\ra 1$.
When $x\gg 1$, $R_S(x)\ra 1/12x$ and $R_F(x)\ra 1/6x $.
Note the fermionic and bosonic contributions have opposite signs, and the charged fermion in the loop is the anti-$b^{(')}$.

Similarly, the contribution from the diagram with leptoquark $D^{-\tth}$ and $b^{(')}$ running in the loop yields
\beqa
d^{\tau \mu}_L = { e N_c  m_\tau \over 32 \pi^2 M_D^2 } && \left\{ \lambda^D_{\tau b'} (\lambda^D_{\mu b'})^* \beta'_D
\left[  +\tth  R_S(\beta'_D) - \ot  R_F(\beta'_D)\right]\right.\nonr\\
&+& \left. \lambda^D_{\tau b} (\lambda^D_{\mu b})^* \beta_D
\left[  +\tth R_S(\beta_D) -\ot  R_F(\beta_D)\right]\; \right\}\,.
\eeqa
As discussed in Sec.\ref{sec:g-2}, we set $\lambda^D_{\mu b'}=0$  in ( Sol-2) to avoid the dangerous $\mu\ra e \gamma$ transition\footnote{It will be clear that this is the case for  fitting neutrino oscillation data successfully.}.
Then, we obtain
\beq
d^{\tau \mu}_L \simeq { e N_c  m_\tau \over 32 \pi^2 M_D^2 }\lambda^D_{\tau b} (\lambda^D_{\mu b})^*\beta_D\left[\tth R_S(\beta_D)-\ot R_F(\beta_D) \right] \simeq 0\,,
\eeq
and due to the accidental cancellation in the squared bracket, it vanishes in the limit of $m_b \ll m_D$ in this case. Therefore, only $d^{\tau \mu}_R$ needs to be taken into account.
From $\tau_\tau=(290.3\pm 0.5)\times 10^{-15}s$\cite{Zyla:2020zbs}, the branching ratio of this rare process is
\beqa
{\cal B}(\tau\ra \mu \gamma)&=& \Gamma(\tau \ra \mu \gamma)/\Gamma_\tau \nonr\\
&\simeq& 5.14\times 10^{-6}
 \left|  \lambda^T_{\tau b'} (\lambda^T_{\mu b'})^* \frac{\beta'_T}{3}[4R_S(\beta'_T)+ R_F(\beta'_T)]    + \frac{\lambda^T_{\tau b} (\lambda^T_{\mu b})^*}{6} \right|^2
  \times \left({\mbox{TeV}\over M_T}\right)^4\,.
\eeqa
The current experimental bound, ${\cal B}(\tau\ra \mu \gamma)< 4.4\times 10^{-8}$\cite{Aubert:2009ag}, sets an upper bound
\beq
\left|  \lambda^T_{\tau b'} (\lambda^T_{\mu b'})^* \frac{4\beta'_T}{3}[R_S(\beta'_T)+ R_F(\beta'_T)]    + \frac{\lambda^T_{\tau b} (\lambda^T_{\mu b})^*}{6} \right|
   < 0.092  \times \left({ M_T \over \mbox{TeV}}\right)^2\,.
\eeq

On the other hand, if both $\lambda^D_{\tau b'}$ and $\lambda^D_{e b'}$ present, we have
\beq
d^{\tau e}_L  \simeq { e N_c  m_\tau \over 16 \pi^2 M_D^2 } {\lambda^D_{\tau b'} (\lambda^D_{\mu b'})^* \beta'_D \over 6}
\left[  2 R_S(\beta'_D) -  R_F(\beta'_D)\right]
\eeq
for $\tau\ra e\gamma$ transition.
From ${\cal B}(\tau\ra e \gamma)< 3.3 \times 10^{-8}$ \cite{Aubert:2009ag}, it gives a weak bound
\beq
\left| \lambda^D_{\tau b'} (\lambda^D_{e b'})^*\right|< 16.93  \times \left({ M_D \over \mtev }\right)^2
\eeq
for $\beta'_D=(1.0/1.5)^2$.

\subsection{$b\ra s \gamma$ }
\label{sec:bsg}
Similar to the previous discussion on $\tau\ra \mu\gamma$, the $b \ra s \gamma (g)$ transition can be induced when both  $\lambda^T_{\tau b}$ and $\lambda^T_{\tau s}$ are nonzero, see Fig.\ref{fig:bsg}(b,c).
Moreover, the fermion masses, $m_\tau$ or $m_{\nu_\tau}$, can be ignored, which corresponds to the $\beta_T\gg 1$ limit.
From Eq.(\ref{eq:mueg_loop}), it is easy to see that  $R_S(x) \ra 1/(12 x)$ and $R_F(x)\ra 1/(6x)$ when $x\gg 1$.
Therefore, by plugging in the electric charges in the loop, the $b\ra s \gamma$ transition amplitude can readily read as
\beq
i{\cal M}^\mu(b\ra s \gamma) \simeq
i {e m_b \over 16\pi^2} {\lambda^T_{\tau b} (\lambda^T_{\tau s})^* \over 12 m_T^2}\left[
-\frac{1}{2}\left(-\ot\right) -\left(-\fth \right) + 2(1)\right] \left[ \bar{s}\left( i \sigma^{\mu\nu}k_\nu \right)\PR b \right]\,.
\eeq
The first one-half factor comes from the $T^{-\ot}$ Yukawa couplings which associate with the $(1/\sqrt{2})$ normalization, see Eq.(\ref{eq:LQYukawa_coupling}), and the factor 2 in the last term comes from the anti-tau contribution.
We also need to consider $b\ra s g$ transition because the RGE running will generate the $b\ra s \gamma$ operator at the low energy.
Because the gluon can only couple to the leptoquark, the amplitude reads
\beq
i{\cal M}^\mu(b\ra s  g) \simeq
i {g_s m_b \over 16\pi^2} {\lambda^T_{\tau b} (\lambda^T_{\tau s})^* \over 12 m_T^2}\left[
-\frac{1}{2}  -\left( 1 \right)\right] \left[ \bar{s}\left( i T^{(a)}\sigma^{\mu\nu}k_\nu \right)\PR b \right]\,.
\eeq
where $T^{(a)}$ is the $SU(3)_c$ generator.

Conventionally, the relevant effective Hamiltonian is given as
\beq
{\cal H}^{b\ra s\gamma}_{eff} = -\frac{4 G_F}{\sqrt{2}} V_{tb} V_{ts}^* ( {\cal C}_7 {\cal O}_7 + {\cal C}_8 {\cal O}_8)\,,
\eeq
with
\beqa
{\cal O}_7 &=& \frac{e}{16\pi^2} m_b \bar{s} \sigma^{\mu\nu}\PR b F_{\mu\nu}\,,\nonr\\
{\cal O}_8 &=& \frac{g_s}{16\pi^2} m_b \bar{s}_\alpha \sigma^{\mu\nu}\PR T^{(a)}_{\alpha\beta} b_\beta G^{(a)}_{\mu\nu}\,,
\eeqa
In our model, the Wilson coefficients for ${\cal O}_7$ and ${\cal O}_8$ can be identified as
\beqa
{\cal C}_7 \simeq {1\over 2\sqrt{2} G_F  V_{tb} V_{ts}^* } {7 \lambda^T_{\tau b} (\lambda^T_{\tau s})^* \over 48 M_T^2}\,,\nonr\\
{\cal C}_8 \simeq -{1\over 2\sqrt{2} G_F  V_{tb} V_{ts}^* } { \lambda^T_{\tau b} (\lambda^T_{\tau s})^* \over 8 M_T^2}\,.
\eeqa

The current experimental measurement, $Br^{Exp}(b\ra s\gamma)=(3.32\pm 0.15)\times 10^{-4}$\cite{Amhis:2016xyh}, and the SM prediction, $Br^{SM}(b\ra s\gamma)=(3.36\pm 0.23)\times 10^{-4}$\cite{Misiak:2015xwa,Misiak:2017woa}, agree with each other and set constraints on the NP contribution. Following Refs.\cite{Arnan:2016cpy,Arnan:2019uhr,Huang:2020ris}, we adopt the $2\sigma$ bound for the NP that
$|{\cal C}_7 +0.19 {\cal C}_8|\lesssim 0.06$, which leads to $|\lambda^T_{\tau b} (\lambda^T_{\tau s})^*|< 0.55$. This limit is much weaker than the one obtained from $\tri M_s$.

\subsection{Neutrino oscillation data }
\label{sec:num_fit}
As discussed before, the vanishing $\lambda^T_{e d_i}$ are preferred by phenomenological consideration.
It is then followed by a robust prediction  that ${\cal M}_{ee}=0$, and the neutrinoless double beta decay mediated by ${\cal M}_{ee}$ vanishes as well. Therefore, the neutrino mass is predicted to be the normal hierarchical(NH).
Later we should discuss the consequence if this vanishing-$\lambda^T_{e d_i}$ assumption is relaxed.

A comprehensive numerical fit to the neutrino data is unnecessary to understand the physics, and it is beyond the scope of this paper as well. For simplicity, we assume all the Yukawa couplings are real, and thus all the CP phases in the neutrino mixings vanish.
However,  this model has no problem to fit the CP violation phases of any values once the requirement of
all the Yukawa couplings being real is lifted.
Moreover, to adhere to the philosophy of using the least number of real parameters,  only two more Yukawa couplings,
 $\lambda^D_{\tau b'}$ and  $\lambda^D_{\tau b}$,  are introduced to  fit the neutrino data\footnote{Note that we have not employed $\lambda^T_{\tau s}$ in the neutrino data fitting yet.}. Together with $\mbox{MinS}_T$, we make use of eleven Yukawa couplings, and the complete minimal set of parameter is
\beq
\mbox{MinS} = \mbox{MinS}_T \cup \left\{ \lambda^D_{e b'}, \lambda^D_{\tau b'}, \lambda^D_{\mu b}, \lambda^D_{\tau b}, \lambda^S_{e b'}, \lambda^S_{\mu b} \right\}\,.
\eeq

Assuming that $M_D\simeq M_T\simeq M_{LQ}$,  the neutrino mass matrix takes the form
\beq
{\cal M}^\nu \simeq N^\nu
\left( \begin{array}{ccc}
 0 &  \lambda^D_{eb'} \lambda^T_{\mu b'}  & \lambda^D_{eb'} \lambda^T_{\tau b'} \\
\lambda^D_{eb'} \lambda^T_{\mu b'} &   2 \rho_b \lambda^D_{\mu b} \lambda^T_{\mu b} &
                          \rho_b(\lambda^D_{\mu b}\lambda^T_{\tau b}+ \lambda^D_{\tau b}\lambda^T_{\mu b})+ \lambda^D_{\mu b'} \lambda^T_{\tau b'} \\
\lambda^D_{eb'} \lambda^T_{\tau b'}  &  \rho_b (\lambda^D_{\mu b}\lambda^T_{\tau b}+ \lambda^D_{\tau b}\lambda^T_{\mu b})+ \lambda^D_{\mu b'} \lambda^T_{\tau b'} &
 2 \rho_b\lambda^D_{\tau b} \lambda^T_{\tau b}+ 2 \lambda^D_{\tau b'} \lambda^T_{\tau b'} \\
 \end{array}  \right)\,,
 \label{eq:nuM_expression}
\eeq
where $\rho_b = m_b/M_{b'}$ and $N^\nu= { 3 \mu_3 v_0 M_{b'}\over 32 \pi^2 M^2_{LQ}} $.
Note that the leptoquark Yukawa couplings are tightly entangled with the neutrino mass matrix.
For instance, $\lambda^T_{\tau b'}/\lambda^T_{\mu b'} = {\cal M}^\nu_{e\tau}/{\cal M}^\nu_{e\mu}$ is required by this minimal assumption.

For illustration, we consider an approximate  neutrino mass matrix\footnote{ It is just a randomly generated example for illustration.  There are infinite  ones with the similar structure. }
\beq
{\cal M}^\nu  \simeq  \left(
                        \begin{array}{ccc}
                          0 & 0.90792& 0.13812 \\
                             0.90792 & -2.4923 & -2.7643 \\
                                0.13812 & -2.7643 & -1.9353 \\
                        \end{array}
                      \right)\times 10^{-2} \mev\,,
                      \label{eq:Mnu_num}
\eeq
 with all elements being real.
It leads to the following mixing angles and mass squared differences
\beqa
&&\theta_{12} \simeq 33.0 ^\circ\,,\;
\theta_{23}\simeq 48.7^\circ\,,\;
\theta_{13}\simeq 8.6^\circ\,,\;
\delta_{CP}=0^\circ\,,\nonr\\
&&\tri m_{21}^2 \sim 7.47 \times10^{-5} \mbox{eV}^2\,,\;
\tri m_{31}^2 \sim 2.53\times 10^{-3} \mbox{eV}^2\,.
\eeqa
Note  all of the above values, except $\delta_{CP}$, are  inside the $1\sigma$  best fit (with SK atmospheric data)  range for the normal ordering given by \cite{Esteban:2020cvm},
\beqa
&&\theta_{12} \in ( 32.7- 34.21)^\circ\,,\;
\theta_{23} \in ( 48.0- 50.1)^\circ\,,\;
\theta_{13} \in ( 8.45- 8.69)^\circ\,,\;
\delta_{CP} \in (173-224)^\circ\,,\nonr\\
&&\tri m_{21}^2 \in (7.22-7.63)\times 10^{-5} \mbox{eV}^2\,,\;
\tri m_{31}^2 \in (2.489-2.543)\times 10^{-3} \mbox{eV}^2\,.
\eeqa

This example neutrino mass matrix captures the essential features of the current neutrino data.
A better fitting to the neutrino oscillation data, including the phase,  by using more(complex) model parameters is expected.

In order to reproduce the neutrino mass matrix, the second solution to $\tri a_{e,\mu}$, Eq.(\ref{eq:g2sol2}), must be adopted.
Because the first solution  $\tri a_{e,\mu}$, Eq.(\ref{eq:g2sol1}), requires $\mu_1 \gg M_{LQ}$ to render  mass matrix elements of about the same order, as shown in Eq.(\ref{eq:Mnu_num}).
All the best fit central values for $b\ra s\mu\mu$, $\tri a_{e,\mu}$, CAA, and the approximate neutrino mass matrix shown in Eq.(\ref{eq:Mnu_num}) can be easily accommodated with the specified non-vanishing parameters in the model.
 Since $(M_{b'}/m_{b})\gg 1$, $ \lambda^D_{\mu b}, \lambda^S_{e b'} >\sqrt{4\pi}$ are required if aiming for explaining the central values of $\tri a_{e,\mu}$.
However, reliable nonperturbative treatment is beyond the scope of this paper.
Instead, we consider the $1\sigma$ ranges of best fit for $b\ra s \mu \bar{\mu}$  and $\tri a_e$.
As an example, below is  one of the viable sets of model parameters  for  $\tri a_e =-5.1[1.8]\times 10^{-13}$ and $C_9=- C_{10} =-0.351$\footnote{The values of $\tri a_e$ correspond  to the 1 $\sigma$ boundaries of $\tri a_e^{Cs}[\tri a_e^{Rb}]$, and
the $C_{9(10)}$ is close to the fitted value by only using the theoretically clean modes in \cite{Altmannshofer:2021qrr}.
All others are taken to be their central values. },
\beqa
M_{LQ}=1.0\, \mtev\,,\, M_{b'}=1.5\, \mtev\,,\, \mu_1=2.298[-0.818]\,\mtev\,,\, \mu_3= 0.513\,\mkev\,,s_1= 0.039\nonr\\
\lambda^T_{\mu b'}=-3.34514 \,,\,
\lambda^T_{\tau b'}=-0.508902\,,\,
\lambda^T_{\mu b}=-0.728492\,,\nonr\\
\lambda^T_{\tau b}=0.929742\,,\,
\,\lambda^T_{\tau s}=-0.076018\,,\nonr\\
\lambda^D_{e b'}=-0.00151\,,\,
\lambda^D_{\mu b}=3.4\,,\,
\lambda^D_{\tau b'}=0.00760\,,\,
\lambda^D_{\tau b}=-0.58293\,,\nonr\\
\lambda^S_{e b'}= 3.4\,,\,
\lambda^S_{\mu b}=-0.088264[0.24798]\,,\,
\label{eq:num_result}
\eeqa
and all the other Yukawa couplings are set to zero.
This specific set of model parameters also predicts $\tri M_s \simeq 1.06\, \tri M_s^{SM}$, $R^\nu_{K^{(*)}}=1.21$,
\beq
{\cal B}(\tau \ra \mu \gamma)=6.0\times 10^{-9}\,,\;
{\cal B}(\tau \ra e \gamma)=1.5 \times 10^{-20}\,,\;
{\cal B}(B^+\ra K^+\mu^+ \tau^-) = 1.4\times 10^{-5}\,,
\eeq
and pass all experimental limits we have considered, the last column in Tables \ref{tab:NCTL_list} and \ref{tab:CCTL_list}.

Note that $\lambda^T_{\mu b'}$, $\lambda^D_{\mu b}$, and $\lambda^S_{e b'}$  in Eq.(\ref{eq:num_result}) are close to but below the nonperturbative limit $\sqrt{4\pi}$.
However, this is because we want to use the minimal number of model parameters.
For instance, if the complex Yukawa is allowed, the degrees of freedom are doubled.
 Lowering $M_{b'}$ and increasing $\mu_1$ can both
make $\lambda^D_{\mu b}$ and $\lambda^S_{e b'}$ smaller as well. We have no doubt that a better fitting can be achieved in this model by using more  (complex) free parameters. But we are content with the demonstration about the ability of this model to accommodate all the observed anomalies and explain the pattern of the observed neutrino data with minimal number of real parameters.

\subsection{$0\nu\beta\beta$ decay}
\begin{figure}
\centering
\includegraphics[width=0.5\textwidth]{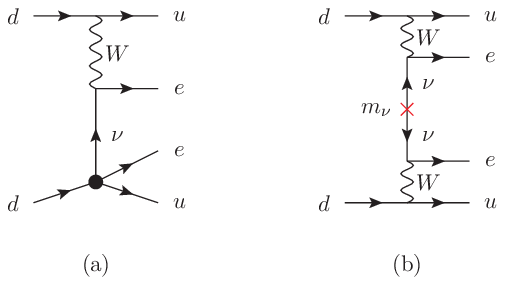}
\caption{ The Feynman diagrams for $0\nu\beta\beta$ decay (a) mediated by the $T\mhyphen D$ mixing, and (b) mediated by the neutrino Majorana mass.  }
\label{fig:0n2b}
\end{figure}
The mixing between $T$ and $D$ breaks the global lepton number, see Appendix \ref{sec:H_eff}. Here, we consider whether the neutrinoless double beta decay can be generated beyond the contribution from the neutrino Majorana mass.
From the low energy effective Hamiltonian, together with the SM CC interaction, the lepton number violating charged current operators can induce the $0\nu\beta\beta$ process via the diagram, see Fig.\ref{fig:0n2b}(a).
By order of magnitude estimation, the absolute value of the amplitude strength relative to the usual one mediated by the neutrino Majorana mass, Fig.\ref{fig:0n2b}(b),  is given by
\beq
\left| { {\cal M}_{TD} \over {\cal M}_{m_\nu} }\right| \simeq
{\lambda_T \lambda_D \mu_3 v_0 \over M_T^2 M_D^2} \frac{M_W^2 \langle p\rangle}{g_2^2 {\cal M}^\nu_{ee}}
\simeq 10^{-5} \left( \frac{1\, \mbox{TeV}}{M_{LQ}}\right)^4 \ll 1\,.
\eeq
In arriving the above value, we take $\mu_3=0.5\mkev$, $\langle p\rangle \sim 1\, \mmev$ for the typical average momentum transfer in the $0\nu\beta\beta$ process and assume ${\cal M}^\nu_{ee} \sim 0.01\mev$ for comparison.
Thus, this tree-level process mediated by $T$ and $D$ can be ignored.

\subsection{ A recap}
After taking into account all the phenomenological limits, we found the following simple assignment with minimal number of real parameters,
\beqa
&& M_{LQ}\simeq 1.0\, \mtev\,,\,M_{b'}= 1.5\, \mtev\,,\,\mu_1 = 2.3[-0.82]\, \mtev\,,\, \mu_3 \simeq 0.5\, \mkev\,, \nonr\\
&& \mu_1 \lambda^S_{e b'}\lambda^D_{e b'} = -12[4] \mgev\,,\;
\mu_1 \lambda^S_{\mu b}\lambda^D_{\mu b}= -690 \mgev\,,\nonr\\
&& \theta_2 =\theta_3=0\,,\;\sin\theta_1 = 0.039\,,\nonr\\
&&\lambda^T_{\mu b'} \simeq  -3.3 \,,\,\lambda^T_{\tau b'} \simeq -0.51 \,,\,\lambda^T_{\mu b} \simeq- 0.72 \,,\,
\lambda^T_{\tau b}\simeq 0.93\,,\,\lambda^T_{\tau s} \simeq  -0.08 \,,\nonr\\
&& \lambda^D_{e b'}\simeq -0.002\,,\, \lambda^D_{\mu b}\simeq  3.4\,,\,\lambda^D_{\tau b'}\simeq 0.008\,,\, \lambda^D_{\tau b}\simeq -0.58\,,
\label{eq:recap_range}
\eeqa
is able to accommodate $\tri a_e^{Cs} [\tri a_e^{Rb} ] $, $\tri a_\mu$, the Cabibbo angle, and $b\ra s \mu\mu$  anomalies
 simultaneously. Moreover, the resulting neutrino mass pattern is very close to the observed one.

\section{Discussion}
\label{sec:discussion}

\subsection{Neutrino mass hierarchy and  neutrinoless double beta decay}
\label{sec:parameter}
Because we set $\lambda^T_{e d_i}=0 (i=d,s,b,b')$, the neutrino mass element $M^\nu_{ee}$ vanishes and the neutrino mass is of the NH type.
Since we have used $\lambda^S_{e b'}$  to explain the observed $\tri a_e$, adding $\lambda^T_{e b'}$ is  the minimal extension to yield a non-zero $M^\nu_{ee}$.
Together with  $\lambda^T_{\mu b}$ and $\lambda^T_{\mu b'}$, the augmentation of $\lambda^T_{e b'}$ leads to an effective Hamiltonian,
\beq
H \supset \frac{{\cal C}_{\mu e}}{4M_T^2}(\bar{u}\gamma^\mu \PL u)(\bar{\mu}\gamma_\mu \PL e) +H.c.\,,
\eeq
where
\beq
{\cal C}_{\mu e} = \lambda^T_{eb'} (\wt{V}_{ub'})^*\left[ \wt{V}_{ub}(\lambda^T_{\mu b})^* +\wt{V}_{ub'}(\lambda^T_{\mu b'})^*\right]\,. \eeq
Numerically,
\beq
{\cal C}_{\mu e} \simeq  1.44\times10^{-3} \lambda^T_{eb'}(\lambda^T_{\mu b'})^*+  1.37\times10^{-4} \lambda^T_{eb'}(\lambda^T_{\mu b})^* \,,
\label{eq:mu-e-con-C}
\eeq
if we set $s_1=0.039$ and $\wt{V}_{ub}=0.00361$.

The Wilson coefficient is severely constrained,  ${\cal C}_{\mu e}<9.61\times 10^{-5}$, from the experimental limit of $\mu\mhyphen e$ conversion rate\cite{Dohmen:1993mp, Carpentier:2010ue}.
Namely,   $\lambda^T_{eb'}(\lambda^T_{\mu b'})^* \lesssim 0.07$ unless the cancellation is arranged in Eq.(\ref{eq:mu-e-con-C}).
Moreover, in this model, the ratio of neutrino mass element $M^\nu_{ee}$ to $M^\nu_{e\mu}$, see Eq.(\ref{eq:nuM_expression}),
\beq
\frac{M^\nu_{ee}}{M^\nu_{e\mu}} ={ 2 M_{b'} \lambda^D_{eb'}\lambda^T_{eb'} \over M_{b'} \lambda^D_{eb'}\lambda^T_{\mu b'} }
= \frac{2 \lambda^T_{eb'}}{\lambda^T_{\mu b'}}\,,
\eeq
should be around $\sim {\cal O}(1)$  and  $\sim {\cal O}(10)$ for the NH and Inverted Hierarchy (IH) type, respectively.
Since we need $\lambda^T_{\mu b'}\sim {\cal O}(1)$ to accommodate the b-anomalies, that implies $\frac{M^\nu_{ee}}{M^\nu_{e\mu}} \lesssim 0.07$. Thus, even if we include a non-zero $\lambda^T_{eb'}$ to generate the $ee$-component of $M_\nu$, the neutrino mass is still of the NH type, and roughly $|M^\nu_{ee}| \lesssim 3\times 10^{-4} \mev$. The  precision required is beyond the capabilities of the near future  experiments\cite{Tornow:2014vta}.

\subsection{Some phenomenological consequences at the colliders}

The smoking gun signature of this model will be the discovery of  $b'$ and the three scalar leptoquarks.
Once their quantum numbers are identified, the gauge invariant allowed Yukawa couplings and the mechanisms to address the anomalies discussed in the paper follow automatically.
The collider physics of leptoquarks  have been extensively studied before, and thus we do not have much to add. The readers interested in this topic are referred to the comprehensive review \cite{Dorsner:2016wpm} and the references therein.
In the paper, we should concentrate on the flavor physics at around or below the Z pole.

However, it is worthy to point out the nontrivial decay branching ratios of the exotic color states.
If we assume the mixings among the leptoquarks are small, their isospin members should be approximately degenerate in mass.
Then,  the decays are dominated by 2-body decay with two SM fermions in the final states.
From the Yukawa couplings shown in Eq.(\ref{eq:num_result}), the decay branching ratio of leptoquarks can be easily read.
For $T^{-\ot}$, its decay branching ratios are
\beqa
&&{\cal B}(T^{-\ot} \ra b \nu_{\mu}) \simeq 18.9\%\,,\;
{\cal B}(T^{-\ot} \ra b \nu_{\tau}) \simeq 30.9\%\,,\;
{\cal B}(T^{-\ot} \ra s \nu_{\tau}) \simeq 2.1\times 10^{-3}\,,\nonr\\
&&{\cal B}(T^{-\ot} \ra \tau t ) \simeq 30.8\%\,,\;
{\cal B}(T^{-\ot} \ra \mu t ) \simeq 18.9\%\,,\;
{\cal B}(T^{-\ot} \ra \tau  c) \simeq 2.0\times 10^{-3}\,.
\eeqa
For $T^{\tth}$ and $T^\ft$, the corresponding  decay branching ratios are
\beqa
&&{\cal B}(T^\tth \ra t \nu_{\mu}) \simeq 37.9\%\,,\;
{\cal B}(T^\tth \ra t \nu_{\tau}) \simeq 61.7\%\,,\;
{\cal B}(T^\tth \ra c \nu_{\tau}) \simeq 0.4\%\,,\nonr\\
&&{\cal B}(T^{-\ft} \ra b \mu^- ) \simeq 37.9\%\,,\;
{\cal B}(T^{-\ft} \ra b \tau^-) \simeq 61.7\%\,,\;
{\cal B}(T^{-\ft} \ra s \tau^-) \simeq 0.4\%\,.
\eeqa
Finally, we have
\beqa
&&{\cal B}(D^{-\ot} \ra b \bar{\nu}_{\mu}) \simeq 97.1\%\,,\;
{\cal B}(D^{-\ot} \ra b \bar{\nu}_{\tau}) \simeq 2.9\%\,,\nonr\\
&&{\cal B}(D^{\tth} \ra b \mu^+ ) \simeq 97.1\%\,,\;{\cal B}(D^{\tth} \ra b \tau^+) \simeq 2.9\%\,,
\nonr\\
&&{\cal B}(S^{\tth} \ra b \mu^+) \simeq 100\%\,,
\eeqa
for $D$ and $S$ leptoquarks.

The dominate decay modes of $b'$ are
$b'\ra LQ + l$, and $ b'\ra W^- u_i (u_i=u,c,t)$  through the mixing of $\wt{V}_{u_i b'}$.
Comparing to $M_{b'}$, the masses of final state particles can be ignored.
The width for $ b'\ra W u_i$ is simply given by
\beq
\Gamma(b'\ra  u_i W^-)\simeq \frac{G_F |\wt{V}_{u_i b'}|^2 M_{b'}^3}{8\sqrt{2} \pi}\,,
\eeq
and
\beqa
\Gamma(b'\ra  \bar{\nu_i} T^{-\ot})\simeq \frac{|\lambda^T_{l_i b'}|^2 }{ 64 \pi}M_{b'}\left(1-\frac{M_T^2}{M_{b'}^2}\right)^2\,,\;
\Gamma(b'\ra  \ell^+_i T^{-\ft})\simeq \frac{|\lambda^T_{l_i b'}|^2}{ 32 \pi}M_{b'}\left(1-\frac{M_T^2}{M_{b'}^2}\right)^2\,,\nonr\\
\Gamma(b'\ra  \nu_i D^{-\ot})\simeq \Gamma(b'\ra  \ell^-_i D^{\tth})\simeq \frac{|\lambda^D_{l_i b'}|^2 }{ 64 \pi}M_{b'}\left(1-\frac{M_D^2}{M_{b'}^2}\right)^2\,,\nonr\\
\Gamma(b'\ra  \ell_i S^{\tth})\simeq \frac{|\lambda^S_{l_i b'}|^2 }{ 32 \pi}M_{b'}\left(1-\frac{M_S^2}{M_{b'}^2}\right)^2\,.\;
\eeqa
The Yukawa coupling between $b'$ and the SM Higgs is through the $b'\mhyphen d$ mixing. So the resulting  Yukawa coupling gets double suppression from the small $\theta_1$ and the ratio of $m_d/v_0$, and so the $b'\ra H d$ decay can be ignored. Similarly the decays of $b'\ra Z^0 d_i$ can be ignored as well.
By plugging in the parameters we found, the total decay width of $b'$ is   $\Gamma_{b'}\simeq 134.0\mgev$, and the branching ratios are
\beqa
{\cal B}( b'\ra u W^- ) \simeq 1.2\%\,,\;
{\cal B}( b'\ra c W^- ) \simeq 6\times 10^{-4}\,,\;
{\cal B}( b'\ra t W^- ) \simeq 9\times 10^{-7}\,,\nonr\\
{\cal B}( b'\ra \bar{\nu} T^{-\ot} ) \simeq 19.7\%\,,\;
{\cal B}( b'\ra \mu^+ T^{-\ft} ) \simeq 38.5\%\,,\nonr\\
{\cal B}( b'\ra \tau^+ T^{-\ft} ) \simeq 0.9\%\,,\;
{\cal B}( b'\ra e  S^{\tth} ) \simeq 39.7\%\,,
\eeqa
for $M_{LQ}=1\mtev$ and $M_{b'}=1.5\mtev$.

We stress that the above decay branching ratios are the result of using the example parameter set given in Eq.(\ref{eq:num_result}).
The decay branching ratios depend strongly on the model parameters, and the branching ratio pattern varies dramatically from one neutrino mass matrix to another\footnote{ In particular, in the minimal setup one has $\lambda^T_{\tau b'}/\lambda^T_{\mu b'} = {\cal M}^\nu_{e\tau}/{\cal M}^\nu_{e\mu}$.}.  However, one can see that the decay pattern of the heavy exotic color states in this example solution is
very different from the working assumption of 100\% $b'\ra t W, Z b, H b$ used for singlet $b'$ and other assumption  used for the leptoquark searches at the colliders.

Before closing this section, we want to point out some potentially interesting FCNC top 3-body decays in this framework.
From the example solution, we have
\beq
{\cal B}(t\ra u  \tau^+ \mu^- ) \simeq {\cal B}(t\ra u \mu^+ \mu^-)\simeq   2\times 10^{-9}\,.
\eeq
With an integrated luminosity of $3\mbox{ab}^{-1}$ and CM energy at $13\mtev$, about $2.5\times 10^9$ top quark pair events will be produced at the LHC. Therefore, only   $\sim 5$  events which include at least one top decaying in these 3-body FCNC are expected.
However,  the 3-body FCNC $t\ra c l_i l_j$ branching ratios change if adopting a different solution.
During our numerical study, we observe that in some cases there are one or two of them
at the ${\cal O}(10^{-7})$ level, which might be detectable at the LHC.
See \cite{Kim:2018oih} for the prospect of studying  these potentially interesting 3-body top decay modes at the LHC.

\subsection{Flavor violating neutral current processes }
A few comments on the data fitting  are in order: From Table \ref{tab:NCTL_list}, one sees that  the fit almost saturates, $\sim 80\%$, of the current limit on decay branching ratio  $B^+\ra K^+ \mu^+\tau^- $. Moreover,  the fit is very close to the $2\sigma$ limit from $B_s-\overline{B}_s$ mixing.
The solution seems to be stretched to the limit, and the discovery of lepton flavor violating signals are around the corners. However, this is the trade-off of using minimal number of parameters to reproduce all the central values. If instead aiming for the $1\sigma$ values, both can be reduced by half. In addition, the Yukawa couplings are tightly connected  with the neutrino mass matrix. During our numerical study, we observe that $D^+\ra \pi^+\mu^+ \mu^-$ and $B^+\ra K^+ \mu^+\tau^- $ can be far below their experimental upper bounds while $\tau \ra \mu \gamma$  close to the current experimental limit if using some different neutrino mass matrix or relaxing the strict relationship that $\wt{V}^*_{cb}\lambda^T_{\mu b}+ \wt{V}^*_{cb'}\lambda^T_{\mu b'} =0$.  Therefore, the model has vast parameter space to accommodate the anomalies with diversified predictions, and we cannot conclusively predict the pattern of the rare process rates at the moment.

However, because we need $ \lambda^T_{\tau b}\lambda^T_{\tau s}\neq 0$ to explain the $b\ra s \mu\mu$ anomaly,  the $b\ra s \tau\tau$ transition will be always generated at the tree-level.
From the example solution, we have
\beqa
{\cal H}^{b\ra s \tau\tau}_{eff}& \simeq&   -\frac{G_F}{\sqrt{2}} \wt{V}_{tb}\wt{V}_{ts}^* \frac{\alpha}{\pi}
{\cal C}^{bs\tau\tau}\left[\bar{s}\gamma^\alpha \PL b \right] \left[\bar{\tau}\gamma_\alpha (1-\gamma^5) \tau\right] +H.c.\,,\nonr\\
{\cal C}^{bs\tau\tau} &\simeq & \frac{\sqrt{2} \pi }{ 4 \alpha} {  \lambda^T_{\tau b} (\lambda^T_{\tau s})^*  \over \wt{V}_{tb}\wt{V}_{ts}^*  G_F M_T^2 } =   23.2  \times \left({\mtev \over M_T}\right)^2\,,
\eeqa
if taking $\wt{V}_{tb}\wt{V}_{ts}^*=-0.03975$.
The additional 1-loop contribution via the box diagram similar to that of $b\ra s\mu\mu$ can be ignored.
This Wilson coefficient is roughly six times larger than the SM prediction that ${\cal C}^{bs\tau\tau}_{SM} \sim -4.3$\cite{Bobeth:1999mk,Huber:2005ig,DescotesGenon:2011yn}, and push the decay branching ratio to $Br(B_s\ra \tau^+\tau^-)\simeq   1.6\times 10^{-5} $.
 Although the above value is still two orders below the relevant experimental upper limit, $\sim {\cal O}(10^{-3})$, for $B_s\ra \tau^+\tau^-$ at LHCb\cite{Aaij:2017xqt} and $B^+\ra K^+\tau^+\tau^-$ at BaBar\cite{TheBaBar:2016xwe},
this interesting $b\ra s \tau\tau$ transition could be potentially studied  at the LHCb and Belle II\cite{Capdevila:2017iqn}, or at the Z-pole\cite{Li:2020bvr}.

\subsection{$B\ra D^{(*)}\tau \bar{\nu}$}

\begin{figure}
\centering
\includegraphics[width=0.35\textwidth]{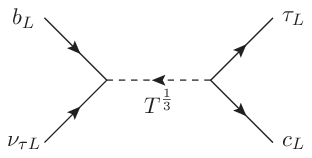}
\caption{ The Feynman diagram for $b\ra c \tau \nu$ transition. }
\label{fig:RD}
\end{figure}

 Alongside the $b\ra s l^+ l^-$ anomaly, the global analysis\cite{Amhis:2016xyh, Murgui:2019czp, Shi:2019gxi, Blanke:2019qrx, Kumbhakar:2019avh}  of the $R(D^{(*)})={\cal B}(B\ra D^{(*)}\tau \nu)/{\cal B}(B\ra D^{(*)}\mu \nu) $  data \cite{Lees:2012xj, Lees:2013uzd, Aaij:2015yra, Aaij:2017deq, Aaij:2017uff, Abdesselam:2019dgh}
also point to the $\tau\mhyphen\mu$ lepton flavor universality violation with a significance of $\geq 3 \sigma$.
In this mode, the $\mbox{MinS}_T$ of $\lambda^T$ contains the needed tree-level $b\ra c\tau \nu$ operator,  Fig.\ref{fig:RD}, to address the $R(D^{(*)})$ problem, and
\beq
{\cal H}^{CC}_{eff} \supset  -\left[{\lambda^T_{\tau s} (\lambda^T_{\tau b})^*\wt{V}_{cs}+ \lambda^T_{\tau b} (\lambda^T_{\tau b'})^*\wt{V}_{c b'} + |\lambda^T_{\tau b}|^2  \wt{V}_{cb} \over 4 M_T^2}\right]\left (\bar{c}  \gamma^\alpha \PL b  \right)\left(\bar{\tau} \gamma_\alpha\PL \nu_\tau \right) + H.c.
\label{eq:CC_bctaunu}
\eeq
This is to compare with the standard  effective Hamiltonian
\beq
{\cal H}_{eff}^{b\ra cl\nu} =  \frac{4 G_F}{\sqrt{2}} V_{cb} [ (1+ {\cal C}_{V_L}){\cal O}_{V_L}
+ {\cal C}_{V_R} {\cal O}_{V_R} +{\cal C}_{S_L} {\cal O}_{S_L} +{\cal C}_{S_R} {\cal O}_{S_R}+{\cal C}_{T} {\cal O}_{T}
] +H.c.
\eeq
where
\beqa
{\cal O}_{V_{L,R}} &=& (\bar{c}\gamma^\mu b_{L,R})(\bar{l}_L\gamma_\mu \nu_{lL})\,,\nonr\\
{\cal O}_{S_{L,R}} &=& (\bar{c}  b_{L,R})(\bar{l}_R \nu_{lL})\,,\;
{\cal O}_{T}  = (\bar{c}\sigma^{\mu\nu} b_{L})(\bar{l}_R \sigma_{\mu\nu}  \nu_{lL})\,.
\eeqa
In this model, the only CC operator can be generated at tree-level is ${\cal O}_{V_L}$, thus ${\cal C}_{V_{R}}=0$, ${\cal C}_{S_{R,L}}=0$, and ${\cal C}_{T}=0$.
From the global fit with single operator\cite{Murgui:2019czp}, the $R(D^{(*)})$ anomaly can be well addressed if ${\cal C}_{V_L}\simeq 0.08$.
However, the requirement to retain the $\mu\mhyphen e$ universality in $b\ra c l \nu$ strictly limits the parameter space.
By using the real MinS of parameters, we found the model can render at most ${\cal C}^{NP}_{V_L}\lesssim 0.01$.
On the other hand, we cannot rule out the possibility that the model has the viable complex number parameter space to accommodate this anomaly with others simultaneously.

On the other hand, this model predicts the lepton universality violation in the $b\ra u l_i \nu_j (i,j=\mu,\tau)$ transition, see the relevant Wilson coefficients in Table \ref{tab:CCTL_list}. Again, there is no electron counter parts if we use the $\mbox{MinS}_T$.
From our example solution, the rate of $b\ra u \mu \nu$ could deviate from the SM one by $\sim 40\%$ due to the smallness of $\wt{V}_{ub}$ and the interference between the NP and the SM weak interaction. More insights of the intriguing flavor problem are expected if the better experimental measurements on the $b\ra u l\nu$ transitions are available\cite{Colangelo:2020jmb}.

\subsection{Origin of the flavor structure}
Not only the subset of parameters are required to explain the anomalies, the nearly vanishing entities in the Yukawa matrices play vital  roles to bypass the strong flavor-changing experimental constraints.
The next question is how to understand the origin of this staggering flavor pattern.
Usually, the flavor pattern is considered within the framework of flavor symmetries.
It is highly nontrivial to embed the flavor pattern we found into a flavor symmetry, and it is beyond the scope of this paper.
Alternatively, we discuss the possible geometric origin of the flavor pattern in the extra-dimensional theories\cite{ArkaniHamed:1998rs, Antoniadis:1998ig,Randall:1999ee,ArkaniHamed:1999dc}.

A comprehensive fitting, including all the SM fermion masses and mixings, like \cite{Chang:2002ww, Chang:2008vx, Chang:2008zx, Chang:2010ic}  is also beyond the scope of this paper.
Instead, here we only consider how to generate the required flavor pattern of the leptoquark Yukawa coupling shown in Eq.(\ref{eq:recap_range}).
To illustrate, we consider a simple split fermion toy model\cite{ArkaniHamed:1999dc}. We assume all the chiral fermion wave functions are Gaussian locating in a small region, but  at  different positions, in the fifth dimension.
Moreover, all the 5-dim Gaussian distributions are assumed to share a universal width, $\sigma_{SF}$.
As for the three leptoquarks, we assume they do not have the zero mode such that their first Kaluza-Klein(KK) mode are naturally heavy.
More importantly, this setup forbids the leptoquark to develop VEV and break the $SU(3)_c$ symmetry.
In addition, the wavefunctions of the first leptoquark KK mode are assumed to be slowly varying in the vicinity of the fermion cluster,
and can be approximated as constants.  On the other hand, the SM Higgs must acquires the zero mode so that it can develop $v_0$ and breaks the SM electroweak symmetry.
Then, the 4-dim effective theory  is obtained after integrating out the fifth dimension.
The effective $\lambda^{s}_{ij}(s=T,D,S)$ Yukawa couplings  is determined by the overlapping of two chiral fermions' 5-dim wave functions  times the product of the scalar-specific 5-dim Yukawa coupling and the scalar's 5-dim wavefunction, denoted as $N_s$, in the vicinity where the fermions locate. The 4-dim Yukawa coupling is given by
\beq
\lambda^{s}_{ij} = N_s\, \mbox{Exp}\left[- \frac{(z_i-z_j)^2}{2 \sigma_{SF}^2 } \right]\,,\; (s= T,D,S)\,,
\eeq
where $z_i$ is the center location in the fifth dimension of the Gaussian wave function of particle$-i$.
It is clear that only the relative distances matter, so we arbitrarily set $z_{\tau_L}=0$ for the LH tau.
Note that  different fermion chiralities are involved for different scalar leptoquark Yukawa couplings.
For example, $\lambda^T_{ij}$ is determined by the separation between the corresponding LH down-quark($z_{d_{jL}}$) and the LH lepton($z_{\ell_{iL}}$), but $z_{l_{iR}}$  and $z_{d_{jL}}$ are involved for  $\lambda^S_{ij}$.

For simplicity, we assume the mass and interaction eigenstates coincide for the down-type quarks and the SM charged leptons.
We found the flavor structure can be excellently  reproduced
if the chiral fermion locations  in the fifth dimension are
\beqa
\{ z_{\tau_L},z_{\mu_L}, z_{e_L}, z_{\mu_R}, z_{e_R}\}&=&\{0,\; -0.49,\; 7.75[7.80],\; -4.99[-4.72],\; -0.89[-0.97]\}\,,\nonr\\
\{ z_{d_L},z_{s_L}, z_{b_L}, z_{b'_L}, z_{b_R}, z_{b'_R}\}
&=&\{-5.27[-5.53],\; 2.86[2.90],\; -2.14[-2.13],\nonr\\
 && -1.90[-1.95],\; -2.06[-2.09],\; 3.66[3.74]\} \,,
\eeqa
in the unit of $\sigma_{SF}$,
and $\{N_T,N_D,N_S\}=\{4.5, 6.2, 5.5 \}$.
The above configuration of split fermion locations results in
\beqa
|\lambda^T_{\mu b'}|=1.66[1.56]\,,\,
|\lambda^T_{\tau b'}|=0.74[0.68]\,,\,
|\lambda^T_{\mu b}|=1.15[1.17]\,,\,
|\lambda^T_{\tau b}|=0.46[0.46]\,,\,
|\lambda^T_{\tau s}|=7.5[6.8]\times 10^{-2}\,,\,\nonr\\
|\lambda^D_{e b'}|= 1.5[1.6]\times 10^{-3}\,,\,
|\lambda^D_{\mu b}|= 1.78[1.71]\,,\,
|\lambda^D_{\tau b'}|= 7.5[5.8]\times 10^{-3}\,,\,
|\lambda^D_{\tau b}|=0.74[0.69]\,,\,
\nonr\\
|\lambda^S_{e b'}|= 3.31[3.42]\,,\,
|\lambda^S_{\mu b}|=9.4[19.2]\times 10^{-2}\,.\nonr
\eeqa
Comparing to Eq.(\ref{eq:num_result}), one can see that all the Yukawa coupling magnitudes agree with the fitted values within $\lesssim 60\%$.
Moreover, the parameters we set to zero to evade the stringent constraints from Koan and muon data are indeed  very small,
\beqa
|\lambda^D_{e b}|= 7.5[3.6] \times 10^{-21}\,,\,
|\lambda^D_{\mu b'}|= 1.1[0.8] \times 10^{-3}\,,\nonr\\
|\lambda^T_{e d}|=7.0[0.13] \times 10^{-37}\,,\,
|\lambda^T_{e s}|=2.9[2.8] \times 10^{-5}\,,\,
|\lambda^T_{e b}|=2.6[1.8] \times 10^{-21}\,,\,\nonr\\
|\lambda^T_{e b'}|=2.7[1.1] \times 10^{-20}\,,\,
|\lambda^T_{\mu d}|=4.8[1.4] \times 10^{-5}\,,\,
|\lambda^T_{\tau d}|= 4.2[1.0] \times 10^{-6}\,,
\eeqa
in this given split fermion location configuration.

Finally, the lepton number symmetry is broken if $\mu_3\neq 0$. The phenomenological solution we found only calls for a very tiny
  $\mu_3\sim {\cal O}(0.5) \mkev$.
The smallness of $\mu_3$ can be arranged by assigning different orbiforlding parities to $T$ and $D$  such that their 5-D wave functions are almost orthogonal to each other and leads to the tiny 4D effective mixing\footnote{For instance, one takes $(+-)$ and the other takes $(-+)$ Kaluza-Klein parity on the $S_1/(Z_2\times Z_2)$ orbiford.}.  Contrarily,   $S$  and $D$ should share the same  orbifolding parities such that the maximal mixing yields a large effective 4D mixing $\mu_1\sim {\cal O}(\mtev)$.
 On the other hand, in terms of flavor symmetry, the smallness of $\mu_3$  seems to indicate  the global/gauged lepton number symmetry is well preserved and only broken very softly or radiatively.

\section{Conclusion}
\label{sec:conclusion}
We proposed a simple scenario with the addition of three scalar leptoquarks $T(3,3,-1/3)$, $D(3,2,1/6)$, $S(3,1,2/3)$, and one pair of down-quark-like vector fermion $b'_{L,R}(3,1,-1/3)$ to the SM particle content. The global baryon number $U(1)_B$ is assumed for simplicity.
This model is able to accommodate the observed $\tri a_e^{Cs[Rb]}$, $a_\mu$, $R(K)$, Cabibbo angle anomalies, and pass all experimental limits simultaneously. Moreover, the right pattern of  neutrino oscillation data can be reproduced as well.
 We have shown the existence of phenomenologically viable model parameter set by furnishing one example configuration with the minimal number of real Yukawa couplings.
For the possible UV origin, we provided a  split fermion toy model to explain the flavor structure embedded in the viable model parameter set. It will be interesting to reproduce the flavor pattern by nontrivial flavor symmetry. However, the tiny lepton number violating parameter, $\mu_3 \sim {\cal O}(0.5) \mkev$, seems to indicate the possible link of global/gauged lepton number and the unknown underling flavor symmetry.

In addition to the smoking gun signatures, the discovery of these new color states, this model robustly predicts the neutrino mass is of the normal hierarchy type with ${\cal M}^\nu_{ee}\lesssim 3\times 10^{-4} \mev$.
The $R(D^{(*)})$ anomaly can only be partially addressed in this model if one employs the minimal number of real Yukawa couplings.
However, we cannot rule out the possibility that could be achieved by using more (complex) parameters.
 From the parameter set example,  more motivated heavy color state decay branching ratios should be taken into account  in their collider searches.

\section*{Acknowledgments}
 WFC thanks Prof. John Ng for his comments on the draft. This work is supported by the Ministry of Science and Technology (MOST) of Taiwan under Grant No.~MOST-109-2112-M-007-012.


\appendix

\section{some properties of the $SU(2)$ triplet}
It is well-known that the Pauli matrices can serve as the generators for the  2-dimensional  $SU(2)$ representations.
Namely, $t^{(2)}_i= \frac{ \sigma_i}{2}\,, (i=1,2,3)$, and they satisfy the relationship $[t^{(2)}_i, t^{(2)}_j]= i \epsilon_{ijk} t^{(2)}_k$ ( $i,j,k=1,2,3$ ).
Any $SU(2)$ doublet $R^{(2)}$ transforms as $R^{(2)}\ra U^{(2)}(\vec{\theta}) R^{(2)}$, where $U^{(2)}(\vec{\theta})=\exp ( i \vec{\theta}\cdot \vec{t^{(2)}} )$ and $\vec{\theta}=(\theta_1,\theta_2,\theta_3)$ is the transformation angle vector.
It is popular to  represent the triplet by a bi-doublet, and  the  3-dimensional representation is less discussed in the literature. Equivalently, one can also construct the $SU(2)$ invariants intuitively using the 3-dimensional representation. This is the notation adopted in this paper, and we think it
 might be useful to collect some facts here for the reader.

It can be checked the following 3-dimensional representation,
\beq
t^{(3)}_1 = \frac{1}{\sqrt{2}}\left( \begin{array}{ccc} 0 &1&0 \\1 &0&1 \\0 &1&0 \\
        \end{array} \right)\,,\,\,
t^{(3)}_2 = \frac{1}{\sqrt{2}} \left( \begin{array}{ccc} 0 &-i &0 \\i &0&-i \\0 &i&0 \\
        \end{array} \right)\,,\,\,
t^{(3)}_3 = \left( \begin{array}{ccc} 1 &0&0 \\0 &0&0 \\0 &0&-1 \\
        \end{array} \right)\,,
\eeq
also satisfy the algebra $[t^{(3)}_i, t^{(3)}_j]= i \epsilon_{ijk} t^{(3)}_k$,  and serve as the group generators.
Therefore, any $SU(2)$ triplet $R^{(3)}$ transforms as $R^{(3)}\ra U^{(3)}(\vec{\theta}) R^{(3)}$, where $U^{(3)}(\vec{\theta})=\exp ( i \vec{\theta}\cdot \vec{t^{(3)}} )$. Because $U^{(3)}(\vec{\theta})$ is unitary, $(R^{(3)})^\dag\cdot R^{(3)}$ is $SU(2)$ invariant. For two given $SU(2)$ doublet $R^{(2)}_1=\begin{pmatrix} u_1\\ d_1 \end{pmatrix}$ and $R^{(2)}_2=\begin{pmatrix} u_2\\ d_2 \end{pmatrix}$, it can be shown that $\{R^{(2)}_1,R^{(2)}_2\}\equiv (u_1 u_2, (u_1d_2+ u_2 d_1)/\sqrt{2}, d_1 d_2 )^T$ is a triplet which transforms according to $U^{(3)}(\vec{\theta})$, and
$[R^{(2)}_1,R^{(2)}_2]\equiv  (u_1d_2- u_2 d_1)/\sqrt{2}$ is a singlet.

In order to construct an $SU(2)$ singlet  from any two $SU(2)$ triplets, $R_1^{(3)}$ and $R_2^{(3)}$,
we define
\beq
t^{(3)}_4= \left( \begin{array}{ccc} 0 &0&1 \\0 &-1&0 \\1 &0&0 \\
        \end{array} \right)\,,\,\,\mbox{ and }\, \left(t^{(3)}_4\right)^2=\mathrm{1}\,.
\eeq
It is  easy to verify  that
\beq
 t^{(3)}_4 t^{(3)}_j= (-)^j t^{(3)}_j t^{(3)}_4\,\,\, (j=1,2,3)\,.
\eeq
Therefore,
\beq
t^{(3)}_4 \left(t^{(3)}_i\right)^* t^{(3)}_4 = - t^{(3)}_i\,,\,\,t^{(3)}_4 \left(t^{(3)}_i\right)^T t^{(3)}_4 = - t^{(3)}_i\,.
\eeq
One can prove that
\beq
R_1^{(3)} \odot R_2^{(3)} \equiv (R_1^{(3)})^T\cdot t^{(3)}_4 \cdot R_2^{(3)}
\eeq
is an $SU(2)$ singlet.

Finally, similar to $\widetilde{ R^{(2)}} \equiv  i \sigma_2 ( R^{(2)})^*$ in the doublet cases, the object
\beq
\widetilde{ R^{(3)}} \equiv  t^{(3)}_4 ( R^{(3)})^*
\eeq
transforms as a triplet but it carries the opposite $U(1)$ charge(s) of $R^{(3)}$.

\section{low energy effective Hamiltonian}
\label{sec:H_eff}
After the electroweak SSB and the mass diagonalization, see Sec., the leptoquark coupling matrix $\lambda$'s are in the charged fermion mass basis. The relevant lagrangian is:
\beqa
{\cal L} &\supset&
 -\lambda_T \left[ \bar{\nu}^c u_L T^{-\tth}+ \left(\bar{\nu}^c d_L +\bar{e}^c u_L\right)\frac{ T^{\ot} }{\sqrt{2}}+ \bar{e}^c d_L T^{\ft}\right]- \lambda_D \frac{ \bar{d}_R}{\sqrt{2}} \left(\nu_L D^{-\ot}-e_L D^{\tth}\right) \nonr\\
&& - \lambda_S  \bar{e}_R d_L S^{-\tth}
-\frac{\mu_3 v_0}{2} D^{\ot} T^{-\ot} -\frac{\mu_3 v_0}{\sqrt{2}} D^{-\tth} T^{\tth} - \frac{\mu_1 v_0}{2} D^{\tth} S^{-\tth}  +H.c.
\eeqa
In the above expression, the fields should be understood as the vectors in the flavor space:  $u=\widetilde{V}^\dag\cdot(u,c,t)^T$, $d=(d,s,b, b')^T$, $e=(e,\mu,\tau)^T$, and $\nu=(\nu_e, \nu_\mu, \nu_\tau)^T$.

If the mixings among  the leptoquarks are small, it can be treated by the triple scalar interaction vertices to the leading approximation. Moreover, masses are nearly degenerated within the leptoquark $SU(2)$ multiplets.
After integrating out the heavy leptoquarks, we arrive the following low energy effective Hamiltonian:
\beqa
{\cal H}_{eff} \simeq &-&\frac{1}{M_T^2}\lambda_T \lambda_T^\dag\left[(\bar{\nu}^cu_L)(\bar{u}_L\nu^c)
+\frac{1}{2}(\bar{\nu}^c d_L+\bar{e}^c u_L)(\bar{d}_L\nu^c+\bar{u}_L e^c)+(\bar{e}^c d_L)(\bar{d}_L e^c)\right]\nonr\\
&-&\frac{1}{2 M_D^2}\lambda_D \lambda_D^\dag\left[(\bar{d}_R \nu_L)(\bar{\nu}_L d_R)+(\bar{d}_R e_L)(\bar{e}_L d_R)\right]
-\frac{1}{ M_S^2} \lambda_S \lambda_S^\dag(\bar{e}_R d_L)(\bar{d}_L e_R)\nonr\\
&-&\left\{ \frac{\mu_1 v_0 }{2\sqrt{2} M_S^2 M_D^2} \lambda_S \lambda_D(\bar{e}_R d_L)(\bar{d}_R e_L)+H.c. \right \}\nonr\\
&-&\left\{ \frac{\mu_3 v_0 }{2 M_T^2 M_D^2}\lambda_T \lambda_D\left[(\bar{\nu}^c u_L)(\bar{d}_R e_L)-\frac{1}{2}(\bar{\nu}^c d_L+\bar{e}^c u_L)(\bar{d}_R \nu_L)\right] +H.c. \right\}
\eeqa
After performing the Fierz transformation and applying the identities associated with charge conjugation, the effective Hamiltonian takes a more familiar form:
\beqa
{\cal H}_{eff} \simeq &-&\frac{1}{2 M_T^2}\lambda_T \lambda_T^\dag\left[
(\bar{u}\gamma^\alpha\PL u)\left(\bar{\nu}\gamma_\alpha\PL \nu +\half \bar{e}\gamma_\alpha\PL e \right)
+(\bar{d}\gamma^\alpha\PL d)\left(\bar{e}\gamma_\alpha\PL e +\half \bar{\nu}\gamma_\alpha\PL \nu \right)
\right]\nonr\\
&-&\frac{1}{4 M_T^2}\lambda_T \lambda_T^\dag\left[
(\bar{u}\gamma^\alpha\PL d) (\bar{e}\gamma_\alpha\PL \nu)  +  (\bar{d}\gamma^\alpha\PL u) (\bar{\nu}\gamma_\alpha\PL e)
\right]\nonr\\
&+&\frac{1}{4 M_D^2}\lambda_D \lambda_D^\dag\left[(\bar{d} \gamma^\alpha \PR d)(\bar{\nu}\gamma_\alpha \PL \nu + \bar{e}\gamma_\alpha \PL e )\right]
+ \frac{1}{2 M_S^2} \lambda_S \lambda_S^\dag(\bar{d}\gamma^\alpha\PL  d)(\bar{e} \gamma_\alpha\PR e)\nonr\\
&+&\left\{ \frac{\mu_1 v_0 }{4\sqrt{2} M_S^2 M_D^2} \lambda_S \lambda_D\left[(\bar{d}\PL d)(\bar{e}\PL e) +\fth(\bar{d}\sigma^{\alpha\beta}\PL d)(\bar{e}\sigma_{\alpha\beta}\PL e) \right] +H.c.\right\} \nonr\\
&+&\left\{ \frac{\mu_3 v_0 }{4 M_T^2 M_D^2}\lambda_T \lambda_D\left[(\bar{d}\PL u)(\bar{\nu}^c\PL e)-(\bar{d}\PL d)(\bar{\nu}^c\PL \nu)
\right] +H.c.\right\} \nonr\\
&+&\left\{ \frac{\mu_3 v_0 }{32 M_T^2 M_D^2}\lambda_T \lambda_D\left[ 3 (\bar{d}\sigma^{\alpha\beta}\PL u)(\bar{\nu}^c\sigma_{\alpha\beta}\PL e)
- (\bar{d}\sigma^{\alpha\beta}\PL d)(\bar{\nu}^c\sigma_{\alpha\beta}\PL \nu)\right] +H.c. \right\}
\label{eq:H_eff}
\eeqa
where $\PL=(1-\gamma^5)/2$ and  $\PR=(1+\gamma^5)/2$.
From the last three terms in the above, it is clear that: (1) the $SU(2)$ symmetry is broken after the electroweak SSB,
(2) the mixing of leptoquarks with different lepton number breaks the global lepton number, as has been discussed in Sec.\ref{sec:model}.

So far, we have suppressed the flavor indices to avoid the unnecessary notational burden, but it is easy to trace and put them back in when needed. For example, the effective Hamiltonian generated by the pair of triplet Yukawa couplings, $\lambda^T_{i d}$ and $\lambda^T_{j D}$,
is given by
\beqa
{\cal H}_{eff}  & \supset &  -{\lambda^T_{i d}(\lambda^T_{j D})^* \over 2 M_T^2 }\sum_{u,U=u,c,t}\left[
 \left(\overline{D} \gamma^\alpha\PL d\right)\left(\bar{e}_j \gamma_\alpha \PL e_i +\frac{1}{2}\bar{\nu}_j \gamma_\alpha \PL \nu_i \right) \right.\nonr\\
&& + \frac{1}{2} \wt{V}_{U D} \left(\overline{U} \gamma^\alpha\PL d\right)\left(\bar{e}_j \gamma_\alpha\PL \nu_i\right)
+ \frac{1}{2} \wt{V}^*_{u d} \left(\bar{D} \gamma^\alpha\PL u\right)\left(\bar{\nu}_j \gamma_\alpha\PL e_i\right)\nonr\\
&&\left.  + \wt{V}^*_{u d} \wt{V}_{U D} \left( \overline{U} \gamma^\alpha\PL u\right)\left(\bar{\nu}_j \gamma_\alpha \PL \nu_i
+ \frac{1}{2} \bar{e}_j \gamma_\alpha \PL e_i \right) \right] + H.c.
 \label{eq:2THeff}
\eeqa
%
%

\bibliography{c1loop_Ref}

\begin{thebibliography}{165}%
\makeatletter
\providecommand \@ifxundefined [1]{%
 \@ifx{#1\undefined}
}%
\providecommand \@ifnum [1]{%
 \ifnum #1\expandafter \@firstoftwo
 \else \expandafter \@secondoftwo
 \fi
}%
\providecommand \@ifx [1]{%
 \ifx #1\expandafter \@firstoftwo
 \else \expandafter \@secondoftwo
 \fi
}%
\providecommand \natexlab [1]{#1}%
\providecommand \enquote  [1]{``#1''}%
\providecommand \bibnamefont  [1]{#1}%
\providecommand \bibfnamefont [1]{#1}%
\providecommand \citenamefont [1]{#1}%
\providecommand \href@noop [0]{\@secondoftwo}%
\providecommand \href [0]{\begingroup \@sanitize@url \@href}%
\providecommand \@href[1]{\@@startlink{#1}\@@href}%
\providecommand \@@href[1]{\endgroup#1\@@endlink}%
\providecommand \@sanitize@url [0]{\catcode `\\12\catcode `\$12\catcode
  `\&12\catcode `\#12\catcode `\^12\catcode `\_12\catcode `\%12\relax}%
\providecommand \@@startlink[1]{}%
\providecommand \@@endlink[0]{}%
\providecommand \url  [0]{\begingroup\@sanitize@url \@url }%
\providecommand \@url [1]{\endgroup\@href {#1}{\urlprefix }}%
\providecommand \urlprefix  [0]{URL }%
\providecommand \Eprint [0]{\href }%
\providecommand \doibase [0]{https://doi.org/}%
\providecommand \selectlanguage [0]{\@gobble}%
\providecommand \bibinfo  [0]{\@secondoftwo}%
\providecommand \bibfield  [0]{\@secondoftwo}%
\providecommand \translation [1]{[#1]}%
\providecommand \BibitemOpen [0]{}%
\providecommand \bibitemStop [0]{}%
\providecommand \bibitemNoStop [0]{.\EOS\space}%
\providecommand \EOS [0]{\spacefactor3000\relax}%
\providecommand \BibitemShut  [1]{\csname bibitem#1\endcsname}%
\let\auto@bib@innerbib\@empty
\bibitem [{\citenamefont {Zyla}\ \emph {et~al.}(2020)\citenamefont {Zyla} \emph
  {et~al.}}]{Zyla:2020zbs}%
  \BibitemOpen
  \bibfield  {author} {\bibinfo {author} {\bibfnamefont {P.~A.}\ \bibnamefont
  {Zyla}} \emph {et~al.} (\bibinfo {collaboration} {Particle Data Group}),\
  }\bibfield  {title} {\bibinfo {title} {{Review of Particle Physics}},\ }\href
  {https://doi.org/10.1093/ptep/ptaa104} {\bibfield  {journal} {\bibinfo
  {journal} {PTEP}\ }\textbf {\bibinfo {volume} {2020}},\ \bibinfo {pages}
  {083C01} (\bibinfo {year} {2020})}\BibitemShut {NoStop}%
\bibitem [{\citenamefont {Bennett}\ \emph {et~al.}(2006)\citenamefont {Bennett}
  \emph {et~al.}}]{Bennett:2006fi}%
  \BibitemOpen
  \bibfield  {author} {\bibinfo {author} {\bibfnamefont {G.~W.}\ \bibnamefont
  {Bennett}} \emph {et~al.} (\bibinfo {collaboration} {Muon g-2}),\ }\bibfield
  {title} {\bibinfo {title} {{Final Report of the Muon E821 Anomalous Magnetic
  Moment Measurement at BNL}},\ }\href
  {https://doi.org/10.1103/PhysRevD.73.072003} {\bibfield  {journal} {\bibinfo
  {journal} {Phys. Rev. D}\ }\textbf {\bibinfo {volume} {73}},\ \bibinfo
  {pages} {072003} (\bibinfo {year} {2006})},\ \Eprint
  {https://arxiv.org/abs/hep-ex/0602035} {arXiv:hep-ex/0602035} \BibitemShut
  {NoStop}%
\bibitem [{\citenamefont {Blum}\ \emph {et~al.}(2018)\citenamefont {Blum},
  \citenamefont {Boyle}, \citenamefont {G\"ulpers}, \citenamefont {Izubuchi},
  \citenamefont {Jin}, \citenamefont {Jung}, \citenamefont {J\"uttner},
  \citenamefont {Lehner}, \citenamefont {Portelli},\ and\ \citenamefont
  {Tsang}}]{Blum:2018mom}%
  \BibitemOpen
  \bibfield  {author} {\bibinfo {author} {\bibfnamefont {T.}~\bibnamefont
  {Blum}}, \bibinfo {author} {\bibfnamefont {P.~A.}\ \bibnamefont {Boyle}},
  \bibinfo {author} {\bibfnamefont {V.}~\bibnamefont {G\"ulpers}}, \bibinfo
  {author} {\bibfnamefont {T.}~\bibnamefont {Izubuchi}}, \bibinfo {author}
  {\bibfnamefont {L.}~\bibnamefont {Jin}}, \bibinfo {author} {\bibfnamefont
  {C.}~\bibnamefont {Jung}}, \bibinfo {author} {\bibfnamefont {A.}~\bibnamefont
  {J\"uttner}}, \bibinfo {author} {\bibfnamefont {C.}~\bibnamefont {Lehner}},
  \bibinfo {author} {\bibfnamefont {A.}~\bibnamefont {Portelli}},\ and\
  \bibinfo {author} {\bibfnamefont {J.~T.}\ \bibnamefont {Tsang}} (\bibinfo
  {collaboration} {RBC, UKQCD}),\ }\bibfield  {title} {\bibinfo {title}
  {{Calculation of the hadronic vacuum polarization contribution to the muon
  anomalous magnetic moment}},\ }\href
  {https://doi.org/10.1103/PhysRevLett.121.022003} {\bibfield  {journal}
  {\bibinfo  {journal} {Phys. Rev. Lett.}\ }\textbf {\bibinfo {volume} {121}},\
  \bibinfo {pages} {022003} (\bibinfo {year} {2018})},\ \Eprint
  {https://arxiv.org/abs/1801.07224} {arXiv:1801.07224 [hep-lat]} \BibitemShut
  {NoStop}%
\bibitem [{\citenamefont {Davier}\ \emph {et~al.}(2020)\citenamefont {Davier},
  \citenamefont {Hoecker}, \citenamefont {Malaescu},\ and\ \citenamefont
  {Zhang}}]{Davier:2019can}%
  \BibitemOpen
  \bibfield  {author} {\bibinfo {author} {\bibfnamefont {M.}~\bibnamefont
  {Davier}}, \bibinfo {author} {\bibfnamefont {A.}~\bibnamefont {Hoecker}},
  \bibinfo {author} {\bibfnamefont {B.}~\bibnamefont {Malaescu}},\ and\
  \bibinfo {author} {\bibfnamefont {Z.}~\bibnamefont {Zhang}},\ }\bibfield
  {title} {\bibinfo {title} {{A new evaluation of the hadronic vacuum
  polarisation contributions to the muon anomalous magnetic moment and to
  $\mathbf{\boldsymbol\alpha(m_Z^2)}$}},\ }\href
  {https://doi.org/10.1140/epjc/s10052-020-7792-2} {\bibfield  {journal}
  {\bibinfo  {journal} {Eur. Phys. J. C}\ }\textbf {\bibinfo {volume} {80}},\
  \bibinfo {pages} {241} (\bibinfo {year} {2020})},\ \bibinfo {note} {[Erratum:
  Eur.Phys.J.C 80, 410 (2020)]},\ \Eprint {https://arxiv.org/abs/1908.00921}
  {arXiv:1908.00921 [hep-ph]} \BibitemShut {NoStop}%
\bibitem [{\citenamefont {Abi}\ \emph {et~al.}(2021)\citenamefont {Abi} \emph
  {et~al.}}]{Abi:2021gix}%
  \BibitemOpen
  \bibfield  {author} {\bibinfo {author} {\bibfnamefont {B.}~\bibnamefont
  {Abi}} \emph {et~al.} (\bibinfo {collaboration} {Muon g-2}),\ }\bibfield
  {title} {\bibinfo {title} {{Measurement of the Positive Muon Anomalous
  Magnetic Moment to 0.46 ppm}},\ }\href
  {https://doi.org/10.1103/PhysRevLett.126.141801} {\bibfield  {journal}
  {\bibinfo  {journal} {Phys. Rev. Lett.}\ }\textbf {\bibinfo {volume} {126}},\
  \bibinfo {pages} {141801} (\bibinfo {year} {2021})},\ \Eprint
  {https://arxiv.org/abs/2104.03281} {arXiv:2104.03281 [hep-ex]} \BibitemShut
  {NoStop}%
\bibitem [{\citenamefont {Saito}(2012)}]{Saito:2012zz}%
  \BibitemOpen
  \bibfield  {author} {\bibinfo {author} {\bibfnamefont {N.}~\bibnamefont
  {Saito}} (\bibinfo {collaboration} {J-PARC g-'2/EDM}),\ }\bibfield  {title}
  {\bibinfo {title} {{A novel precision measurement of muon g-2 and EDM at
  J-PARC}},\ }\href {https://doi.org/10.1063/1.4742078} {\bibfield  {journal}
  {\bibinfo  {journal} {AIP Conf. Proc.}\ }\textbf {\bibinfo {volume} {1467}},\
  \bibinfo {pages} {45} (\bibinfo {year} {2012})}\BibitemShut {NoStop}%
\bibitem [{\citenamefont {Parker}\ \emph {et~al.}(2018)\citenamefont {Parker},
  \citenamefont {Yu}, \citenamefont {Zhong}, \citenamefont {Estey},\ and\
  \citenamefont {M\"uller}}]{Parker:2018vye}%
  \BibitemOpen
  \bibfield  {author} {\bibinfo {author} {\bibfnamefont {R.~H.}\ \bibnamefont
  {Parker}}, \bibinfo {author} {\bibfnamefont {C.}~\bibnamefont {Yu}}, \bibinfo
  {author} {\bibfnamefont {W.}~\bibnamefont {Zhong}}, \bibinfo {author}
  {\bibfnamefont {B.}~\bibnamefont {Estey}},\ and\ \bibinfo {author}
  {\bibfnamefont {H.}~\bibnamefont {M\"uller}},\ }\bibfield  {title} {\bibinfo
  {title} {{Measurement of the fine-structure constant as a test of the
  Standard Model}},\ }\href {https://doi.org/10.1126/science.aap7706}
  {\bibfield  {journal} {\bibinfo  {journal} {Science}\ }\textbf {\bibinfo
  {volume} {360}},\ \bibinfo {pages} {191} (\bibinfo {year} {2018})},\ \Eprint
  {https://arxiv.org/abs/1812.04130} {arXiv:1812.04130 [physics.atom-ph]}
  \BibitemShut {NoStop}%
\bibitem [{\citenamefont {Hanneke}\ \emph {et~al.}(2011)\citenamefont
  {Hanneke}, \citenamefont {Hoogerheide},\ and\ \citenamefont
  {Gabrielse}}]{Hanneke:2010au}%
  \BibitemOpen
  \bibfield  {author} {\bibinfo {author} {\bibfnamefont {D.}~\bibnamefont
  {Hanneke}}, \bibinfo {author} {\bibfnamefont {S.~F.}\ \bibnamefont
  {Hoogerheide}},\ and\ \bibinfo {author} {\bibfnamefont {G.}~\bibnamefont
  {Gabrielse}},\ }\bibfield  {title} {\bibinfo {title} {{Cavity Control of a
  Single-Electron Quantum Cyclotron: Measuring the Electron Magnetic Moment}},\
  }\href {https://doi.org/10.1103/PhysRevA.83.052122} {\bibfield  {journal}
  {\bibinfo  {journal} {Phys. Rev. A}\ }\textbf {\bibinfo {volume} {83}},\
  \bibinfo {pages} {052122} (\bibinfo {year} {2011})},\ \Eprint
  {https://arxiv.org/abs/1009.4831} {arXiv:1009.4831 [physics.atom-ph]}
  \BibitemShut {NoStop}%
\bibitem [{\citenamefont {Aoyama}\ \emph {et~al.}(2018)\citenamefont {Aoyama},
  \citenamefont {Kinoshita},\ and\ \citenamefont {Nio}}]{Aoyama:2017uqe}%
  \BibitemOpen
  \bibfield  {author} {\bibinfo {author} {\bibfnamefont {T.}~\bibnamefont
  {Aoyama}}, \bibinfo {author} {\bibfnamefont {T.}~\bibnamefont {Kinoshita}},\
  and\ \bibinfo {author} {\bibfnamefont {M.}~\bibnamefont {Nio}},\ }\bibfield
  {title} {\bibinfo {title} {{Revised and Improved Value of the QED Tenth-Order
  Electron Anomalous Magnetic Moment}},\ }\href
  {https://doi.org/10.1103/PhysRevD.97.036001} {\bibfield  {journal} {\bibinfo
  {journal} {Phys. Rev. D}\ }\textbf {\bibinfo {volume} {97}},\ \bibinfo
  {pages} {036001} (\bibinfo {year} {2018})},\ \Eprint
  {https://arxiv.org/abs/1712.06060} {arXiv:1712.06060 [hep-ph]} \BibitemShut
  {NoStop}%
\bibitem [{\citenamefont {Liu}\ \emph {et~al.}(2019)\citenamefont {Liu},
  \citenamefont {Wagner},\ and\ \citenamefont {Wang}}]{Liu:2018xkx}%
  \BibitemOpen
  \bibfield  {author} {\bibinfo {author} {\bibfnamefont {J.}~\bibnamefont
  {Liu}}, \bibinfo {author} {\bibfnamefont {C.~E.~M.}\ \bibnamefont {Wagner}},\
  and\ \bibinfo {author} {\bibfnamefont {X.-P.}\ \bibnamefont {Wang}},\
  }\bibfield  {title} {\bibinfo {title} {{A light complex scalar for the
  electron and muon anomalous magnetic moments}},\ }\href
  {https://doi.org/10.1007/JHEP03(2019)008} {\bibfield  {journal} {\bibinfo
  {journal} {JHEP}\ }\textbf {\bibinfo {volume} {03}},\ \bibinfo {pages}
  {008}},\ \Eprint {https://arxiv.org/abs/1810.11028} {arXiv:1810.11028
  [hep-ph]} \BibitemShut {NoStop}%
\bibitem [{\citenamefont {Crivellin}\ \emph {et~al.}(2018)\citenamefont
  {Crivellin}, \citenamefont {Hoferichter},\ and\ \citenamefont
  {Schmidt-Wellenburg}}]{Crivellin:2018qmi}%
  \BibitemOpen
  \bibfield  {author} {\bibinfo {author} {\bibfnamefont {A.}~\bibnamefont
  {Crivellin}}, \bibinfo {author} {\bibfnamefont {M.}~\bibnamefont
  {Hoferichter}},\ and\ \bibinfo {author} {\bibfnamefont {P.}~\bibnamefont
  {Schmidt-Wellenburg}},\ }\bibfield  {title} {\bibinfo {title} {{Combined
  explanations of $(g-2)_{\mu,e}$ and implications for a large muon EDM}},\
  }\href {https://doi.org/10.1103/PhysRevD.98.113002} {\bibfield  {journal}
  {\bibinfo  {journal} {Phys. Rev. D}\ }\textbf {\bibinfo {volume} {98}},\
  \bibinfo {pages} {113002} (\bibinfo {year} {2018})},\ \Eprint
  {https://arxiv.org/abs/1807.11484} {arXiv:1807.11484 [hep-ph]} \BibitemShut
  {NoStop}%
\bibitem [{\citenamefont {Endo}\ and\ \citenamefont
  {Yin}(2019)}]{Endo:2019bcj}%
  \BibitemOpen
  \bibfield  {author} {\bibinfo {author} {\bibfnamefont {M.}~\bibnamefont
  {Endo}}\ and\ \bibinfo {author} {\bibfnamefont {W.}~\bibnamefont {Yin}},\
  }\bibfield  {title} {\bibinfo {title} {{Explaining electron and muon $g-2$
  anomaly in SUSY without lepton-flavor mixings}},\ }\href
  {https://doi.org/10.1007/JHEP08(2019)122} {\bibfield  {journal} {\bibinfo
  {journal} {JHEP}\ }\textbf {\bibinfo {volume} {08}},\ \bibinfo {pages}
  {122}},\ \Eprint {https://arxiv.org/abs/1906.08768} {arXiv:1906.08768
  [hep-ph]} \BibitemShut {NoStop}%
\bibitem [{\citenamefont {Bauer}\ \emph {et~al.}(2020)\citenamefont {Bauer},
  \citenamefont {Neubert}, \citenamefont {Renner}, \citenamefont {Schnubel},\
  and\ \citenamefont {Thamm}}]{Bauer:2019gfk}%
  \BibitemOpen
  \bibfield  {author} {\bibinfo {author} {\bibfnamefont {M.}~\bibnamefont
  {Bauer}}, \bibinfo {author} {\bibfnamefont {M.}~\bibnamefont {Neubert}},
  \bibinfo {author} {\bibfnamefont {S.}~\bibnamefont {Renner}}, \bibinfo
  {author} {\bibfnamefont {M.}~\bibnamefont {Schnubel}},\ and\ \bibinfo
  {author} {\bibfnamefont {A.}~\bibnamefont {Thamm}},\ }\bibfield  {title}
  {\bibinfo {title} {{Axionlike Particles, Lepton-Flavor Violation, and a New
  Explanation of $a_\mu$ and $a_e$}},\ }\href
  {https://doi.org/10.1103/PhysRevLett.124.211803} {\bibfield  {journal}
  {\bibinfo  {journal} {Phys. Rev. Lett.}\ }\textbf {\bibinfo {volume} {124}},\
  \bibinfo {pages} {211803} (\bibinfo {year} {2020})},\ \Eprint
  {https://arxiv.org/abs/1908.00008} {arXiv:1908.00008 [hep-ph]} \BibitemShut
  {NoStop}%
\bibitem [{\citenamefont {Badziak}\ and\ \citenamefont
  {Sakurai}(2019)}]{Badziak:2019gaf}%
  \BibitemOpen
  \bibfield  {author} {\bibinfo {author} {\bibfnamefont {M.}~\bibnamefont
  {Badziak}}\ and\ \bibinfo {author} {\bibfnamefont {K.}~\bibnamefont
  {Sakurai}},\ }\bibfield  {title} {\bibinfo {title} {{Explanation of electron
  and muon g \ensuremath{-} 2 anomalies in the MSSM}},\ }\href
  {https://doi.org/10.1007/JHEP10(2019)024} {\bibfield  {journal} {\bibinfo
  {journal} {JHEP}\ }\textbf {\bibinfo {volume} {10}},\ \bibinfo {pages}
  {024}},\ \Eprint {https://arxiv.org/abs/1908.03607} {arXiv:1908.03607
  [hep-ph]} \BibitemShut {NoStop}%
\bibitem [{\citenamefont {Abdullah}\ \emph {et~al.}(2019)\citenamefont
  {Abdullah}, \citenamefont {Dutta}, \citenamefont {Ghosh},\ and\ \citenamefont
  {Li}}]{Abdullah:2019ofw}%
  \BibitemOpen
  \bibfield  {author} {\bibinfo {author} {\bibfnamefont {M.}~\bibnamefont
  {Abdullah}}, \bibinfo {author} {\bibfnamefont {B.}~\bibnamefont {Dutta}},
  \bibinfo {author} {\bibfnamefont {S.}~\bibnamefont {Ghosh}},\ and\ \bibinfo
  {author} {\bibfnamefont {T.}~\bibnamefont {Li}},\ }\bibfield  {title}
  {\bibinfo {title} {{$(g-2)_{\mu,e}$ and the ANITA anomalous events in a
  three-loop neutrino mass model}},\ }\href
  {https://doi.org/10.1103/PhysRevD.100.115006} {\bibfield  {journal} {\bibinfo
   {journal} {Phys. Rev. D}\ }\textbf {\bibinfo {volume} {100}},\ \bibinfo
  {pages} {115006} (\bibinfo {year} {2019})},\ \Eprint
  {https://arxiv.org/abs/1907.08109} {arXiv:1907.08109 [hep-ph]} \BibitemShut
  {NoStop}%
\bibitem [{\citenamefont {Hiller}\ \emph {et~al.}(2020)\citenamefont {Hiller},
  \citenamefont {Hormigos-Feliu}, \citenamefont {Litim},\ and\ \citenamefont
  {Steudtner}}]{Hiller:2019mou}%
  \BibitemOpen
  \bibfield  {author} {\bibinfo {author} {\bibfnamefont {G.}~\bibnamefont
  {Hiller}}, \bibinfo {author} {\bibfnamefont {C.}~\bibnamefont
  {Hormigos-Feliu}}, \bibinfo {author} {\bibfnamefont {D.~F.}\ \bibnamefont
  {Litim}},\ and\ \bibinfo {author} {\bibfnamefont {T.}~\bibnamefont
  {Steudtner}},\ }\bibfield  {title} {\bibinfo {title} {{Anomalous magnetic
  moments from asymptotic safety}},\ }\href
  {https://doi.org/10.1103/PhysRevD.102.071901} {\bibfield  {journal} {\bibinfo
   {journal} {Phys. Rev. D}\ }\textbf {\bibinfo {volume} {102}},\ \bibinfo
  {pages} {071901} (\bibinfo {year} {2020})},\ \Eprint
  {https://arxiv.org/abs/1910.14062} {arXiv:1910.14062 [hep-ph]} \BibitemShut
  {NoStop}%
\bibitem [{\citenamefont {Cornella}\ \emph {et~al.}(2020)\citenamefont
  {Cornella}, \citenamefont {Paradisi},\ and\ \citenamefont
  {Sumensari}}]{Cornella:2019uxs}%
  \BibitemOpen
  \bibfield  {author} {\bibinfo {author} {\bibfnamefont {C.}~\bibnamefont
  {Cornella}}, \bibinfo {author} {\bibfnamefont {P.}~\bibnamefont {Paradisi}},\
  and\ \bibinfo {author} {\bibfnamefont {O.}~\bibnamefont {Sumensari}},\
  }\bibfield  {title} {\bibinfo {title} {{Hunting for ALPs with Lepton Flavor
  Violation}},\ }\href {https://doi.org/10.1007/JHEP01(2020)158} {\bibfield
  {journal} {\bibinfo  {journal} {JHEP}\ }\textbf {\bibinfo {volume} {01}},\
  \bibinfo {pages} {158}},\ \Eprint {https://arxiv.org/abs/1911.06279}
  {arXiv:1911.06279 [hep-ph]} \BibitemShut {NoStop}%
\bibitem [{\citenamefont {Haba}\ \emph {et~al.}(2020)\citenamefont {Haba},
  \citenamefont {Shimizu},\ and\ \citenamefont {Yamada}}]{Haba:2020gkr}%
  \BibitemOpen
  \bibfield  {author} {\bibinfo {author} {\bibfnamefont {N.}~\bibnamefont
  {Haba}}, \bibinfo {author} {\bibfnamefont {Y.}~\bibnamefont {Shimizu}},\ and\
  \bibinfo {author} {\bibfnamefont {T.}~\bibnamefont {Yamada}},\ }\bibfield
  {title} {\bibinfo {title} {{Muon and electron $g-2$ and the origin of the
  fermion mass hierarchy}},\ }\href {https://doi.org/10.1093/ptep/ptaa098}
  {\bibfield  {journal} {\bibinfo  {journal} {PTEP}\ }\textbf {\bibinfo
  {volume} {2020}},\ \bibinfo {pages} {093B05} (\bibinfo {year} {2020})},\
  \Eprint {https://arxiv.org/abs/2002.10230} {arXiv:2002.10230 [hep-ph]}
  \BibitemShut {NoStop}%
\bibitem [{\citenamefont {Bigaran}\ and\ \citenamefont
  {Volkas}(2020)}]{Bigaran:2020jil}%
  \BibitemOpen
  \bibfield  {author} {\bibinfo {author} {\bibfnamefont {I.}~\bibnamefont
  {Bigaran}}\ and\ \bibinfo {author} {\bibfnamefont {R.~R.}\ \bibnamefont
  {Volkas}},\ }\bibfield  {title} {\bibinfo {title} {{Getting chirality right:
  Single scalar leptoquark solutions to the $(g-2)_{e,\mu}$ puzzle}},\ }\href
  {https://doi.org/10.1103/PhysRevD.102.075037} {\bibfield  {journal} {\bibinfo
   {journal} {Phys. Rev. D}\ }\textbf {\bibinfo {volume} {102}},\ \bibinfo
  {pages} {075037} (\bibinfo {year} {2020})},\ \Eprint
  {https://arxiv.org/abs/2002.12544} {arXiv:2002.12544 [hep-ph]} \BibitemShut
  {NoStop}%
\bibitem [{\citenamefont {Jana}\ \emph
  {et~al.}(2020{\natexlab{a}})\citenamefont {Jana}, \citenamefont {K.},\ and\
  \citenamefont {Saad}}]{Jana:2020pxx}%
  \BibitemOpen
  \bibfield  {author} {\bibinfo {author} {\bibfnamefont {S.}~\bibnamefont
  {Jana}}, \bibinfo {author} {\bibfnamefont {V.~P.}\ \bibnamefont {K.}},\ and\
  \bibinfo {author} {\bibfnamefont {S.}~\bibnamefont {Saad}},\ }\bibfield
  {title} {\bibinfo {title} {{Resolving electron and muon $g-2$ within the
  2HDM}},\ }\href {https://doi.org/10.1103/PhysRevD.101.115037} {\bibfield
  {journal} {\bibinfo  {journal} {Phys. Rev. D}\ }\textbf {\bibinfo {volume}
  {101}},\ \bibinfo {pages} {115037} (\bibinfo {year} {2020}{\natexlab{a}})},\
  \Eprint {https://arxiv.org/abs/2003.03386} {arXiv:2003.03386 [hep-ph]}
  \BibitemShut {NoStop}%
\bibitem [{\citenamefont {Calibbi}\ \emph {et~al.}(2020)\citenamefont
  {Calibbi}, \citenamefont {L\'opez-Ib\'a\~nez}, \citenamefont {Melis},\ and\
  \citenamefont {Vives}}]{Calibbi:2020emz}%
  \BibitemOpen
  \bibfield  {author} {\bibinfo {author} {\bibfnamefont {L.}~\bibnamefont
  {Calibbi}}, \bibinfo {author} {\bibfnamefont {M.~L.}\ \bibnamefont
  {L\'opez-Ib\'a\~nez}}, \bibinfo {author} {\bibfnamefont {A.}~\bibnamefont
  {Melis}},\ and\ \bibinfo {author} {\bibfnamefont {O.}~\bibnamefont {Vives}},\
  }\bibfield  {title} {\bibinfo {title} {{Muon and electron $g-2$ and lepton
  masses in flavor models}},\ }\href {https://doi.org/10.1007/JHEP06(2020)087}
  {\bibfield  {journal} {\bibinfo  {journal} {JHEP}\ }\textbf {\bibinfo
  {volume} {06}},\ \bibinfo {pages} {087}},\ \Eprint
  {https://arxiv.org/abs/2003.06633} {arXiv:2003.06633 [hep-ph]} \BibitemShut
  {NoStop}%
\bibitem [{\citenamefont {Yang}\ \emph {et~al.}(2020)\citenamefont {Yang},
  \citenamefont {Feng},\ and\ \citenamefont {Zhang}}]{Yang:2020bmh}%
  \BibitemOpen
  \bibfield  {author} {\bibinfo {author} {\bibfnamefont {J.-L.}\ \bibnamefont
  {Yang}}, \bibinfo {author} {\bibfnamefont {T.-F.}\ \bibnamefont {Feng}},\
  and\ \bibinfo {author} {\bibfnamefont {H.-B.}\ \bibnamefont {Zhang}},\
  }\bibfield  {title} {\bibinfo {title} {{Electron and muon $(g-2)$ in the
  B-LSSM}},\ }\href {https://doi.org/10.1088/1361-6471/ab7986} {\bibfield
  {journal} {\bibinfo  {journal} {J. Phys. G}\ }\textbf {\bibinfo {volume}
  {47}},\ \bibinfo {pages} {055004} (\bibinfo {year} {2020})},\ \Eprint
  {https://arxiv.org/abs/2003.09781} {arXiv:2003.09781 [hep-ph]} \BibitemShut
  {NoStop}%
\bibitem [{\citenamefont {Chen}\ and\ \citenamefont
  {Nomura}(2021)}]{Chen:2020jvl}%
  \BibitemOpen
  \bibfield  {author} {\bibinfo {author} {\bibfnamefont {C.-H.}\ \bibnamefont
  {Chen}}\ and\ \bibinfo {author} {\bibfnamefont {T.}~\bibnamefont {Nomura}},\
  }\bibfield  {title} {\bibinfo {title} {{Electron and muon $g-2$, radiative
  neutrino mass, and $\ell' \to \ell \gamma$ in a $U(1)_{e-\mu}$ model}},\
  }\href {https://doi.org/10.1016/j.nuclphysb.2021.115314} {\bibfield
  {journal} {\bibinfo  {journal} {Nucl. Phys. B}\ }\textbf {\bibinfo {volume}
  {964}},\ \bibinfo {pages} {115314} (\bibinfo {year} {2021})},\ \Eprint
  {https://arxiv.org/abs/2003.07638} {arXiv:2003.07638 [hep-ph]} \BibitemShut
  {NoStop}%
\bibitem [{\citenamefont {Hati}\ \emph {et~al.}(2020)\citenamefont {Hati},
  \citenamefont {Kriewald}, \citenamefont {Orloff},\ and\ \citenamefont
  {Teixeira}}]{Hati:2020fzp}%
  \BibitemOpen
  \bibfield  {author} {\bibinfo {author} {\bibfnamefont {C.}~\bibnamefont
  {Hati}}, \bibinfo {author} {\bibfnamefont {J.}~\bibnamefont {Kriewald}},
  \bibinfo {author} {\bibfnamefont {J.}~\bibnamefont {Orloff}},\ and\ \bibinfo
  {author} {\bibfnamefont {A.~M.}\ \bibnamefont {Teixeira}},\ }\bibfield
  {title} {\bibinfo {title} {{Anomalies in $^8$Be nuclear transitions and
  $(g-2)_{e,\mu}$: towards a minimal combined explanation}},\ }\href
  {https://doi.org/10.1007/JHEP07(2020)235} {\bibfield  {journal} {\bibinfo
  {journal} {JHEP}\ }\textbf {\bibinfo {volume} {07}},\ \bibinfo {pages}
  {235}},\ \Eprint {https://arxiv.org/abs/2005.00028} {arXiv:2005.00028
  [hep-ph]} \BibitemShut {NoStop}%
\bibitem [{\citenamefont {Dutta}\ \emph {et~al.}(2020)\citenamefont {Dutta},
  \citenamefont {Ghosh},\ and\ \citenamefont {Li}}]{Dutta:2020scq}%
  \BibitemOpen
  \bibfield  {author} {\bibinfo {author} {\bibfnamefont {B.}~\bibnamefont
  {Dutta}}, \bibinfo {author} {\bibfnamefont {S.}~\bibnamefont {Ghosh}},\ and\
  \bibinfo {author} {\bibfnamefont {T.}~\bibnamefont {Li}},\ }\bibfield
  {title} {\bibinfo {title} {{Explaining $(g-2)_{\mu,e}$, the KOTO anomaly and
  the MiniBooNE excess in an extended Higgs model with sterile neutrinos}},\
  }\href {https://doi.org/10.1103/PhysRevD.102.055017} {\bibfield  {journal}
  {\bibinfo  {journal} {Phys. Rev. D}\ }\textbf {\bibinfo {volume} {102}},\
  \bibinfo {pages} {055017} (\bibinfo {year} {2020})},\ \Eprint
  {https://arxiv.org/abs/2006.01319} {arXiv:2006.01319 [hep-ph]} \BibitemShut
  {NoStop}%
\bibitem [{\citenamefont {Chen}\ \emph {et~al.}(2020)\citenamefont {Chen},
  \citenamefont {Chiang},\ and\ \citenamefont {Yagyu}}]{Chen:2020tfr}%
  \BibitemOpen
  \bibfield  {author} {\bibinfo {author} {\bibfnamefont {K.-F.}\ \bibnamefont
  {Chen}}, \bibinfo {author} {\bibfnamefont {C.-W.}\ \bibnamefont {Chiang}},\
  and\ \bibinfo {author} {\bibfnamefont {K.}~\bibnamefont {Yagyu}},\ }\bibfield
   {title} {\bibinfo {title} {{An explanation for the muon and electron $g-2$
  anomalies and dark matter}},\ }\href
  {https://doi.org/10.1007/JHEP09(2020)119} {\bibfield  {journal} {\bibinfo
  {journal} {JHEP}\ }\textbf {\bibinfo {volume} {09}},\ \bibinfo {pages}
  {119}},\ \Eprint {https://arxiv.org/abs/2006.07929} {arXiv:2006.07929
  [hep-ph]} \BibitemShut {NoStop}%
\bibitem [{\citenamefont {Chun}\ and\ \citenamefont
  {Mondal}(2020)}]{Chun:2020uzw}%
  \BibitemOpen
  \bibfield  {author} {\bibinfo {author} {\bibfnamefont {E.~J.}\ \bibnamefont
  {Chun}}\ and\ \bibinfo {author} {\bibfnamefont {T.}~\bibnamefont {Mondal}},\
  }\bibfield  {title} {\bibinfo {title} {{Explaining $g-2$ anomalies in two
  Higgs doublet model with vector-like leptons}},\ }\href
  {https://doi.org/10.1007/JHEP11(2020)077} {\bibfield  {journal} {\bibinfo
  {journal} {JHEP}\ }\textbf {\bibinfo {volume} {11}},\ \bibinfo {pages}
  {077}},\ \Eprint {https://arxiv.org/abs/2009.08314} {arXiv:2009.08314
  [hep-ph]} \BibitemShut {NoStop}%
\bibitem [{\citenamefont {Li}\ \emph {et~al.}(2021)\citenamefont {Li},
  \citenamefont {Li}, \citenamefont {Li}, \citenamefont {Yang},\ and\
  \citenamefont {Zhang}}]{Li:2020dbg}%
  \BibitemOpen
  \bibfield  {author} {\bibinfo {author} {\bibfnamefont {S.-P.}\ \bibnamefont
  {Li}}, \bibinfo {author} {\bibfnamefont {X.-Q.}\ \bibnamefont {Li}}, \bibinfo
  {author} {\bibfnamefont {Y.-Y.}\ \bibnamefont {Li}}, \bibinfo {author}
  {\bibfnamefont {Y.-D.}\ \bibnamefont {Yang}},\ and\ \bibinfo {author}
  {\bibfnamefont {X.}~\bibnamefont {Zhang}},\ }\bibfield  {title} {\bibinfo
  {title} {{Power-aligned 2HDM: a correlative perspective on
  $(g-2)_{e,\mu}$}},\ }\href {https://doi.org/10.1007/JHEP01(2021)034}
  {\bibfield  {journal} {\bibinfo  {journal} {JHEP}\ }\textbf {\bibinfo
  {volume} {01}},\ \bibinfo {pages} {034}},\ \Eprint
  {https://arxiv.org/abs/2010.02799} {arXiv:2010.02799 [hep-ph]} \BibitemShut
  {NoStop}%
\bibitem [{\citenamefont {Dor\v{s}ner}\ \emph {et~al.}(2020)\citenamefont
  {Dor\v{s}ner}, \citenamefont {Fajfer},\ and\ \citenamefont
  {Saad}}]{Dorsner:2020aaz}%
  \BibitemOpen
  \bibfield  {author} {\bibinfo {author} {\bibfnamefont {I.}~\bibnamefont
  {Dor\v{s}ner}}, \bibinfo {author} {\bibfnamefont {S.}~\bibnamefont
  {Fajfer}},\ and\ \bibinfo {author} {\bibfnamefont {S.}~\bibnamefont {Saad}},\
  }\bibfield  {title} {\bibinfo {title} {{$\mu \to e \gamma$ selecting scalar
  leptoquark solutions for the $(g-2)_{e,\mu}$ puzzles}},\ }\href
  {https://doi.org/10.1103/PhysRevD.102.075007} {\bibfield  {journal} {\bibinfo
   {journal} {Phys. Rev. D}\ }\textbf {\bibinfo {volume} {102}},\ \bibinfo
  {pages} {075007} (\bibinfo {year} {2020})},\ \Eprint
  {https://arxiv.org/abs/2006.11624} {arXiv:2006.11624 [hep-ph]} \BibitemShut
  {NoStop}%
\bibitem [{\citenamefont {Keung}\ \emph {et~al.}(2021)\citenamefont {Keung},
  \citenamefont {Marfatia},\ and\ \citenamefont {Tseng}}]{Keung:2021rps}%
  \BibitemOpen
  \bibfield  {author} {\bibinfo {author} {\bibfnamefont {W.-Y.}\ \bibnamefont
  {Keung}}, \bibinfo {author} {\bibfnamefont {D.}~\bibnamefont {Marfatia}},\
  and\ \bibinfo {author} {\bibfnamefont {P.-Y.}\ \bibnamefont {Tseng}},\
  }\bibfield  {title} {\bibinfo {title} {{Axion-like particles,
  two-Higgs-doublet models, leptoquarks, and the electron and muon $g-2$}},\
  }\href@noop {} {\  (\bibinfo {year} {2021})},\ \Eprint
  {https://arxiv.org/abs/2104.03341} {arXiv:2104.03341 [hep-ph]} \BibitemShut
  {NoStop}%
\bibitem [{\citenamefont {Arbel\'aez}\ \emph {et~al.}(2020)\citenamefont
  {Arbel\'aez}, \citenamefont {Cepedello}, \citenamefont {Fonseca},\ and\
  \citenamefont {Hirsch}}]{Arbelaez:2020rbq}%
  \BibitemOpen
  \bibfield  {author} {\bibinfo {author} {\bibfnamefont {C.}~\bibnamefont
  {Arbel\'aez}}, \bibinfo {author} {\bibfnamefont {R.}~\bibnamefont
  {Cepedello}}, \bibinfo {author} {\bibfnamefont {R.~M.}\ \bibnamefont
  {Fonseca}},\ and\ \bibinfo {author} {\bibfnamefont {M.}~\bibnamefont
  {Hirsch}},\ }\bibfield  {title} {\bibinfo {title} {{$(g-2)$ anomalies and
  neutrino mass}},\ }\href {https://doi.org/10.1103/PhysRevD.102.075005}
  {\bibfield  {journal} {\bibinfo  {journal} {Phys. Rev. D}\ }\textbf {\bibinfo
  {volume} {102}},\ \bibinfo {pages} {075005} (\bibinfo {year} {2020})},\
  \Eprint {https://arxiv.org/abs/2007.11007} {arXiv:2007.11007 [hep-ph]}
  \BibitemShut {NoStop}%
\bibitem [{\citenamefont {Jana}\ \emph
  {et~al.}(2020{\natexlab{b}})\citenamefont {Jana}, \citenamefont {Vishnu},
  \citenamefont {Rodejohann},\ and\ \citenamefont {Saad}}]{Jana:2020joi}%
  \BibitemOpen
  \bibfield  {author} {\bibinfo {author} {\bibfnamefont {S.}~\bibnamefont
  {Jana}}, \bibinfo {author} {\bibfnamefont {P.~K.}\ \bibnamefont {Vishnu}},
  \bibinfo {author} {\bibfnamefont {W.}~\bibnamefont {Rodejohann}},\ and\
  \bibinfo {author} {\bibfnamefont {S.}~\bibnamefont {Saad}},\ }\bibfield
  {title} {\bibinfo {title} {{Dark matter assisted lepton anomalous magnetic
  moments and neutrino masses}},\ }\href
  {https://doi.org/10.1103/PhysRevD.102.075003} {\bibfield  {journal} {\bibinfo
   {journal} {Phys. Rev. D}\ }\textbf {\bibinfo {volume} {102}},\ \bibinfo
  {pages} {075003} (\bibinfo {year} {2020}{\natexlab{b}})},\ \Eprint
  {https://arxiv.org/abs/2008.02377} {arXiv:2008.02377 [hep-ph]} \BibitemShut
  {NoStop}%
\bibitem [{\citenamefont {Escribano}\ \emph {et~al.}(2021)\citenamefont
  {Escribano}, \citenamefont {Terol-Calvo},\ and\ \citenamefont
  {Vicente}}]{Escribano:2021css}%
  \BibitemOpen
  \bibfield  {author} {\bibinfo {author} {\bibfnamefont {P.}~\bibnamefont
  {Escribano}}, \bibinfo {author} {\bibfnamefont {J.}~\bibnamefont
  {Terol-Calvo}},\ and\ \bibinfo {author} {\bibfnamefont {A.}~\bibnamefont
  {Vicente}},\ }\bibfield  {title} {\bibinfo {title}
  {{$\boldsymbol{(g-2)_{e,\mu}}$ in an extended inverse type-III seesaw}},\
  }\href@noop {} {\  (\bibinfo {year} {2021})},\ \Eprint
  {https://arxiv.org/abs/2104.03705} {arXiv:2104.03705 [hep-ph]} \BibitemShut
  {NoStop}%
\bibitem [{\citenamefont {Morel}\ \emph {et~al.}(2020)\citenamefont {Morel},
  \citenamefont {Yao}, \citenamefont {Clad\'e},\ and\ \citenamefont
  {Guellati-Kh\'elifa}}]{Morel:2020dww}%
  \BibitemOpen
  \bibfield  {author} {\bibinfo {author} {\bibfnamefont {L.}~\bibnamefont
  {Morel}}, \bibinfo {author} {\bibfnamefont {Z.}~\bibnamefont {Yao}}, \bibinfo
  {author} {\bibfnamefont {P.}~\bibnamefont {Clad\'e}},\ and\ \bibinfo {author}
  {\bibfnamefont {S.}~\bibnamefont {Guellati-Kh\'elifa}},\ }\bibfield  {title}
  {\bibinfo {title} {{Determination of the fine-structure constant with an
  accuracy of 81 parts per trillion}},\ }\href
  {https://doi.org/10.1038/s41586-020-2964-7} {\bibfield  {journal} {\bibinfo
  {journal} {Nature}\ }\textbf {\bibinfo {volume} {588}},\ \bibinfo {pages}
  {61} (\bibinfo {year} {2020})}\BibitemShut {NoStop}%
\bibitem [{\citenamefont {Capdevila}\ \emph
  {et~al.}(2018{\natexlab{a}})\citenamefont {Capdevila}, \citenamefont
  {Crivellin}, \citenamefont {Descotes-Genon}, \citenamefont {Matias},\ and\
  \citenamefont {Virto}}]{Capdevila:2017bsm}%
  \BibitemOpen
  \bibfield  {author} {\bibinfo {author} {\bibfnamefont {B.}~\bibnamefont
  {Capdevila}}, \bibinfo {author} {\bibfnamefont {A.}~\bibnamefont
  {Crivellin}}, \bibinfo {author} {\bibfnamefont {S.}~\bibnamefont
  {Descotes-Genon}}, \bibinfo {author} {\bibfnamefont {J.}~\bibnamefont
  {Matias}},\ and\ \bibinfo {author} {\bibfnamefont {J.}~\bibnamefont
  {Virto}},\ }\bibfield  {title} {\bibinfo {title} {{Patterns of New Physics in
  $b\to s\ell^+\ell^-$ transitions in the light of recent data}},\ }\href
  {https://doi.org/10.1007/JHEP01(2018)093} {\bibfield  {journal} {\bibinfo
  {journal} {JHEP}\ }\textbf {\bibinfo {volume} {01}},\ \bibinfo {pages}
  {093}},\ \Eprint {https://arxiv.org/abs/1704.05340} {arXiv:1704.05340
  [hep-ph]} \BibitemShut {NoStop}%
\bibitem [{\citenamefont {Altmannshofer}\ \emph {et~al.}(2017)\citenamefont
  {Altmannshofer}, \citenamefont {Stangl},\ and\ \citenamefont
  {Straub}}]{Altmannshofer:2017yso}%
  \BibitemOpen
  \bibfield  {author} {\bibinfo {author} {\bibfnamefont {W.}~\bibnamefont
  {Altmannshofer}}, \bibinfo {author} {\bibfnamefont {P.}~\bibnamefont
  {Stangl}},\ and\ \bibinfo {author} {\bibfnamefont {D.~M.}\ \bibnamefont
  {Straub}},\ }\bibfield  {title} {\bibinfo {title} {{Interpreting Hints for
  Lepton Flavor Universality Violation}},\ }\href
  {https://doi.org/10.1103/PhysRevD.96.055008} {\bibfield  {journal} {\bibinfo
  {journal} {Phys. Rev. D}\ }\textbf {\bibinfo {volume} {96}},\ \bibinfo
  {pages} {055008} (\bibinfo {year} {2017})},\ \Eprint
  {https://arxiv.org/abs/1704.05435} {arXiv:1704.05435 [hep-ph]} \BibitemShut
  {NoStop}%
\bibitem [{\citenamefont {D'Amico}\ \emph {et~al.}(2017)\citenamefont
  {D'Amico}, \citenamefont {Nardecchia}, \citenamefont {Panci}, \citenamefont
  {Sannino}, \citenamefont {Strumia}, \citenamefont {Torre},\ and\
  \citenamefont {Urbano}}]{DAmico:2017mtc}%
  \BibitemOpen
  \bibfield  {author} {\bibinfo {author} {\bibfnamefont {G.}~\bibnamefont
  {D'Amico}}, \bibinfo {author} {\bibfnamefont {M.}~\bibnamefont {Nardecchia}},
  \bibinfo {author} {\bibfnamefont {P.}~\bibnamefont {Panci}}, \bibinfo
  {author} {\bibfnamefont {F.}~\bibnamefont {Sannino}}, \bibinfo {author}
  {\bibfnamefont {A.}~\bibnamefont {Strumia}}, \bibinfo {author} {\bibfnamefont
  {R.}~\bibnamefont {Torre}},\ and\ \bibinfo {author} {\bibfnamefont
  {A.}~\bibnamefont {Urbano}},\ }\bibfield  {title} {\bibinfo {title} {{Flavour
  anomalies after the $R_{K^*}$ measurement}},\ }\href
  {https://doi.org/10.1007/JHEP09(2017)010} {\bibfield  {journal} {\bibinfo
  {journal} {JHEP}\ }\textbf {\bibinfo {volume} {09}},\ \bibinfo {pages}
  {010}},\ \Eprint {https://arxiv.org/abs/1704.05438} {arXiv:1704.05438
  [hep-ph]} \BibitemShut {NoStop}%
\bibitem [{\citenamefont {Hiller}\ and\ \citenamefont
  {Nisandzic}(2017)}]{Hiller:2017bzc}%
  \BibitemOpen
  \bibfield  {author} {\bibinfo {author} {\bibfnamefont {G.}~\bibnamefont
  {Hiller}}\ and\ \bibinfo {author} {\bibfnamefont {I.}~\bibnamefont
  {Nisandzic}},\ }\bibfield  {title} {\bibinfo {title} {{$R_K$ and
  $R_{K^{\ast}}$ beyond the standard model}},\ }\href
  {https://doi.org/10.1103/PhysRevD.96.035003} {\bibfield  {journal} {\bibinfo
  {journal} {Phys. Rev. D}\ }\textbf {\bibinfo {volume} {96}},\ \bibinfo
  {pages} {035003} (\bibinfo {year} {2017})},\ \Eprint
  {https://arxiv.org/abs/1704.05444} {arXiv:1704.05444 [hep-ph]} \BibitemShut
  {NoStop}%
\bibitem [{\citenamefont {Ciuchini}\ \emph {et~al.}(2017)\citenamefont
  {Ciuchini}, \citenamefont {Coutinho}, \citenamefont {Fedele}, \citenamefont
  {Franco}, \citenamefont {Paul}, \citenamefont {Silvestrini},\ and\
  \citenamefont {Valli}}]{Ciuchini:2017mik}%
  \BibitemOpen
  \bibfield  {author} {\bibinfo {author} {\bibfnamefont {M.}~\bibnamefont
  {Ciuchini}}, \bibinfo {author} {\bibfnamefont {A.~M.}\ \bibnamefont
  {Coutinho}}, \bibinfo {author} {\bibfnamefont {M.}~\bibnamefont {Fedele}},
  \bibinfo {author} {\bibfnamefont {E.}~\bibnamefont {Franco}}, \bibinfo
  {author} {\bibfnamefont {A.}~\bibnamefont {Paul}}, \bibinfo {author}
  {\bibfnamefont {L.}~\bibnamefont {Silvestrini}},\ and\ \bibinfo {author}
  {\bibfnamefont {M.}~\bibnamefont {Valli}},\ }\bibfield  {title} {\bibinfo
  {title} {{On Flavourful Easter eggs for New Physics hunger and Lepton Flavour
  Universality violation}},\ }\href
  {https://doi.org/10.1140/epjc/s10052-017-5270-2} {\bibfield  {journal}
  {\bibinfo  {journal} {Eur. Phys. J. C}\ }\textbf {\bibinfo {volume} {77}},\
  \bibinfo {pages} {688} (\bibinfo {year} {2017})},\ \Eprint
  {https://arxiv.org/abs/1704.05447} {arXiv:1704.05447 [hep-ph]} \BibitemShut
  {NoStop}%
\bibitem [{\citenamefont {Geng}\ \emph {et~al.}(2017)\citenamefont {Geng},
  \citenamefont {Grinstein}, \citenamefont {J\"ager}, \citenamefont
  {Martin~Camalich}, \citenamefont {Ren},\ and\ \citenamefont
  {Shi}}]{Geng:2017svp}%
  \BibitemOpen
  \bibfield  {author} {\bibinfo {author} {\bibfnamefont {L.-S.}\ \bibnamefont
  {Geng}}, \bibinfo {author} {\bibfnamefont {B.}~\bibnamefont {Grinstein}},
  \bibinfo {author} {\bibfnamefont {S.}~\bibnamefont {J\"ager}}, \bibinfo
  {author} {\bibfnamefont {J.}~\bibnamefont {Martin~Camalich}}, \bibinfo
  {author} {\bibfnamefont {X.-L.}\ \bibnamefont {Ren}},\ and\ \bibinfo {author}
  {\bibfnamefont {R.-X.}\ \bibnamefont {Shi}},\ }\bibfield  {title} {\bibinfo
  {title} {{Towards the discovery of new physics with lepton-universality
  ratios of $b\to s\ell\ell$ decays}},\ }\href
  {https://doi.org/10.1103/PhysRevD.96.093006} {\bibfield  {journal} {\bibinfo
  {journal} {Phys. Rev. D}\ }\textbf {\bibinfo {volume} {96}},\ \bibinfo
  {pages} {093006} (\bibinfo {year} {2017})},\ \Eprint
  {https://arxiv.org/abs/1704.05446} {arXiv:1704.05446 [hep-ph]} \BibitemShut
  {NoStop}%
\bibitem [{\citenamefont {Hurth}\ \emph {et~al.}(2017)\citenamefont {Hurth},
  \citenamefont {Mahmoudi}, \citenamefont {Martinez~Santos},\ and\
  \citenamefont {Neshatpour}}]{Hurth:2017hxg}%
  \BibitemOpen
  \bibfield  {author} {\bibinfo {author} {\bibfnamefont {T.}~\bibnamefont
  {Hurth}}, \bibinfo {author} {\bibfnamefont {F.}~\bibnamefont {Mahmoudi}},
  \bibinfo {author} {\bibfnamefont {D.}~\bibnamefont {Martinez~Santos}},\ and\
  \bibinfo {author} {\bibfnamefont {S.}~\bibnamefont {Neshatpour}},\ }\bibfield
   {title} {\bibinfo {title} {{Lepton nonuniversality in exclusive
  $b{\rightarrow}s{\ell}{\ell}$ decays}},\ }\href
  {https://doi.org/10.1103/PhysRevD.96.095034} {\bibfield  {journal} {\bibinfo
  {journal} {Phys. Rev. D}\ }\textbf {\bibinfo {volume} {96}},\ \bibinfo
  {pages} {095034} (\bibinfo {year} {2017})},\ \Eprint
  {https://arxiv.org/abs/1705.06274} {arXiv:1705.06274 [hep-ph]} \BibitemShut
  {NoStop}%
\bibitem [{\citenamefont {Alok}\ \emph {et~al.}(2017)\citenamefont {Alok},
  \citenamefont {Bhattacharya}, \citenamefont {Datta}, \citenamefont {Kumar},
  \citenamefont {Kumar},\ and\ \citenamefont {London}}]{Alok:2017sui}%
  \BibitemOpen
  \bibfield  {author} {\bibinfo {author} {\bibfnamefont {A.~K.}\ \bibnamefont
  {Alok}}, \bibinfo {author} {\bibfnamefont {B.}~\bibnamefont {Bhattacharya}},
  \bibinfo {author} {\bibfnamefont {A.}~\bibnamefont {Datta}}, \bibinfo
  {author} {\bibfnamefont {D.}~\bibnamefont {Kumar}}, \bibinfo {author}
  {\bibfnamefont {J.}~\bibnamefont {Kumar}},\ and\ \bibinfo {author}
  {\bibfnamefont {D.}~\bibnamefont {London}},\ }\bibfield  {title} {\bibinfo
  {title} {{New Physics in $b \to s \mu^+ \mu^-$ after the Measurement of
  $R_{K^*}$}},\ }\href {https://doi.org/10.1103/PhysRevD.96.095009} {\bibfield
  {journal} {\bibinfo  {journal} {Phys. Rev. D}\ }\textbf {\bibinfo {volume}
  {96}},\ \bibinfo {pages} {095009} (\bibinfo {year} {2017})},\ \Eprint
  {https://arxiv.org/abs/1704.07397} {arXiv:1704.07397 [hep-ph]} \BibitemShut
  {NoStop}%
\bibitem [{\citenamefont {Alguer\'o}\ \emph {et~al.}(2019)\citenamefont
  {Alguer\'o}, \citenamefont {Capdevila}, \citenamefont {Crivellin},
  \citenamefont {Descotes-Genon}, \citenamefont {Masjuan}, \citenamefont
  {Matias}, \citenamefont {Novoa~Brunet},\ and\ \citenamefont
  {Virto}}]{Alguero:2019ptt}%
  \BibitemOpen
  \bibfield  {author} {\bibinfo {author} {\bibfnamefont {M.}~\bibnamefont
  {Alguer\'o}}, \bibinfo {author} {\bibfnamefont {B.}~\bibnamefont
  {Capdevila}}, \bibinfo {author} {\bibfnamefont {A.}~\bibnamefont
  {Crivellin}}, \bibinfo {author} {\bibfnamefont {S.}~\bibnamefont
  {Descotes-Genon}}, \bibinfo {author} {\bibfnamefont {P.}~\bibnamefont
  {Masjuan}}, \bibinfo {author} {\bibfnamefont {J.}~\bibnamefont {Matias}},
  \bibinfo {author} {\bibfnamefont {M.}~\bibnamefont {Novoa~Brunet}},\ and\
  \bibinfo {author} {\bibfnamefont {J.}~\bibnamefont {Virto}},\ }\bibfield
  {title} {\bibinfo {title} {{Emerging patterns of New Physics with and without
  Lepton Flavour Universal contributions}},\ }\href
  {https://doi.org/10.1140/epjc/s10052-019-7216-3} {\bibfield  {journal}
  {\bibinfo  {journal} {Eur. Phys. J. C}\ }\textbf {\bibinfo {volume} {79}},\
  \bibinfo {pages} {714} (\bibinfo {year} {2019})},\ \bibinfo {note}
  {[Addendum: Eur.Phys.J.C 80, 511 (2020)]},\ \Eprint
  {https://arxiv.org/abs/1903.09578} {arXiv:1903.09578 [hep-ph]} \BibitemShut
  {NoStop}%
\bibitem [{\citenamefont {Aebischer}\ \emph {et~al.}(2020)\citenamefont
  {Aebischer}, \citenamefont {Altmannshofer}, \citenamefont {Guadagnoli},
  \citenamefont {Reboud}, \citenamefont {Stangl},\ and\ \citenamefont
  {Straub}}]{Aebischer:2019mlg}%
  \BibitemOpen
  \bibfield  {author} {\bibinfo {author} {\bibfnamefont {J.}~\bibnamefont
  {Aebischer}}, \bibinfo {author} {\bibfnamefont {W.}~\bibnamefont
  {Altmannshofer}}, \bibinfo {author} {\bibfnamefont {D.}~\bibnamefont
  {Guadagnoli}}, \bibinfo {author} {\bibfnamefont {M.}~\bibnamefont {Reboud}},
  \bibinfo {author} {\bibfnamefont {P.}~\bibnamefont {Stangl}},\ and\ \bibinfo
  {author} {\bibfnamefont {D.~M.}\ \bibnamefont {Straub}},\ }\bibfield  {title}
  {\bibinfo {title} {{$B$-decay discrepancies after Moriond 2019}},\ }\href
  {https://doi.org/10.1140/epjc/s10052-020-7817-x} {\bibfield  {journal}
  {\bibinfo  {journal} {Eur. Phys. J. C}\ }\textbf {\bibinfo {volume} {80}},\
  \bibinfo {pages} {252} (\bibinfo {year} {2020})},\ \Eprint
  {https://arxiv.org/abs/1903.10434} {arXiv:1903.10434 [hep-ph]} \BibitemShut
  {NoStop}%
\bibitem [{\citenamefont {Ciuchini}\ \emph {et~al.}(2019)\citenamefont
  {Ciuchini}, \citenamefont {Coutinho}, \citenamefont {Fedele}, \citenamefont
  {Franco}, \citenamefont {Paul}, \citenamefont {Silvestrini},\ and\
  \citenamefont {Valli}}]{Ciuchini:2019usw}%
  \BibitemOpen
  \bibfield  {author} {\bibinfo {author} {\bibfnamefont {M.}~\bibnamefont
  {Ciuchini}}, \bibinfo {author} {\bibfnamefont {A.~M.}\ \bibnamefont
  {Coutinho}}, \bibinfo {author} {\bibfnamefont {M.}~\bibnamefont {Fedele}},
  \bibinfo {author} {\bibfnamefont {E.}~\bibnamefont {Franco}}, \bibinfo
  {author} {\bibfnamefont {A.}~\bibnamefont {Paul}}, \bibinfo {author}
  {\bibfnamefont {L.}~\bibnamefont {Silvestrini}},\ and\ \bibinfo {author}
  {\bibfnamefont {M.}~\bibnamefont {Valli}},\ }\bibfield  {title} {\bibinfo
  {title} {{New Physics in $b \to s \ell^+ \ell^-$ confronts new data on Lepton
  Universality}},\ }\href {https://doi.org/10.1140/epjc/s10052-019-7210-9}
  {\bibfield  {journal} {\bibinfo  {journal} {Eur. Phys. J. C}\ }\textbf
  {\bibinfo {volume} {79}},\ \bibinfo {pages} {719} (\bibinfo {year} {2019})},\
  \Eprint {https://arxiv.org/abs/1903.09632} {arXiv:1903.09632 [hep-ph]}
  \BibitemShut {NoStop}%
\bibitem [{\citenamefont {Datta}\ \emph {et~al.}(2019)\citenamefont {Datta},
  \citenamefont {Kumar},\ and\ \citenamefont {London}}]{Datta:2019zca}%
  \BibitemOpen
  \bibfield  {author} {\bibinfo {author} {\bibfnamefont {A.}~\bibnamefont
  {Datta}}, \bibinfo {author} {\bibfnamefont {J.}~\bibnamefont {Kumar}},\ and\
  \bibinfo {author} {\bibfnamefont {D.}~\bibnamefont {London}},\ }\bibfield
  {title} {\bibinfo {title} {{The $B$ anomalies and new physics in $b \to s e^+
  e^-$}},\ }\href {https://doi.org/10.1016/j.physletb.2019.134858} {\bibfield
  {journal} {\bibinfo  {journal} {Phys. Lett. B}\ }\textbf {\bibinfo {volume}
  {797}},\ \bibinfo {pages} {134858} (\bibinfo {year} {2019})},\ \Eprint
  {https://arxiv.org/abs/1903.10086} {arXiv:1903.10086 [hep-ph]} \BibitemShut
  {NoStop}%
\bibitem [{\citenamefont {Aaij}\ \emph {et~al.}(2019)\citenamefont {Aaij} \emph
  {et~al.}}]{Aaij:2019wad}%
  \BibitemOpen
  \bibfield  {author} {\bibinfo {author} {\bibfnamefont {R.}~\bibnamefont
  {Aaij}} \emph {et~al.} (\bibinfo {collaboration} {LHCb}),\ }\bibfield
  {title} {\bibinfo {title} {{Search for lepton-universality violation in
  $B^+\to K^+\ell^+\ell^-$ decays}},\ }\href
  {https://doi.org/10.1103/PhysRevLett.122.191801} {\bibfield  {journal}
  {\bibinfo  {journal} {Phys. Rev. Lett.}\ }\textbf {\bibinfo {volume} {122}},\
  \bibinfo {pages} {191801} (\bibinfo {year} {2019})},\ \Eprint
  {https://arxiv.org/abs/1903.09252} {arXiv:1903.09252 [hep-ex]} \BibitemShut
  {NoStop}%
\bibitem [{\citenamefont {Aaij}\ \emph
  {et~al.}(2017{\natexlab{a}})\citenamefont {Aaij} \emph
  {et~al.}}]{Aaij:2017vbb}%
  \BibitemOpen
  \bibfield  {author} {\bibinfo {author} {\bibfnamefont {R.}~\bibnamefont
  {Aaij}} \emph {et~al.} (\bibinfo {collaboration} {LHCb}),\ }\bibfield
  {title} {\bibinfo {title} {{Test of lepton universality with $B^{0}
  \rightarrow K^{*0}\ell^{+}\ell^{-}$ decays}},\ }\href
  {https://doi.org/10.1007/JHEP08(2017)055} {\bibfield  {journal} {\bibinfo
  {journal} {JHEP}\ }\textbf {\bibinfo {volume} {08}},\ \bibinfo {pages}
  {055}},\ \Eprint {https://arxiv.org/abs/1705.05802} {arXiv:1705.05802
  [hep-ex]} \BibitemShut {NoStop}%
\bibitem [{\citenamefont {Abdesselam}\ \emph
  {et~al.}(2019{\natexlab{a}})\citenamefont {Abdesselam} \emph
  {et~al.}}]{Abdesselam:2019wac}%
  \BibitemOpen
  \bibfield  {author} {\bibinfo {author} {\bibfnamefont {A.}~\bibnamefont
  {Abdesselam}} \emph {et~al.} (\bibinfo {collaboration} {Belle}),\ }\bibfield
  {title} {\bibinfo {title} {{Test of lepton flavor universality in ${B\to
  K^\ast\ell^+\ell^-}$ decays at Belle}},\ }\href@noop {} {\  (\bibinfo {year}
  {2019}{\natexlab{a}})},\ \Eprint {https://arxiv.org/abs/1904.02440}
  {arXiv:1904.02440 [hep-ex]} \BibitemShut {NoStop}%
\bibitem [{\citenamefont {Abdesselam}\ \emph
  {et~al.}(2019{\natexlab{b}})\citenamefont {Abdesselam} \emph
  {et~al.}}]{Abdesselam:2019lab}%
  \BibitemOpen
  \bibfield  {author} {\bibinfo {author} {\bibfnamefont {A.}~\bibnamefont
  {Abdesselam}} \emph {et~al.} (\bibinfo {collaboration} {Belle}),\ }\bibfield
  {title} {\bibinfo {title} {{Test of lepton flavor universality in $B \to K
  \ell^{+}\ell^{-}$ decays}},\ }\href@noop {} {\  (\bibinfo {year}
  {2019}{\natexlab{b}})},\ \Eprint {https://arxiv.org/abs/1908.01848}
  {arXiv:1908.01848 [hep-ex]} \BibitemShut {NoStop}%
\bibitem [{\citenamefont {Aad}\ \emph {et~al.}(2014)\citenamefont {Aad} \emph
  {et~al.}}]{Aad:2014fwa}%
  \BibitemOpen
  \bibfield  {author} {\bibinfo {author} {\bibfnamefont {G.}~\bibnamefont
  {Aad}} \emph {et~al.} (\bibinfo {collaboration} {ATLAS}),\ }\bibfield
  {title} {\bibinfo {title} {{Comprehensive measurements of $t$-channel single
  top-quark production cross sections at $\sqrt{s} = 7$ TeV with the ATLAS
  detector}},\ }\href {https://doi.org/10.1103/PhysRevD.90.112006} {\bibfield
  {journal} {\bibinfo  {journal} {Phys. Rev. D}\ }\textbf {\bibinfo {volume}
  {90}},\ \bibinfo {pages} {112006} (\bibinfo {year} {2014})},\ \Eprint
  {https://arxiv.org/abs/1406.7844} {arXiv:1406.7844 [hep-ex]} \BibitemShut
  {NoStop}%
\bibitem [{\citenamefont {Aaij}\ \emph
  {et~al.}(2015{\natexlab{a}})\citenamefont {Aaij} \emph
  {et~al.}}]{Aaij:2015esa}%
  \BibitemOpen
  \bibfield  {author} {\bibinfo {author} {\bibfnamefont {R.}~\bibnamefont
  {Aaij}} \emph {et~al.} (\bibinfo {collaboration} {LHCb}),\ }\bibfield
  {title} {\bibinfo {title} {{Angular analysis and differential branching
  fraction of the decay $B^0_s\to\phi\mu^+\mu^-$}},\ }\href
  {https://doi.org/10.1007/JHEP09(2015)179} {\bibfield  {journal} {\bibinfo
  {journal} {JHEP}\ }\textbf {\bibinfo {volume} {09}},\ \bibinfo {pages}
  {179}},\ \Eprint {https://arxiv.org/abs/1506.08777} {arXiv:1506.08777
  [hep-ex]} \BibitemShut {NoStop}%
\bibitem [{\citenamefont {Wei}\ \emph {et~al.}(2009)\citenamefont {Wei} \emph
  {et~al.}}]{Wei:2009zv}%
  \BibitemOpen
  \bibfield  {author} {\bibinfo {author} {\bibfnamefont {J.~T.}\ \bibnamefont
  {Wei}} \emph {et~al.} (\bibinfo {collaboration} {Belle}),\ }\bibfield
  {title} {\bibinfo {title} {{Measurement of the Differential Branching
  Fraction and Forward-Backword Asymmetry for $B \to K^{(*)}\ell^+\ell^-$}},\
  }\href {https://doi.org/10.1103/PhysRevLett.103.171801} {\bibfield  {journal}
  {\bibinfo  {journal} {Phys. Rev. Lett.}\ }\textbf {\bibinfo {volume} {103}},\
  \bibinfo {pages} {171801} (\bibinfo {year} {2009})},\ \Eprint
  {https://arxiv.org/abs/0904.0770} {arXiv:0904.0770 [hep-ex]} \BibitemShut
  {NoStop}%
\bibitem [{\citenamefont {Aaltonen}\ \emph {et~al.}(2012)\citenamefont
  {Aaltonen} \emph {et~al.}}]{Aaltonen:2011ja}%
  \BibitemOpen
  \bibfield  {author} {\bibinfo {author} {\bibfnamefont {T.}~\bibnamefont
  {Aaltonen}} \emph {et~al.} (\bibinfo {collaboration} {CDF}),\ }\bibfield
  {title} {\bibinfo {title} {{Measurements of the Angular Distributions in the
  Decays $B \to K^{(*)} \mu^+ \mu^-$ at CDF}},\ }\href
  {https://doi.org/10.1103/PhysRevLett.108.081807} {\bibfield  {journal}
  {\bibinfo  {journal} {Phys. Rev. Lett.}\ }\textbf {\bibinfo {volume} {108}},\
  \bibinfo {pages} {081807} (\bibinfo {year} {2012})},\ \Eprint
  {https://arxiv.org/abs/1108.0695} {arXiv:1108.0695 [hep-ex]} \BibitemShut
  {NoStop}%
\bibitem [{\citenamefont {Khachatryan}\ \emph {et~al.}(2016)\citenamefont
  {Khachatryan} \emph {et~al.}}]{Khachatryan:2015isa}%
  \BibitemOpen
  \bibfield  {author} {\bibinfo {author} {\bibfnamefont {V.}~\bibnamefont
  {Khachatryan}} \emph {et~al.} (\bibinfo {collaboration} {CMS}),\ }\bibfield
  {title} {\bibinfo {title} {{Angular analysis of the decay $B^0 \to K^{*0}
  \mu^+ \mu^-$ from pp collisions at $\sqrt s = 8$ TeV}},\ }\href
  {https://doi.org/10.1016/j.physletb.2015.12.020} {\bibfield  {journal}
  {\bibinfo  {journal} {Phys. Lett. B}\ }\textbf {\bibinfo {volume} {753}},\
  \bibinfo {pages} {424} (\bibinfo {year} {2016})},\ \Eprint
  {https://arxiv.org/abs/1507.08126} {arXiv:1507.08126 [hep-ex]} \BibitemShut
  {NoStop}%
\bibitem [{\citenamefont {Abdesselam}\ \emph {et~al.}(2016)\citenamefont
  {Abdesselam} \emph {et~al.}}]{Abdesselam:2016llu}%
  \BibitemOpen
  \bibfield  {author} {\bibinfo {author} {\bibfnamefont {A.}~\bibnamefont
  {Abdesselam}} \emph {et~al.} (\bibinfo {collaboration} {Belle}),\ }\bibfield
  {title} {\bibinfo {title} {{Angular analysis of $B^0 \to (K^*(892))^0 \ell^+
  \ell^-$}},\ }in\ \href@noop {} {\emph {\bibinfo {booktitle} {{LHC Ski 2016}:
  {A First Discussion of 13 TeV Results}}}}\ (\bibinfo {year} {2016})\ \Eprint
  {https://arxiv.org/abs/1604.04042} {arXiv:1604.04042 [hep-ex]} \BibitemShut
  {NoStop}%
\bibitem [{\citenamefont {Aaij}\ \emph {et~al.}(2016)\citenamefont {Aaij} \emph
  {et~al.}}]{Aaij:2015oid}%
  \BibitemOpen
  \bibfield  {author} {\bibinfo {author} {\bibfnamefont {R.}~\bibnamefont
  {Aaij}} \emph {et~al.} (\bibinfo {collaboration} {LHCb}),\ }\bibfield
  {title} {\bibinfo {title} {{Angular analysis of the $B^{0} \to K^{*0} \mu^{+}
  \mu^{-}$ decay using 3 fb$^{-1}$ of integrated luminosity}},\ }\href
  {https://doi.org/10.1007/JHEP02(2016)104} {\bibfield  {journal} {\bibinfo
  {journal} {JHEP}\ }\textbf {\bibinfo {volume} {02}},\ \bibinfo {pages}
  {104}},\ \Eprint {https://arxiv.org/abs/1512.04442} {arXiv:1512.04442
  [hep-ex]} \BibitemShut {NoStop}%
\bibitem [{\citenamefont {Wehle}\ \emph {et~al.}(2017)\citenamefont {Wehle}
  \emph {et~al.}}]{Wehle:2016yoi}%
  \BibitemOpen
  \bibfield  {author} {\bibinfo {author} {\bibfnamefont {S.}~\bibnamefont
  {Wehle}} \emph {et~al.} (\bibinfo {collaboration} {Belle}),\ }\bibfield
  {title} {\bibinfo {title} {{Lepton-Flavor-Dependent Angular Analysis of $B\to
  K^\ast \ell^+\ell^-$}},\ }\href
  {https://doi.org/10.1103/PhysRevLett.118.111801} {\bibfield  {journal}
  {\bibinfo  {journal} {Phys. Rev. Lett.}\ }\textbf {\bibinfo {volume} {118}},\
  \bibinfo {pages} {111801} (\bibinfo {year} {2017})},\ \Eprint
  {https://arxiv.org/abs/1612.05014} {arXiv:1612.05014 [hep-ex]} \BibitemShut
  {NoStop}%
\bibitem [{\citenamefont {Sirunyan}\ \emph {et~al.}(2018)\citenamefont
  {Sirunyan} \emph {et~al.}}]{Sirunyan:2017dhj}%
  \BibitemOpen
  \bibfield  {author} {\bibinfo {author} {\bibfnamefont {A.~M.}\ \bibnamefont
  {Sirunyan}} \emph {et~al.} (\bibinfo {collaboration} {CMS}),\ }\bibfield
  {title} {\bibinfo {title} {{Measurement of angular parameters from the decay
  $\mathrm{B}^0 \to \mathrm{K}^{*0} \mu^+ \mu^-$ in proton-proton collisions at
  $\sqrt{s} = $ 8 TeV}},\ }\href
  {https://doi.org/10.1016/j.physletb.2018.04.030} {\bibfield  {journal}
  {\bibinfo  {journal} {Phys. Lett. B}\ }\textbf {\bibinfo {volume} {781}},\
  \bibinfo {pages} {517} (\bibinfo {year} {2018})},\ \Eprint
  {https://arxiv.org/abs/1710.02846} {arXiv:1710.02846 [hep-ex]} \BibitemShut
  {NoStop}%
\bibitem [{\citenamefont {Aaboud}\ \emph
  {et~al.}(2018{\natexlab{a}})\citenamefont {Aaboud} \emph
  {et~al.}}]{Aaboud:2018krd}%
  \BibitemOpen
  \bibfield  {author} {\bibinfo {author} {\bibfnamefont {M.}~\bibnamefont
  {Aaboud}} \emph {et~al.} (\bibinfo {collaboration} {ATLAS}),\ }\bibfield
  {title} {\bibinfo {title} {{Angular analysis of $B^0_d \rightarrow
  K^{*}\mu^+\mu^-$ decays in $pp$ collisions at $\sqrt{s}= 8$ TeV with the
  ATLAS detector}},\ }\href {https://doi.org/10.1007/JHEP10(2018)047}
  {\bibfield  {journal} {\bibinfo  {journal} {JHEP}\ }\textbf {\bibinfo
  {volume} {10}},\ \bibinfo {pages} {047}},\ \Eprint
  {https://arxiv.org/abs/1805.04000} {arXiv:1805.04000 [hep-ex]} \BibitemShut
  {NoStop}%
\bibitem [{\citenamefont {Lees}\ \emph {et~al.}(2016)\citenamefont {Lees} \emph
  {et~al.}}]{Lees:2015ymt}%
  \BibitemOpen
  \bibfield  {author} {\bibinfo {author} {\bibfnamefont {J.~P.}\ \bibnamefont
  {Lees}} \emph {et~al.} (\bibinfo {collaboration} {BaBar}),\ }\bibfield
  {title} {\bibinfo {title} {{Measurement of angular asymmetries in the decays
  $B \rightarrow K^* l^+ l^-$}},\ }\href
  {https://doi.org/10.1103/PhysRevD.93.052015} {\bibfield  {journal} {\bibinfo
  {journal} {Phys. Rev. D}\ }\textbf {\bibinfo {volume} {93}},\ \bibinfo
  {pages} {052015} (\bibinfo {year} {2016})},\ \Eprint
  {https://arxiv.org/abs/1508.07960} {arXiv:1508.07960 [hep-ex]} \BibitemShut
  {NoStop}%
\bibitem [{\citenamefont {Aaij}\ \emph {et~al.}(2021)\citenamefont {Aaij} \emph
  {et~al.}}]{Aaij:2021vac}%
  \BibitemOpen
  \bibfield  {author} {\bibinfo {author} {\bibfnamefont {R.}~\bibnamefont
  {Aaij}} \emph {et~al.} (\bibinfo {collaboration} {LHCb}),\ }\bibfield
  {title} {\bibinfo {title} {{Test of lepton universality in beauty-quark
  decays}},\ }\href@noop {} {\  (\bibinfo {year} {2021})},\ \Eprint
  {https://arxiv.org/abs/2103.11769} {arXiv:2103.11769 [hep-ex]} \BibitemShut
  {NoStop}%
\bibitem [{\citenamefont {Altmannshofer}\ and\ \citenamefont
  {Stangl}(2021)}]{Altmannshofer:2021qrr}%
  \BibitemOpen
  \bibfield  {author} {\bibinfo {author} {\bibfnamefont {W.}~\bibnamefont
  {Altmannshofer}}\ and\ \bibinfo {author} {\bibfnamefont {P.}~\bibnamefont
  {Stangl}},\ }\bibfield  {title} {\bibinfo {title} {{New Physics in Rare B
  Decays after Moriond 2021}},\ }\href@noop {} {\  (\bibinfo {year} {2021})},\
  \Eprint {https://arxiv.org/abs/2103.13370} {arXiv:2103.13370 [hep-ph]}
  \BibitemShut {NoStop}%
\bibitem [{\citenamefont {Geng}\ \emph {et~al.}(2021)\citenamefont {Geng},
  \citenamefont {Grinstein}, \citenamefont {J\"ager}, \citenamefont {Li},
  \citenamefont {Martin~Camalich},\ and\ \citenamefont {Shi}}]{Geng:2021nhg}%
  \BibitemOpen
  \bibfield  {author} {\bibinfo {author} {\bibfnamefont {L.-S.}\ \bibnamefont
  {Geng}}, \bibinfo {author} {\bibfnamefont {B.}~\bibnamefont {Grinstein}},
  \bibinfo {author} {\bibfnamefont {S.}~\bibnamefont {J\"ager}}, \bibinfo
  {author} {\bibfnamefont {S.-Y.}\ \bibnamefont {Li}}, \bibinfo {author}
  {\bibfnamefont {J.}~\bibnamefont {Martin~Camalich}},\ and\ \bibinfo {author}
  {\bibfnamefont {R.-X.}\ \bibnamefont {Shi}},\ }\bibfield  {title} {\bibinfo
  {title} {{Implications of new evidence for lepton-universality violation in
  $b\to s\ell^+\ell^-$ decays}},\ }\href@noop {} {\  (\bibinfo {year}
  {2021})},\ \Eprint {https://arxiv.org/abs/2103.12738} {arXiv:2103.12738
  [hep-ph]} \BibitemShut {NoStop}%
\bibitem [{\citenamefont {Capdevila}\ \emph {et~al.}(2021)\citenamefont
  {Capdevila}, \citenamefont {Crivellin}, \citenamefont {Manzari},\ and\
  \citenamefont {Montull}}]{Capdevila:2020rrl}%
  \BibitemOpen
  \bibfield  {author} {\bibinfo {author} {\bibfnamefont {B.}~\bibnamefont
  {Capdevila}}, \bibinfo {author} {\bibfnamefont {A.}~\bibnamefont
  {Crivellin}}, \bibinfo {author} {\bibfnamefont {C.~A.}\ \bibnamefont
  {Manzari}},\ and\ \bibinfo {author} {\bibfnamefont {M.}~\bibnamefont
  {Montull}},\ }\bibfield  {title} {\bibinfo {title} {{Explaining $b\to
  s\ell^+\ell^-$ and the Cabibbo angle anomaly with a vector triplet}},\ }\href
  {https://doi.org/10.1103/PhysRevD.103.015032} {\bibfield  {journal} {\bibinfo
   {journal} {Phys. Rev. D}\ }\textbf {\bibinfo {volume} {103}},\ \bibinfo
  {pages} {015032} (\bibinfo {year} {2021})},\ \Eprint
  {https://arxiv.org/abs/2005.13542} {arXiv:2005.13542 [hep-ph]} \BibitemShut
  {NoStop}%
\bibitem [{\citenamefont {Altmannshofer}\ \emph
  {et~al.}(2020{\natexlab{a}})\citenamefont {Altmannshofer}, \citenamefont
  {Davighi},\ and\ \citenamefont {Nardecchia}}]{Altmannshofer:2019xda}%
  \BibitemOpen
  \bibfield  {author} {\bibinfo {author} {\bibfnamefont {W.}~\bibnamefont
  {Altmannshofer}}, \bibinfo {author} {\bibfnamefont {J.}~\bibnamefont
  {Davighi}},\ and\ \bibinfo {author} {\bibfnamefont {M.}~\bibnamefont
  {Nardecchia}},\ }\bibfield  {title} {\bibinfo {title} {{Gauging the
  accidental symmetries of the standard model, and implications for the flavor
  anomalies}},\ }\href {https://doi.org/10.1103/PhysRevD.101.015004} {\bibfield
   {journal} {\bibinfo  {journal} {Phys. Rev. D}\ }\textbf {\bibinfo {volume}
  {101}},\ \bibinfo {pages} {015004} (\bibinfo {year} {2020}{\natexlab{a}})},\
  \Eprint {https://arxiv.org/abs/1909.02021} {arXiv:1909.02021 [hep-ph]}
  \BibitemShut {NoStop}%
\bibitem [{\citenamefont {Gauld}\ \emph {et~al.}(2014)\citenamefont {Gauld},
  \citenamefont {Goertz},\ and\ \citenamefont {Haisch}}]{Gauld:2013qja}%
  \BibitemOpen
  \bibfield  {author} {\bibinfo {author} {\bibfnamefont {R.}~\bibnamefont
  {Gauld}}, \bibinfo {author} {\bibfnamefont {F.}~\bibnamefont {Goertz}},\ and\
  \bibinfo {author} {\bibfnamefont {U.}~\bibnamefont {Haisch}},\ }\bibfield
  {title} {\bibinfo {title} {{An explicit Z'-boson explanation of the $B \to
  K^* \mu^+ \mu^-$ anomaly}},\ }\href {https://doi.org/10.1007/JHEP01(2014)069}
  {\bibfield  {journal} {\bibinfo  {journal} {JHEP}\ }\textbf {\bibinfo
  {volume} {01}},\ \bibinfo {pages} {069}},\ \Eprint
  {https://arxiv.org/abs/1310.1082} {arXiv:1310.1082 [hep-ph]} \BibitemShut
  {NoStop}%
\bibitem [{\citenamefont {Bauer}\ and\ \citenamefont
  {Neubert}(2016)}]{Bauer:2015knc}%
  \BibitemOpen
  \bibfield  {author} {\bibinfo {author} {\bibfnamefont {M.}~\bibnamefont
  {Bauer}}\ and\ \bibinfo {author} {\bibfnamefont {M.}~\bibnamefont
  {Neubert}},\ }\bibfield  {title} {\bibinfo {title} {{Minimal Leptoquark
  Explanation for the R$_{D^{(*)}}$ , R$_K$ , and $(g-2)_g$ Anomalies}},\
  }\href {https://doi.org/10.1103/PhysRevLett.116.141802} {\bibfield  {journal}
  {\bibinfo  {journal} {Phys. Rev. Lett.}\ }\textbf {\bibinfo {volume} {116}},\
  \bibinfo {pages} {141802} (\bibinfo {year} {2016})},\ \Eprint
  {https://arxiv.org/abs/1511.01900} {arXiv:1511.01900 [hep-ph]} \BibitemShut
  {NoStop}%
\bibitem [{\citenamefont {Coluccio~Leskow}\ \emph {et~al.}(2017)\citenamefont
  {Coluccio~Leskow}, \citenamefont {D'Ambrosio}, \citenamefont {Crivellin},\
  and\ \citenamefont {M\"uller}}]{ColuccioLeskow:2016dox}%
  \BibitemOpen
  \bibfield  {author} {\bibinfo {author} {\bibfnamefont {E.}~\bibnamefont
  {Coluccio~Leskow}}, \bibinfo {author} {\bibfnamefont {G.}~\bibnamefont
  {D'Ambrosio}}, \bibinfo {author} {\bibfnamefont {A.}~\bibnamefont
  {Crivellin}},\ and\ \bibinfo {author} {\bibfnamefont {D.}~\bibnamefont
  {M\"uller}},\ }\bibfield  {title} {\bibinfo {title} {{$(g-2)\mu$, lepton
  flavor violation, and $Z$ decays with leptoquarks: Correlations and future
  prospects}},\ }\href {https://doi.org/10.1103/PhysRevD.95.055018} {\bibfield
  {journal} {\bibinfo  {journal} {Phys. Rev. D}\ }\textbf {\bibinfo {volume}
  {95}},\ \bibinfo {pages} {055018} (\bibinfo {year} {2017})},\ \Eprint
  {https://arxiv.org/abs/1612.06858} {arXiv:1612.06858 [hep-ph]} \BibitemShut
  {NoStop}%
\bibitem [{\citenamefont {Angelescu}\ \emph {et~al.}(2018)\citenamefont
  {Angelescu}, \citenamefont {Be\v{c}irevi\'c}, \citenamefont {Faroughy},\ and\
  \citenamefont {Sumensari}}]{Angelescu:2018tyl}%
  \BibitemOpen
  \bibfield  {author} {\bibinfo {author} {\bibfnamefont {A.}~\bibnamefont
  {Angelescu}}, \bibinfo {author} {\bibfnamefont {D.}~\bibnamefont
  {Be\v{c}irevi\'c}}, \bibinfo {author} {\bibfnamefont {D.~A.}\ \bibnamefont
  {Faroughy}},\ and\ \bibinfo {author} {\bibfnamefont {O.}~\bibnamefont
  {Sumensari}},\ }\bibfield  {title} {\bibinfo {title} {{Closing the window on
  single leptoquark solutions to the $B$-physics anomalies}},\ }\href
  {https://doi.org/10.1007/JHEP10(2018)183} {\bibfield  {journal} {\bibinfo
  {journal} {JHEP}\ }\textbf {\bibinfo {volume} {10}},\ \bibinfo {pages}
  {183}},\ \Eprint {https://arxiv.org/abs/1808.08179} {arXiv:1808.08179
  [hep-ph]} \BibitemShut {NoStop}%
\bibitem [{\citenamefont {Crivellin}\ \emph
  {et~al.}(2020{\natexlab{a}})\citenamefont {Crivellin}, \citenamefont
  {M\"uller},\ and\ \citenamefont {Saturnino}}]{Crivellin:2019dwb}%
  \BibitemOpen
  \bibfield  {author} {\bibinfo {author} {\bibfnamefont {A.}~\bibnamefont
  {Crivellin}}, \bibinfo {author} {\bibfnamefont {D.}~\bibnamefont
  {M\"uller}},\ and\ \bibinfo {author} {\bibfnamefont {F.}~\bibnamefont
  {Saturnino}},\ }\bibfield  {title} {\bibinfo {title} {{Flavor Phenomenology
  of the Leptoquark Singlet-Triplet Model}},\ }\href
  {https://doi.org/10.1007/JHEP06(2020)020} {\bibfield  {journal} {\bibinfo
  {journal} {JHEP}\ }\textbf {\bibinfo {volume} {06}},\ \bibinfo {pages}
  {020}},\ \Eprint {https://arxiv.org/abs/1912.04224} {arXiv:1912.04224
  [hep-ph]} \BibitemShut {NoStop}%
\bibitem [{\citenamefont {Fuentes-Mart\'\i{}n}\ and\ \citenamefont
  {Stangl}(2020)}]{Fuentes-Martin:2020bnh}%
  \BibitemOpen
  \bibfield  {author} {\bibinfo {author} {\bibfnamefont {J.}~\bibnamefont
  {Fuentes-Mart\'\i{}n}}\ and\ \bibinfo {author} {\bibfnamefont
  {P.}~\bibnamefont {Stangl}},\ }\bibfield  {title} {\bibinfo {title}
  {{Third-family quark-lepton unification with a fundamental composite
  Higgs}},\ }\href {https://doi.org/10.1016/j.physletb.2020.135953} {\bibfield
  {journal} {\bibinfo  {journal} {Phys. Lett. B}\ }\textbf {\bibinfo {volume}
  {811}},\ \bibinfo {pages} {135953} (\bibinfo {year} {2020})},\ \Eprint
  {https://arxiv.org/abs/2004.11376} {arXiv:2004.11376 [hep-ph]} \BibitemShut
  {NoStop}%
\bibitem [{\citenamefont {Saad}\ and\ \citenamefont
  {Thapa}(2020)}]{Saad:2020ucl}%
  \BibitemOpen
  \bibfield  {author} {\bibinfo {author} {\bibfnamefont {S.}~\bibnamefont
  {Saad}}\ and\ \bibinfo {author} {\bibfnamefont {A.}~\bibnamefont {Thapa}},\
  }\bibfield  {title} {\bibinfo {title} {{Common origin of neutrino masses and
  $R_{D^{(\ast)}}$, $R_{K^{(\ast)}}$ anomalies}},\ }\href
  {https://doi.org/10.1103/PhysRevD.102.015014} {\bibfield  {journal} {\bibinfo
   {journal} {Phys. Rev. D}\ }\textbf {\bibinfo {volume} {102}},\ \bibinfo
  {pages} {015014} (\bibinfo {year} {2020})},\ \Eprint
  {https://arxiv.org/abs/2004.07880} {arXiv:2004.07880 [hep-ph]} \BibitemShut
  {NoStop}%
\bibitem [{\citenamefont {Balaji}\ and\ \citenamefont
  {Schmidt}(2020)}]{Balaji:2019kwe}%
  \BibitemOpen
  \bibfield  {author} {\bibinfo {author} {\bibfnamefont {S.}~\bibnamefont
  {Balaji}}\ and\ \bibinfo {author} {\bibfnamefont {M.~A.}\ \bibnamefont
  {Schmidt}},\ }\bibfield  {title} {\bibinfo {title} {{Unified SU(4) theory for
  the $R_{D^{(*)}}$ and $R_{K^{(*)}}$ anomalies}},\ }\href
  {https://doi.org/10.1103/PhysRevD.101.015026} {\bibfield  {journal} {\bibinfo
   {journal} {Phys. Rev. D}\ }\textbf {\bibinfo {volume} {101}},\ \bibinfo
  {pages} {015026} (\bibinfo {year} {2020})},\ \Eprint
  {https://arxiv.org/abs/1911.08873} {arXiv:1911.08873 [hep-ph]} \BibitemShut
  {NoStop}%
\bibitem [{\citenamefont {Babu}\ \emph {et~al.}(2021)\citenamefont {Babu},
  \citenamefont {Dev}, \citenamefont {Jana},\ and\ \citenamefont
  {Thapa}}]{Babu:2020hun}%
  \BibitemOpen
  \bibfield  {author} {\bibinfo {author} {\bibfnamefont {K.~S.}\ \bibnamefont
  {Babu}}, \bibinfo {author} {\bibfnamefont {P.~S.~B.}\ \bibnamefont {Dev}},
  \bibinfo {author} {\bibfnamefont {S.}~\bibnamefont {Jana}},\ and\ \bibinfo
  {author} {\bibfnamefont {A.}~\bibnamefont {Thapa}},\ }\bibfield  {title}
  {\bibinfo {title} {{Unified framework for $B$-anomalies, muon $g-2$ and
  neutrino masses}},\ }\href {https://doi.org/10.1007/JHEP03(2021)179}
  {\bibfield  {journal} {\bibinfo  {journal} {JHEP}\ }\textbf {\bibinfo
  {volume} {03}},\ \bibinfo {pages} {179}},\ \Eprint
  {https://arxiv.org/abs/2009.01771} {arXiv:2009.01771 [hep-ph]} \BibitemShut
  {NoStop}%
\bibitem [{\citenamefont {Bhupal~Dev}\ \emph {et~al.}(2020)\citenamefont
  {Bhupal~Dev}, \citenamefont {Mohanta}, \citenamefont {Patra},\ and\
  \citenamefont {Sahoo}}]{Dev:2020qet}%
  \BibitemOpen
  \bibfield  {author} {\bibinfo {author} {\bibfnamefont {P.~S.}\ \bibnamefont
  {Bhupal~Dev}}, \bibinfo {author} {\bibfnamefont {R.}~\bibnamefont {Mohanta}},
  \bibinfo {author} {\bibfnamefont {S.}~\bibnamefont {Patra}},\ and\ \bibinfo
  {author} {\bibfnamefont {S.}~\bibnamefont {Sahoo}},\ }\bibfield  {title}
  {\bibinfo {title} {{Unified explanation of flavor anomalies, radiative
  neutrino masses, and ANITA anomalous events in a vector leptoquark model}},\
  }\href {https://doi.org/10.1103/PhysRevD.102.095012} {\bibfield  {journal}
  {\bibinfo  {journal} {Phys. Rev. D}\ }\textbf {\bibinfo {volume} {102}},\
  \bibinfo {pages} {095012} (\bibinfo {year} {2020})},\ \Eprint
  {https://arxiv.org/abs/2004.09464} {arXiv:2004.09464 [hep-ph]} \BibitemShut
  {NoStop}%
\bibitem [{\citenamefont {Gripaios}\ \emph {et~al.}(2016)\citenamefont
  {Gripaios}, \citenamefont {Nardecchia},\ and\ \citenamefont
  {Renner}}]{Gripaios:2015gra}%
  \BibitemOpen
  \bibfield  {author} {\bibinfo {author} {\bibfnamefont {B.}~\bibnamefont
  {Gripaios}}, \bibinfo {author} {\bibfnamefont {M.}~\bibnamefont
  {Nardecchia}},\ and\ \bibinfo {author} {\bibfnamefont {S.~A.}\ \bibnamefont
  {Renner}},\ }\bibfield  {title} {\bibinfo {title} {{Linear flavour violation
  and anomalies in B physics}},\ }\href
  {https://doi.org/10.1007/JHEP06(2016)083} {\bibfield  {journal} {\bibinfo
  {journal} {JHEP}\ }\textbf {\bibinfo {volume} {06}},\ \bibinfo {pages}
  {083}},\ \Eprint {https://arxiv.org/abs/1509.05020} {arXiv:1509.05020
  [hep-ph]} \BibitemShut {NoStop}%
\bibitem [{\citenamefont {Arnan}\ \emph {et~al.}(2017)\citenamefont {Arnan},
  \citenamefont {Hofer}, \citenamefont {Mescia},\ and\ \citenamefont
  {Crivellin}}]{Arnan:2016cpy}%
  \BibitemOpen
  \bibfield  {author} {\bibinfo {author} {\bibfnamefont {P.}~\bibnamefont
  {Arnan}}, \bibinfo {author} {\bibfnamefont {L.}~\bibnamefont {Hofer}},
  \bibinfo {author} {\bibfnamefont {F.}~\bibnamefont {Mescia}},\ and\ \bibinfo
  {author} {\bibfnamefont {A.}~\bibnamefont {Crivellin}},\ }\bibfield  {title}
  {\bibinfo {title} {{Loop effects of heavy new scalars and fermions in $b\to
  s\mu^+\mu^-$}},\ }\href {https://doi.org/10.1007/JHEP04(2017)043} {\bibfield
  {journal} {\bibinfo  {journal} {JHEP}\ }\textbf {\bibinfo {volume} {04}},\
  \bibinfo {pages} {043}},\ \Eprint {https://arxiv.org/abs/1608.07832}
  {arXiv:1608.07832 [hep-ph]} \BibitemShut {NoStop}%
\bibitem [{\citenamefont {Li}\ \emph {et~al.}(2018)\citenamefont {Li},
  \citenamefont {Li}, \citenamefont {Yang},\ and\ \citenamefont
  {Zhang}}]{Li:2018rax}%
  \BibitemOpen
  \bibfield  {author} {\bibinfo {author} {\bibfnamefont {S.-P.}\ \bibnamefont
  {Li}}, \bibinfo {author} {\bibfnamefont {X.-Q.}\ \bibnamefont {Li}}, \bibinfo
  {author} {\bibfnamefont {Y.-D.}\ \bibnamefont {Yang}},\ and\ \bibinfo
  {author} {\bibfnamefont {X.}~\bibnamefont {Zhang}},\ }\bibfield  {title}
  {\bibinfo {title} {{$ {R}_{D^{\left(*\right)}},{R}_{K^{\left(*\right)}} $ and
  neutrino mass in the 2HDM-III with right-handed neutrinos}},\ }\href
  {https://doi.org/10.1007/JHEP09(2018)149} {\bibfield  {journal} {\bibinfo
  {journal} {JHEP}\ }\textbf {\bibinfo {volume} {09}},\ \bibinfo {pages}
  {149}},\ \Eprint {https://arxiv.org/abs/1807.08530} {arXiv:1807.08530
  [hep-ph]} \BibitemShut {NoStop}%
\bibitem [{\citenamefont {Hu}\ and\ \citenamefont {Huang}(2020)}]{Hu:2019ahp}%
  \BibitemOpen
  \bibfield  {author} {\bibinfo {author} {\bibfnamefont {Q.-Y.}\ \bibnamefont
  {Hu}}\ and\ \bibinfo {author} {\bibfnamefont {L.-L.}\ \bibnamefont {Huang}},\
  }\bibfield  {title} {\bibinfo {title} {{Explaining $b\to s \ell^+ \ell^-$
  data by sneutrinos in the $R$ -parity violating MSSM}},\ }\href
  {https://doi.org/10.1103/PhysRevD.101.035030} {\bibfield  {journal} {\bibinfo
   {journal} {Phys. Rev. D}\ }\textbf {\bibinfo {volume} {101}},\ \bibinfo
  {pages} {035030} (\bibinfo {year} {2020})},\ \Eprint
  {https://arxiv.org/abs/1912.03676} {arXiv:1912.03676 [hep-ph]} \BibitemShut
  {NoStop}%
\bibitem [{\citenamefont {Arnan}\ \emph {et~al.}(2019)\citenamefont {Arnan},
  \citenamefont {Crivellin}, \citenamefont {Fedele},\ and\ \citenamefont
  {Mescia}}]{Arnan:2019uhr}%
  \BibitemOpen
  \bibfield  {author} {\bibinfo {author} {\bibfnamefont {P.}~\bibnamefont
  {Arnan}}, \bibinfo {author} {\bibfnamefont {A.}~\bibnamefont {Crivellin}},
  \bibinfo {author} {\bibfnamefont {M.}~\bibnamefont {Fedele}},\ and\ \bibinfo
  {author} {\bibfnamefont {F.}~\bibnamefont {Mescia}},\ }\bibfield  {title}
  {\bibinfo {title} {{Generic loop effects of new scalars and fermions in $b\to
  s\ell^+\ell^-$ and a vector-like $4^{\rm th}$ generation}},\ }\href
  {https://doi.org/10.1007/JHEP06(2019)118} {\bibfield  {journal} {\bibinfo
  {journal} {JHEP}\ }\textbf {\bibinfo {volume} {06}},\ \bibinfo {pages}
  {118}},\ \Eprint {https://arxiv.org/abs/1904.05890} {arXiv:1904.05890
  [hep-ph]} \BibitemShut {NoStop}%
\bibitem [{\citenamefont {Li}\ and\ \citenamefont {L\"u}(2018)}]{Li:2018lxi}%
  \BibitemOpen
  \bibfield  {author} {\bibinfo {author} {\bibfnamefont {Y.}~\bibnamefont
  {Li}}\ and\ \bibinfo {author} {\bibfnamefont {C.-D.}\ \bibnamefont {L\"u}},\
  }\bibfield  {title} {\bibinfo {title} {{Recent Anomalies in B Physics}},\
  }\href {https://doi.org/10.1016/j.scib.2018.02.003} {\bibfield  {journal}
  {\bibinfo  {journal} {Sci. Bull.}\ }\textbf {\bibinfo {volume} {63}},\
  \bibinfo {pages} {267} (\bibinfo {year} {2018})},\ \Eprint
  {https://arxiv.org/abs/1808.02990} {arXiv:1808.02990 [hep-ph]} \BibitemShut
  {NoStop}%
\bibitem [{\citenamefont {Bifani}\ \emph {et~al.}(2019)\citenamefont {Bifani},
  \citenamefont {Descotes-Genon}, \citenamefont {Romero~Vidal},\ and\
  \citenamefont {Schune}}]{Bifani:2018zmi}%
  \BibitemOpen
  \bibfield  {author} {\bibinfo {author} {\bibfnamefont {S.}~\bibnamefont
  {Bifani}}, \bibinfo {author} {\bibfnamefont {S.}~\bibnamefont
  {Descotes-Genon}}, \bibinfo {author} {\bibfnamefont {A.}~\bibnamefont
  {Romero~Vidal}},\ and\ \bibinfo {author} {\bibfnamefont {M.-H.}\ \bibnamefont
  {Schune}},\ }\bibfield  {title} {\bibinfo {title} {{Review of Lepton
  Universality tests in $B$ decays}},\ }\href
  {https://doi.org/10.1088/1361-6471/aaf5de} {\bibfield  {journal} {\bibinfo
  {journal} {J. Phys. G}\ }\textbf {\bibinfo {volume} {46}},\ \bibinfo {pages}
  {023001} (\bibinfo {year} {2019})},\ \Eprint
  {https://arxiv.org/abs/1809.06229} {arXiv:1809.06229 [hep-ex]} \BibitemShut
  {NoStop}%
\bibitem [{\citenamefont {Grossman}\ \emph {et~al.}(2020)\citenamefont
  {Grossman}, \citenamefont {Passemar},\ and\ \citenamefont
  {Schacht}}]{Grossman:2019bzp}%
  \BibitemOpen
  \bibfield  {author} {\bibinfo {author} {\bibfnamefont {Y.}~\bibnamefont
  {Grossman}}, \bibinfo {author} {\bibfnamefont {E.}~\bibnamefont {Passemar}},\
  and\ \bibinfo {author} {\bibfnamefont {S.}~\bibnamefont {Schacht}},\
  }\bibfield  {title} {\bibinfo {title} {{On the Statistical Treatment of the
  Cabibbo Angle Anomaly}},\ }\href {https://doi.org/10.1007/JHEP07(2020)068}
  {\bibfield  {journal} {\bibinfo  {journal} {JHEP}\ }\textbf {\bibinfo
  {volume} {07}},\ \bibinfo {pages} {068}},\ \Eprint
  {https://arxiv.org/abs/1911.07821} {arXiv:1911.07821 [hep-ph]} \BibitemShut
  {NoStop}%
\bibitem [{\citenamefont {Seng}\ \emph {et~al.}(2020)\citenamefont {Seng},
  \citenamefont {Feng}, \citenamefont {Gorchtein},\ and\ \citenamefont
  {Jin}}]{Seng:2020wjq}%
  \BibitemOpen
  \bibfield  {author} {\bibinfo {author} {\bibfnamefont {C.-Y.}\ \bibnamefont
  {Seng}}, \bibinfo {author} {\bibfnamefont {X.}~\bibnamefont {Feng}}, \bibinfo
  {author} {\bibfnamefont {M.}~\bibnamefont {Gorchtein}},\ and\ \bibinfo
  {author} {\bibfnamefont {L.-C.}\ \bibnamefont {Jin}},\ }\bibfield  {title}
  {\bibinfo {title} {{Joint lattice QCD\textendash{}dispersion theory analysis
  confirms the quark-mixing top-row unitarity deficit}},\ }\href
  {https://doi.org/10.1103/PhysRevD.101.111301} {\bibfield  {journal} {\bibinfo
   {journal} {Phys. Rev. D}\ }\textbf {\bibinfo {volume} {101}},\ \bibinfo
  {pages} {111301} (\bibinfo {year} {2020})},\ \Eprint
  {https://arxiv.org/abs/2003.11264} {arXiv:2003.11264 [hep-ph]} \BibitemShut
  {NoStop}%
\bibitem [{\citenamefont {Amhis}\ \emph {et~al.}(2019)\citenamefont {Amhis}
  \emph {et~al.}}]{Amhis:2019ckw}%
  \BibitemOpen
  \bibfield  {author} {\bibinfo {author} {\bibfnamefont {Y.~S.}\ \bibnamefont
  {Amhis}} \emph {et~al.} (\bibinfo {collaboration} {HFLAV}),\ }\bibfield
  {title} {\bibinfo {title} {{Averages of $b$-hadron, $c$-hadron, and
  $\tau$-lepton properties as of 2018}},\ }\href@noop {} {\  (\bibinfo {year}
  {2019})},\ \Eprint {https://arxiv.org/abs/1909.12524} {arXiv:1909.12524
  [hep-ex]} \BibitemShut {NoStop}%
\bibitem [{\citenamefont {Aoki}\ \emph {et~al.}(2020)\citenamefont {Aoki} \emph
  {et~al.}}]{Aoki:2019cca}%
  \BibitemOpen
  \bibfield  {author} {\bibinfo {author} {\bibfnamefont {S.}~\bibnamefont
  {Aoki}} \emph {et~al.} (\bibinfo {collaboration} {Flavour Lattice Averaging
  Group}),\ }\bibfield  {title} {\bibinfo {title} {{FLAG Review 2019: Flavour
  Lattice Averaging Group (FLAG)}},\ }\href
  {https://doi.org/10.1140/epjc/s10052-019-7354-7} {\bibfield  {journal}
  {\bibinfo  {journal} {Eur. Phys. J. C}\ }\textbf {\bibinfo {volume} {80}},\
  \bibinfo {pages} {113} (\bibinfo {year} {2020})},\ \Eprint
  {https://arxiv.org/abs/1902.08191} {arXiv:1902.08191 [hep-lat]} \BibitemShut
  {NoStop}%
\bibitem [{\citenamefont {Seng}\ \emph {et~al.}(2018)\citenamefont {Seng},
  \citenamefont {Gorchtein}, \citenamefont {Patel},\ and\ \citenamefont
  {Ramsey-Musolf}}]{Seng:2018yzq}%
  \BibitemOpen
  \bibfield  {author} {\bibinfo {author} {\bibfnamefont {C.-Y.}\ \bibnamefont
  {Seng}}, \bibinfo {author} {\bibfnamefont {M.}~\bibnamefont {Gorchtein}},
  \bibinfo {author} {\bibfnamefont {H.~H.}\ \bibnamefont {Patel}},\ and\
  \bibinfo {author} {\bibfnamefont {M.~J.}\ \bibnamefont {Ramsey-Musolf}},\
  }\bibfield  {title} {\bibinfo {title} {{Reduced Hadronic Uncertainty in the
  Determination of $V_{ud}$}},\ }\href
  {https://doi.org/10.1103/PhysRevLett.121.241804} {\bibfield  {journal}
  {\bibinfo  {journal} {Phys. Rev. Lett.}\ }\textbf {\bibinfo {volume} {121}},\
  \bibinfo {pages} {241804} (\bibinfo {year} {2018})},\ \Eprint
  {https://arxiv.org/abs/1807.10197} {arXiv:1807.10197 [hep-ph]} \BibitemShut
  {NoStop}%
\bibitem [{\citenamefont {Czarnecki}\ \emph {et~al.}(2019)\citenamefont
  {Czarnecki}, \citenamefont {Marciano},\ and\ \citenamefont
  {Sirlin}}]{Czarnecki:2019mwq}%
  \BibitemOpen
  \bibfield  {author} {\bibinfo {author} {\bibfnamefont {A.}~\bibnamefont
  {Czarnecki}}, \bibinfo {author} {\bibfnamefont {W.~J.}\ \bibnamefont
  {Marciano}},\ and\ \bibinfo {author} {\bibfnamefont {A.}~\bibnamefont
  {Sirlin}},\ }\bibfield  {title} {\bibinfo {title} {{Radiative Corrections to
  Neutron and Nuclear Beta Decays Revisited}},\ }\href
  {https://doi.org/10.1103/PhysRevD.100.073008} {\bibfield  {journal} {\bibinfo
   {journal} {Phys. Rev. D}\ }\textbf {\bibinfo {volume} {100}},\ \bibinfo
  {pages} {073008} (\bibinfo {year} {2019})},\ \Eprint
  {https://arxiv.org/abs/1907.06737} {arXiv:1907.06737 [hep-ph]} \BibitemShut
  {NoStop}%
\bibitem [{\citenamefont {Belfatto}\ \emph {et~al.}(2020)\citenamefont
  {Belfatto}, \citenamefont {Beradze},\ and\ \citenamefont
  {Berezhiani}}]{Belfatto:2019swo}%
  \BibitemOpen
  \bibfield  {author} {\bibinfo {author} {\bibfnamefont {B.}~\bibnamefont
  {Belfatto}}, \bibinfo {author} {\bibfnamefont {R.}~\bibnamefont {Beradze}},\
  and\ \bibinfo {author} {\bibfnamefont {Z.}~\bibnamefont {Berezhiani}},\
  }\bibfield  {title} {\bibinfo {title} {{The CKM unitarity problem: A trace of
  new physics at the TeV scale?}},\ }\href
  {https://doi.org/10.1140/epjc/s10052-020-7691-6} {\bibfield  {journal}
  {\bibinfo  {journal} {Eur. Phys. J. C}\ }\textbf {\bibinfo {volume} {80}},\
  \bibinfo {pages} {149} (\bibinfo {year} {2020})},\ \Eprint
  {https://arxiv.org/abs/1906.02714} {arXiv:1906.02714 [hep-ph]} \BibitemShut
  {NoStop}%
\bibitem [{\citenamefont {Cheung}\ \emph {et~al.}(2020)\citenamefont {Cheung},
  \citenamefont {Keung}, \citenamefont {Lu},\ and\ \citenamefont
  {Tseng}}]{Cheung:2020vqm}%
  \BibitemOpen
  \bibfield  {author} {\bibinfo {author} {\bibfnamefont {K.}~\bibnamefont
  {Cheung}}, \bibinfo {author} {\bibfnamefont {W.-Y.}\ \bibnamefont {Keung}},
  \bibinfo {author} {\bibfnamefont {C.-T.}\ \bibnamefont {Lu}},\ and\ \bibinfo
  {author} {\bibfnamefont {P.-Y.}\ \bibnamefont {Tseng}},\ }\bibfield  {title}
  {\bibinfo {title} {{Vector-like Quark Interpretation for the CKM Unitarity
  Violation, Excess in Higgs Signal Strength, and Bottom Quark Forward-Backward
  Asymmetry}},\ }\href {https://doi.org/10.1007/JHEP05(2020)117} {\bibfield
  {journal} {\bibinfo  {journal} {JHEP}\ }\textbf {\bibinfo {volume} {05}},\
  \bibinfo {pages} {117}},\ \Eprint {https://arxiv.org/abs/2001.02853}
  {arXiv:2001.02853 [hep-ph]} \BibitemShut {NoStop}%
\bibitem [{\citenamefont {Crivellin}\ \emph
  {et~al.}(2020{\natexlab{b}})\citenamefont {Crivellin}, \citenamefont
  {Manzari}, \citenamefont {Alguero},\ and\ \citenamefont
  {Matias}}]{Crivellin:2020oup}%
  \BibitemOpen
  \bibfield  {author} {\bibinfo {author} {\bibfnamefont {A.}~\bibnamefont
  {Crivellin}}, \bibinfo {author} {\bibfnamefont {C.~A.}\ \bibnamefont
  {Manzari}}, \bibinfo {author} {\bibfnamefont {M.}~\bibnamefont {Alguero}},\
  and\ \bibinfo {author} {\bibfnamefont {J.}~\bibnamefont {Matias}},\
  }\bibfield  {title} {\bibinfo {title} {{Combined Explanation of the $Z\to
  b\bar b$ Forward-Backward Asymmetry, the Cabibbo Angle Anomaly,
  $\tau\to\mu\nu\nu$ and $b\to s\ell^+\ell^-$ Data}},\ }\href@noop {} {\
  (\bibinfo {year} {2020}{\natexlab{b}})},\ \Eprint
  {https://arxiv.org/abs/2010.14504} {arXiv:2010.14504 [hep-ph]} \BibitemShut
  {NoStop}%
\bibitem [{\citenamefont {Coutinho}\ \emph {et~al.}(2020)\citenamefont
  {Coutinho}, \citenamefont {Crivellin},\ and\ \citenamefont
  {Manzari}}]{Coutinho:2019aiy}%
  \BibitemOpen
  \bibfield  {author} {\bibinfo {author} {\bibfnamefont {A.~M.}\ \bibnamefont
  {Coutinho}}, \bibinfo {author} {\bibfnamefont {A.}~\bibnamefont
  {Crivellin}},\ and\ \bibinfo {author} {\bibfnamefont {C.~A.}\ \bibnamefont
  {Manzari}},\ }\bibfield  {title} {\bibinfo {title} {{Global Fit to Modified
  Neutrino Couplings and the Cabibbo-Angle Anomaly}},\ }\href
  {https://doi.org/10.1103/PhysRevLett.125.071802} {\bibfield  {journal}
  {\bibinfo  {journal} {Phys. Rev. Lett.}\ }\textbf {\bibinfo {volume} {125}},\
  \bibinfo {pages} {071802} (\bibinfo {year} {2020})},\ \Eprint
  {https://arxiv.org/abs/1912.08823} {arXiv:1912.08823 [hep-ph]} \BibitemShut
  {NoStop}%
\bibitem [{\citenamefont {Crivellin}\ and\ \citenamefont
  {Hoferichter}(2020)}]{Crivellin:2020lzu}%
  \BibitemOpen
  \bibfield  {author} {\bibinfo {author} {\bibfnamefont {A.}~\bibnamefont
  {Crivellin}}\ and\ \bibinfo {author} {\bibfnamefont {M.}~\bibnamefont
  {Hoferichter}},\ }\bibfield  {title} {\bibinfo {title} {{\ensuremath{\beta}
  Decays as Sensitive Probes of Lepton Flavor Universality}},\ }\href
  {https://doi.org/10.1103/PhysRevLett.125.111801} {\bibfield  {journal}
  {\bibinfo  {journal} {Phys. Rev. Lett.}\ }\textbf {\bibinfo {volume} {125}},\
  \bibinfo {pages} {111801} (\bibinfo {year} {2020})},\ \Eprint
  {https://arxiv.org/abs/2002.07184} {arXiv:2002.07184 [hep-ph]} \BibitemShut
  {NoStop}%
\bibitem [{\citenamefont {Crivellin}\ \emph
  {et~al.}(2020{\natexlab{c}})\citenamefont {Crivellin}, \citenamefont {Kirk},
  \citenamefont {Manzari},\ and\ \citenamefont {Montull}}]{Crivellin:2020ebi}%
  \BibitemOpen
  \bibfield  {author} {\bibinfo {author} {\bibfnamefont {A.}~\bibnamefont
  {Crivellin}}, \bibinfo {author} {\bibfnamefont {F.}~\bibnamefont {Kirk}},
  \bibinfo {author} {\bibfnamefont {C.~A.}\ \bibnamefont {Manzari}},\ and\
  \bibinfo {author} {\bibfnamefont {M.}~\bibnamefont {Montull}},\ }\bibfield
  {title} {\bibinfo {title} {{Global Electroweak Fit and Vector-Like Leptons in
  Light of the Cabibbo Angle Anomaly}}\ }\href
  {https://doi.org/10.1007/JHEP12(2020)166} {10.1007/JHEP12(2020)166} (\bibinfo
  {year} {2020}{\natexlab{c}}),\ \Eprint {https://arxiv.org/abs/2008.01113}
  {arXiv:2008.01113 [hep-ph]} \BibitemShut {NoStop}%
\bibitem [{\citenamefont {Kirk}(2021)}]{Kirk:2020wdk}%
  \BibitemOpen
  \bibfield  {author} {\bibinfo {author} {\bibfnamefont {M.}~\bibnamefont
  {Kirk}},\ }\bibfield  {title} {\bibinfo {title} {{Cabibbo anomaly versus
  electroweak precision tests: An exploration of extensions of the standard
  model}},\ }\href {https://doi.org/10.1103/PhysRevD.103.035004} {\bibfield
  {journal} {\bibinfo  {journal} {Phys. Rev. D}\ }\textbf {\bibinfo {volume}
  {103}},\ \bibinfo {pages} {035004} (\bibinfo {year} {2021})},\ \Eprint
  {https://arxiv.org/abs/2008.03261} {arXiv:2008.03261 [hep-ph]} \BibitemShut
  {NoStop}%
\bibitem [{\citenamefont {Alok}\ \emph {et~al.}(2020)\citenamefont {Alok},
  \citenamefont {Dighe}, \citenamefont {Gangal},\ and\ \citenamefont
  {Kumar}}]{Alok:2020jod}%
  \BibitemOpen
  \bibfield  {author} {\bibinfo {author} {\bibfnamefont {A.~K.}\ \bibnamefont
  {Alok}}, \bibinfo {author} {\bibfnamefont {A.}~\bibnamefont {Dighe}},
  \bibinfo {author} {\bibfnamefont {S.}~\bibnamefont {Gangal}},\ and\ \bibinfo
  {author} {\bibfnamefont {J.}~\bibnamefont {Kumar}},\ }\bibfield  {title}
  {\bibinfo {title} {{The role of non-universal $Z$ couplings in explaining the
  $V_{us}$ anomaly}},\ }\href@noop {} {\  (\bibinfo {year} {2020})},\ \Eprint
  {https://arxiv.org/abs/2010.12009} {arXiv:2010.12009 [hep-ph]} \BibitemShut
  {NoStop}%
\bibitem [{\citenamefont {Belfatto}\ and\ \citenamefont
  {Berezhiani}(2021)}]{Belfatto:2021jhf}%
  \BibitemOpen
  \bibfield  {author} {\bibinfo {author} {\bibfnamefont {B.}~\bibnamefont
  {Belfatto}}\ and\ \bibinfo {author} {\bibfnamefont {Z.}~\bibnamefont
  {Berezhiani}},\ }\bibfield  {title} {\bibinfo {title} {{Are the CKM anomalies
  induced by vector-like quarks? Limits from flavor changing and Standard Model
  precision tests}},\ }\href@noop {} {\  (\bibinfo {year} {2021})},\ \Eprint
  {https://arxiv.org/abs/2103.05549} {arXiv:2103.05549 [hep-ph]} \BibitemShut
  {NoStop}%
\bibitem [{\citenamefont {Borsanyi}\ \emph {et~al.}(2021)\citenamefont
  {Borsanyi} \emph {et~al.}}]{Borsanyi:2020mff}%
  \BibitemOpen
  \bibfield  {author} {\bibinfo {author} {\bibfnamefont {S.}~\bibnamefont
  {Borsanyi}} \emph {et~al.},\ }\bibfield  {title} {\bibinfo {title} {{Leading
  hadronic contribution to the muon magnetic moment from lattice QCD}},\ }\href
  {https://doi.org/10.1038/s41586-021-03418-1} {\bibfield  {journal} {\bibinfo
  {journal} {Nature}\ }\textbf {\bibinfo {volume} {593}},\ \bibinfo {pages}
  {51} (\bibinfo {year} {2021})},\ \Eprint {https://arxiv.org/abs/2002.12347}
  {arXiv:2002.12347 [hep-lat]} \BibitemShut {NoStop}%
\bibitem [{\citenamefont {Buchmuller}\ \emph {et~al.}(1987)\citenamefont
  {Buchmuller}, \citenamefont {Ruckl},\ and\ \citenamefont
  {Wyler}}]{Buchmuller:1986zs}%
  \BibitemOpen
  \bibfield  {author} {\bibinfo {author} {\bibfnamefont {W.}~\bibnamefont
  {Buchmuller}}, \bibinfo {author} {\bibfnamefont {R.}~\bibnamefont {Ruckl}},\
  and\ \bibinfo {author} {\bibfnamefont {D.}~\bibnamefont {Wyler}},\ }\bibfield
   {title} {\bibinfo {title} {{Leptoquarks in Lepton - Quark Collisions}},\
  }\href {https://doi.org/10.1016/0370-2693(87)90637-X} {\bibfield  {journal}
  {\bibinfo  {journal} {Phys. Lett. B}\ }\textbf {\bibinfo {volume} {191}},\
  \bibinfo {pages} {442} (\bibinfo {year} {1987})},\ \bibinfo {note} {[Erratum:
  Phys.Lett.B 448, 320--320 (1999)]}\BibitemShut {NoStop}%
\bibitem [{\citenamefont {Aaboud}\ \emph
  {et~al.}(2018{\natexlab{b}})\citenamefont {Aaboud} \emph
  {et~al.}}]{Aaboud:2018pii}%
  \BibitemOpen
  \bibfield  {author} {\bibinfo {author} {\bibfnamefont {M.}~\bibnamefont
  {Aaboud}} \emph {et~al.} (\bibinfo {collaboration} {ATLAS}),\ }\bibfield
  {title} {\bibinfo {title} {{Combination of the searches for pair-produced
  vector-like partners of the third-generation quarks at $\sqrt{s} =$ 13 TeV
  with the ATLAS detector}},\ }\href
  {https://doi.org/10.1103/PhysRevLett.121.211801} {\bibfield  {journal}
  {\bibinfo  {journal} {Phys. Rev. Lett.}\ }\textbf {\bibinfo {volume} {121}},\
  \bibinfo {pages} {211801} (\bibinfo {year} {2018}{\natexlab{b}})},\ \Eprint
  {https://arxiv.org/abs/1808.02343} {arXiv:1808.02343 [hep-ex]} \BibitemShut
  {NoStop}%
\bibitem [{\citenamefont {Sirunyan}\ \emph
  {et~al.}(2019{\natexlab{a}})\citenamefont {Sirunyan} \emph
  {et~al.}}]{Sirunyan:2018qau}%
  \BibitemOpen
  \bibfield  {author} {\bibinfo {author} {\bibfnamefont {A.~M.}\ \bibnamefont
  {Sirunyan}} \emph {et~al.} (\bibinfo {collaboration} {CMS}),\ }\bibfield
  {title} {\bibinfo {title} {{Search for vector-like quarks in events with two
  oppositely charged leptons and jets in proton-proton collisions at $\sqrt{s}
  =$ 13 TeV}},\ }\href {https://doi.org/10.1140/epjc/s10052-019-6855-8}
  {\bibfield  {journal} {\bibinfo  {journal} {Eur. Phys. J. C}\ }\textbf
  {\bibinfo {volume} {79}},\ \bibinfo {pages} {364} (\bibinfo {year}
  {2019}{\natexlab{a}})},\ \Eprint {https://arxiv.org/abs/1812.09768}
  {arXiv:1812.09768 [hep-ex]} \BibitemShut {NoStop}%
\bibitem [{\citenamefont {Sirunyan}\ \emph
  {et~al.}(2019{\natexlab{b}})\citenamefont {Sirunyan} \emph
  {et~al.}}]{Sirunyan:2019sza}%
  \BibitemOpen
  \bibfield  {author} {\bibinfo {author} {\bibfnamefont {A.~M.}\ \bibnamefont
  {Sirunyan}} \emph {et~al.} (\bibinfo {collaboration} {CMS}),\ }\bibfield
  {title} {\bibinfo {title} {{Search for pair production of vectorlike quarks
  in the fully hadronic final state}},\ }\href
  {https://doi.org/10.1103/PhysRevD.100.072001} {\bibfield  {journal} {\bibinfo
   {journal} {Phys. Rev. D}\ }\textbf {\bibinfo {volume} {100}},\ \bibinfo
  {pages} {072001} (\bibinfo {year} {2019}{\natexlab{b}})},\ \Eprint
  {https://arxiv.org/abs/1906.11903} {arXiv:1906.11903 [hep-ex]} \BibitemShut
  {NoStop}%
\bibitem [{\citenamefont {Chang}\ \emph {et~al.}(2016)\citenamefont {Chang},
  \citenamefont {Liou}, \citenamefont {Wong},\ and\ \citenamefont
  {Xu}}]{Chang:2016zll}%
  \BibitemOpen
  \bibfield  {author} {\bibinfo {author} {\bibfnamefont {W.-F.}\ \bibnamefont
  {Chang}}, \bibinfo {author} {\bibfnamefont {S.-C.}\ \bibnamefont {Liou}},
  \bibinfo {author} {\bibfnamefont {C.-F.}\ \bibnamefont {Wong}},\ and\
  \bibinfo {author} {\bibfnamefont {F.}~\bibnamefont {Xu}},\ }\bibfield
  {title} {\bibinfo {title} {{Charged Lepton Flavor Violating Processes and
  Scalar Leptoquark Decay Branching Ratios in the Colored Zee-Babu Model}},\
  }\href {https://doi.org/10.1007/JHEP10(2016)106} {\bibfield  {journal}
  {\bibinfo  {journal} {JHEP}\ }\textbf {\bibinfo {volume} {10}},\ \bibinfo
  {pages} {106}},\ \Eprint {https://arxiv.org/abs/1608.05511} {arXiv:1608.05511
  [hep-ph]} \BibitemShut {NoStop}%
\bibitem [{\citenamefont {Diaz}\ \emph {et~al.}(2017)\citenamefont {Diaz},
  \citenamefont {Schmaltz},\ and\ \citenamefont {Zhong}}]{Diaz:2017lit}%
  \BibitemOpen
  \bibfield  {author} {\bibinfo {author} {\bibfnamefont {B.}~\bibnamefont
  {Diaz}}, \bibinfo {author} {\bibfnamefont {M.}~\bibnamefont {Schmaltz}},\
  and\ \bibinfo {author} {\bibfnamefont {Y.-M.}\ \bibnamefont {Zhong}},\
  }\bibfield  {title} {\bibinfo {title} {{The leptoquark
  Hunter\textquoteright{}s guide: Pair production}},\ }\href
  {https://doi.org/10.1007/JHEP10(2017)097} {\bibfield  {journal} {\bibinfo
  {journal} {JHEP}\ }\textbf {\bibinfo {volume} {10}},\ \bibinfo {pages}
  {097}},\ \Eprint {https://arxiv.org/abs/1706.05033} {arXiv:1706.05033
  [hep-ph]} \BibitemShut {NoStop}%
\bibitem [{\citenamefont {Schmaltz}\ and\ \citenamefont
  {Zhong}(2019)}]{Schmaltz:2018nls}%
  \BibitemOpen
  \bibfield  {author} {\bibinfo {author} {\bibfnamefont {M.}~\bibnamefont
  {Schmaltz}}\ and\ \bibinfo {author} {\bibfnamefont {Y.-M.}\ \bibnamefont
  {Zhong}},\ }\bibfield  {title} {\bibinfo {title} {{The leptoquark
  Hunter\textquoteright{}s guide: large coupling}},\ }\href
  {https://doi.org/10.1007/JHEP01(2019)132} {\bibfield  {journal} {\bibinfo
  {journal} {JHEP}\ }\textbf {\bibinfo {volume} {01}},\ \bibinfo {pages}
  {132}},\ \Eprint {https://arxiv.org/abs/1810.10017} {arXiv:1810.10017
  [hep-ph]} \BibitemShut {NoStop}%
\bibitem [{\citenamefont {Zee}(1980)}]{Zee:1980ai}%
  \BibitemOpen
  \bibfield  {author} {\bibinfo {author} {\bibfnamefont {A.}~\bibnamefont
  {Zee}},\ }\bibfield  {title} {\bibinfo {title} {{A Theory of Lepton Number
  Violation, Neutrino Majorana Mass, and Oscillation}},\ }\href
  {https://doi.org/10.1016/0370-2693(80)90349-4} {\bibfield  {journal}
  {\bibinfo  {journal} {Phys. Lett. B}\ }\textbf {\bibinfo {volume} {93}},\
  \bibinfo {pages} {389} (\bibinfo {year} {1980})},\ \bibinfo {note} {[Erratum:
  Phys.Lett.B 95, 461 (1980)]}\BibitemShut {NoStop}%
\bibitem [{\citenamefont {Schwaller}\ \emph {et~al.}(2013)\citenamefont
  {Schwaller}, \citenamefont {Tait},\ and\ \citenamefont
  {Vega-Morales}}]{Schwaller:2013hqa}%
  \BibitemOpen
  \bibfield  {author} {\bibinfo {author} {\bibfnamefont {P.}~\bibnamefont
  {Schwaller}}, \bibinfo {author} {\bibfnamefont {T.~M.~P.}\ \bibnamefont
  {Tait}},\ and\ \bibinfo {author} {\bibfnamefont {R.}~\bibnamefont
  {Vega-Morales}},\ }\bibfield  {title} {\bibinfo {title} {{Dark Matter and
  Vectorlike Leptons from Gauged Lepton Number}},\ }\href
  {https://doi.org/10.1103/PhysRevD.88.035001} {\bibfield  {journal} {\bibinfo
  {journal} {Phys. Rev. D}\ }\textbf {\bibinfo {volume} {88}},\ \bibinfo
  {pages} {035001} (\bibinfo {year} {2013})},\ \Eprint
  {https://arxiv.org/abs/1305.1108} {arXiv:1305.1108 [hep-ph]} \BibitemShut
  {NoStop}%
\bibitem [{\citenamefont {Chao}(2011)}]{Chao:2010mp}%
  \BibitemOpen
  \bibfield  {author} {\bibinfo {author} {\bibfnamefont {W.}~\bibnamefont
  {Chao}},\ }\bibfield  {title} {\bibinfo {title} {{Pure Leptonic Gauge
  Symmetry, Neutrino Masses and Dark Matter}},\ }\href
  {https://doi.org/10.1016/j.physletb.2010.10.056} {\bibfield  {journal}
  {\bibinfo  {journal} {Phys. Lett. B}\ }\textbf {\bibinfo {volume} {695}},\
  \bibinfo {pages} {157} (\bibinfo {year} {2011})},\ \Eprint
  {https://arxiv.org/abs/1005.1024} {arXiv:1005.1024 [hep-ph]} \BibitemShut
  {NoStop}%
\bibitem [{\citenamefont {Chang}\ and\ \citenamefont
  {Ng}(2019)}]{Chang:2018nid}%
  \BibitemOpen
  \bibfield  {author} {\bibinfo {author} {\bibfnamefont {W.-F.}\ \bibnamefont
  {Chang}}\ and\ \bibinfo {author} {\bibfnamefont {J.~N.}\ \bibnamefont {Ng}},\
  }\bibfield  {title} {\bibinfo {title} {{Alternative Perspective on Gauged
  Lepton Number and Implications for Collider Physics}},\ }\href
  {https://doi.org/10.1103/PhysRevD.99.075025} {\bibfield  {journal} {\bibinfo
  {journal} {Phys. Rev. D}\ }\textbf {\bibinfo {volume} {99}},\ \bibinfo
  {pages} {075025} (\bibinfo {year} {2019})},\ \Eprint
  {https://arxiv.org/abs/1808.08188} {arXiv:1808.08188 [hep-ph]} \BibitemShut
  {NoStop}%
\bibitem [{\citenamefont {Chang}\ and\ \citenamefont
  {Ng}(2018{\natexlab{a}})}]{Chang:2018wsw}%
  \BibitemOpen
  \bibfield  {author} {\bibinfo {author} {\bibfnamefont {W.-F.}\ \bibnamefont
  {Chang}}\ and\ \bibinfo {author} {\bibfnamefont {J.~N.}\ \bibnamefont {Ng}},\
  }\bibfield  {title} {\bibinfo {title} {{Neutrino masses and gauged
  $U(1)_\ell$ lepton number}},\ }\href
  {https://doi.org/10.1007/JHEP10(2018)015} {\bibfield  {journal} {\bibinfo
  {journal} {JHEP}\ }\textbf {\bibinfo {volume} {10}},\ \bibinfo {pages}
  {015}},\ \Eprint {https://arxiv.org/abs/1807.09439} {arXiv:1807.09439
  [hep-ph]} \BibitemShut {NoStop}%
\bibitem [{\citenamefont {Chang}\ and\ \citenamefont
  {Ng}(2018{\natexlab{b}})}]{Chang:2018vdd}%
  \BibitemOpen
  \bibfield  {author} {\bibinfo {author} {\bibfnamefont {W.-F.}\ \bibnamefont
  {Chang}}\ and\ \bibinfo {author} {\bibfnamefont {J.~N.}\ \bibnamefont {Ng}},\
  }\bibfield  {title} {\bibinfo {title} {{Study of Gauged Lepton Symmetry
  Signatures at Colliders}},\ }\href
  {https://doi.org/10.1103/PhysRevD.98.035015} {\bibfield  {journal} {\bibinfo
  {journal} {Phys. Rev. D}\ }\textbf {\bibinfo {volume} {98}},\ \bibinfo
  {pages} {035015} (\bibinfo {year} {2018}{\natexlab{b}})},\ \Eprint
  {https://arxiv.org/abs/1805.10382} {arXiv:1805.10382 [hep-ph]} \BibitemShut
  {NoStop}%
\bibitem [{\citenamefont {Weinberg}(1979)}]{Weinberg:1979sa}%
  \BibitemOpen
  \bibfield  {author} {\bibinfo {author} {\bibfnamefont {S.}~\bibnamefont
  {Weinberg}},\ }\bibfield  {title} {\bibinfo {title} {{Baryon and Lepton
  Nonconserving Processes}},\ }\href
  {https://doi.org/10.1103/PhysRevLett.43.1566} {\bibfield  {journal} {\bibinfo
   {journal} {Phys. Rev. Lett.}\ }\textbf {\bibinfo {volume} {43}},\ \bibinfo
  {pages} {1566} (\bibinfo {year} {1979})}\BibitemShut {NoStop}%
\bibitem [{\citenamefont {Chang}\ and\ \citenamefont
  {Ng}(2005)}]{Chang:2005ag}%
  \BibitemOpen
  \bibfield  {author} {\bibinfo {author} {\bibfnamefont {W.-F.}\ \bibnamefont
  {Chang}}\ and\ \bibinfo {author} {\bibfnamefont {J.~N.}\ \bibnamefont {Ng}},\
  }\bibfield  {title} {\bibinfo {title} {{Lepton flavor violation in extra
  dimension models}},\ }\href {https://doi.org/10.1103/PhysRevD.71.053003}
  {\bibfield  {journal} {\bibinfo  {journal} {Phys. Rev. D}\ }\textbf {\bibinfo
  {volume} {71}},\ \bibinfo {pages} {053003} (\bibinfo {year} {2005})},\
  \Eprint {https://arxiv.org/abs/hep-ph/0501161} {arXiv:hep-ph/0501161}
  \BibitemShut {NoStop}%
\bibitem [{\citenamefont {Baldini}\ \emph {et~al.}(2016)\citenamefont {Baldini}
  \emph {et~al.}}]{TheMEG:2016wtm}%
  \BibitemOpen
  \bibfield  {author} {\bibinfo {author} {\bibfnamefont {A.~M.}\ \bibnamefont
  {Baldini}} \emph {et~al.} (\bibinfo {collaboration} {MEG}),\ }\bibfield
  {title} {\bibinfo {title} {{Search for the lepton flavour violating decay
  $\mu ^+ \rightarrow \mathrm {e}^+ \gamma $ with the full dataset of the MEG
  experiment}},\ }\href {https://doi.org/10.1140/epjc/s10052-016-4271-x}
  {\bibfield  {journal} {\bibinfo  {journal} {Eur. Phys. J. C}\ }\textbf
  {\bibinfo {volume} {76}},\ \bibinfo {pages} {434} (\bibinfo {year} {2016})},\
  \Eprint {https://arxiv.org/abs/1605.05081} {arXiv:1605.05081 [hep-ex]}
  \BibitemShut {NoStop}%
\bibitem [{\citenamefont {Carpentier}\ and\ \citenamefont
  {Davidson}(2010)}]{Carpentier:2010ue}%
  \BibitemOpen
  \bibfield  {author} {\bibinfo {author} {\bibfnamefont {M.}~\bibnamefont
  {Carpentier}}\ and\ \bibinfo {author} {\bibfnamefont {S.}~\bibnamefont
  {Davidson}},\ }\bibfield  {title} {\bibinfo {title} {{Constraints on
  two-lepton, two quark operators}},\ }\href
  {https://doi.org/10.1140/epjc/s10052-010-1482-4} {\bibfield  {journal}
  {\bibinfo  {journal} {Eur. Phys. J. C}\ }\textbf {\bibinfo {volume} {70}},\
  \bibinfo {pages} {1071} (\bibinfo {year} {2010})},\ \Eprint
  {https://arxiv.org/abs/1008.0280} {arXiv:1008.0280 [hep-ph]} \BibitemShut
  {NoStop}%
\bibitem [{ATL(2018)}]{ATLAS:2018avw}%
  \BibitemOpen
  \bibfield  {title} {\bibinfo {title} {{Search for charged lepton-flavour
  violation in top-quark decays at the LHC with the ATLAS detector}},\
  }\href@noop {} {\  (\bibinfo {year} {2018})}\BibitemShut {NoStop}%
\bibitem [{\citenamefont {Jung}\ and\ \citenamefont
  {Straub}(2019)}]{Jung:2018lfu}%
  \BibitemOpen
  \bibfield  {author} {\bibinfo {author} {\bibfnamefont {M.}~\bibnamefont
  {Jung}}\ and\ \bibinfo {author} {\bibfnamefont {D.~M.}\ \bibnamefont
  {Straub}},\ }\bibfield  {title} {\bibinfo {title} {{Constraining new physics
  in $b\to c\ell\nu$ transitions}},\ }\href
  {https://doi.org/10.1007/JHEP01(2019)009} {\bibfield  {journal} {\bibinfo
  {journal} {JHEP}\ }\textbf {\bibinfo {volume} {01}},\ \bibinfo {pages}
  {009}},\ \Eprint {https://arxiv.org/abs/1801.01112} {arXiv:1801.01112
  [hep-ph]} \BibitemShut {NoStop}%
\bibitem [{\citenamefont {Aebischer}\ \emph {et~al.}(2019)\citenamefont
  {Aebischer}, \citenamefont {Crivellin},\ and\ \citenamefont
  {Greub}}]{Aebischer:2018acj}%
  \BibitemOpen
  \bibfield  {author} {\bibinfo {author} {\bibfnamefont {J.}~\bibnamefont
  {Aebischer}}, \bibinfo {author} {\bibfnamefont {A.}~\bibnamefont
  {Crivellin}},\ and\ \bibinfo {author} {\bibfnamefont {C.}~\bibnamefont
  {Greub}},\ }\bibfield  {title} {\bibinfo {title} {{QCD improved matching for
  semileptonic B decays with leptoquarks}},\ }\href
  {https://doi.org/10.1103/PhysRevD.99.055002} {\bibfield  {journal} {\bibinfo
  {journal} {Phys. Rev. D}\ }\textbf {\bibinfo {volume} {99}},\ \bibinfo
  {pages} {055002} (\bibinfo {year} {2019})},\ \Eprint
  {https://arxiv.org/abs/1811.08907} {arXiv:1811.08907 [hep-ph]} \BibitemShut
  {NoStop}%
\bibitem [{\citenamefont {Abada}\ \emph {et~al.}(2019)\citenamefont {Abada}
  \emph {et~al.}}]{Abada:2019zxq}%
  \BibitemOpen
  \bibfield  {author} {\bibinfo {author} {\bibfnamefont {A.}~\bibnamefont
  {Abada}} \emph {et~al.} (\bibinfo {collaboration} {FCC}),\ }\bibfield
  {title} {\bibinfo {title} {{FCC-ee: The Lepton Collider}: {Future Circular
  Collider Conceptual Design Report Volume 2}},\ }\href
  {https://doi.org/10.1140/epjst/e2019-900045-4} {\bibfield  {journal}
  {\bibinfo  {journal} {Eur. Phys. J. ST}\ }\textbf {\bibinfo {volume} {228}},\
  \bibinfo {pages} {261} (\bibinfo {year} {2019})}\BibitemShut {NoStop}%
\bibitem [{Bae(2013)}]{Baer:2013cma}%
  \BibitemOpen
  \bibfield  {title} {\bibinfo {title} {{The International Linear Collider
  Technical Design Report - Volume 2: Physics}},\ }\href@noop {} {\  (\bibinfo
  {year} {2013})},\ \Eprint {https://arxiv.org/abs/1306.6352} {arXiv:1306.6352
  [hep-ph]} \BibitemShut {NoStop}%
\bibitem [{\citenamefont {Dong}\ \emph {et~al.}(2018)\citenamefont {Dong} \emph
  {et~al.}}]{CEPCStudyGroup:2018ghi}%
  \BibitemOpen
  \bibfield  {author} {\bibinfo {author} {\bibfnamefont {M.}~\bibnamefont
  {Dong}} \emph {et~al.} (\bibinfo {collaboration} {CEPC Study Group}),\
  }\bibfield  {title} {\bibinfo {title} {{CEPC Conceptual Design Report: Volume
  2 - Physics \& Detector}},\ }\href@noop {} {\  (\bibinfo {year} {2018})},\
  \Eprint {https://arxiv.org/abs/1811.10545} {arXiv:1811.10545 [hep-ex]}
  \BibitemShut {NoStop}%
\bibitem [{\citenamefont {Chang}\ \emph {et~al.}(2000)\citenamefont {Chang},
  \citenamefont {Chang},\ and\ \citenamefont {Ma}}]{Chang:1999zc}%
  \BibitemOpen
  \bibfield  {author} {\bibinfo {author} {\bibfnamefont {D.}~\bibnamefont
  {Chang}}, \bibinfo {author} {\bibfnamefont {W.-F.}\ \bibnamefont {Chang}},\
  and\ \bibinfo {author} {\bibfnamefont {E.}~\bibnamefont {Ma}},\ }\bibfield
  {title} {\bibinfo {title} {{Fitting precision electroweak data with exotic
  heavy quarks}},\ }\href {https://doi.org/10.1103/PhysRevD.61.037301}
  {\bibfield  {journal} {\bibinfo  {journal} {Phys. Rev. D}\ }\textbf {\bibinfo
  {volume} {61}},\ \bibinfo {pages} {037301} (\bibinfo {year} {2000})},\
  \Eprint {https://arxiv.org/abs/hep-ph/9909537} {arXiv:hep-ph/9909537}
  \BibitemShut {NoStop}%
\bibitem [{\citenamefont {Chang}\ \emph {et~al.}(1999)\citenamefont {Chang},
  \citenamefont {Chang},\ and\ \citenamefont {Ma}}]{Chang:1998pt}%
  \BibitemOpen
  \bibfield  {author} {\bibinfo {author} {\bibfnamefont {D.}~\bibnamefont
  {Chang}}, \bibinfo {author} {\bibfnamefont {W.-F.}\ \bibnamefont {Chang}},\
  and\ \bibinfo {author} {\bibfnamefont {E.}~\bibnamefont {Ma}},\ }\bibfield
  {title} {\bibinfo {title} {{Alternative interpretation of the Tevatron top
  events}},\ }\href {https://doi.org/10.1103/PhysRevD.59.091503} {\bibfield
  {journal} {\bibinfo  {journal} {Phys. Rev. D}\ }\textbf {\bibinfo {volume}
  {59}},\ \bibinfo {pages} {091503} (\bibinfo {year} {1999})},\ \Eprint
  {https://arxiv.org/abs/hep-ph/9810531} {arXiv:hep-ph/9810531} \BibitemShut
  {NoStop}%
\bibitem [{\citenamefont {Choudhury}\ \emph {et~al.}(2002)\citenamefont
  {Choudhury}, \citenamefont {Tait},\ and\ \citenamefont
  {Wagner}}]{Choudhury:2001hs}%
  \BibitemOpen
  \bibfield  {author} {\bibinfo {author} {\bibfnamefont {D.}~\bibnamefont
  {Choudhury}}, \bibinfo {author} {\bibfnamefont {T.~M.~P.}\ \bibnamefont
  {Tait}},\ and\ \bibinfo {author} {\bibfnamefont {C.~E.~M.}\ \bibnamefont
  {Wagner}},\ }\bibfield  {title} {\bibinfo {title} {{Beautiful mirrors and
  precision electroweak data}},\ }\href
  {https://doi.org/10.1103/PhysRevD.65.053002} {\bibfield  {journal} {\bibinfo
  {journal} {Phys. Rev. D}\ }\textbf {\bibinfo {volume} {65}},\ \bibinfo
  {pages} {053002} (\bibinfo {year} {2002})},\ \Eprint
  {https://arxiv.org/abs/hep-ph/0109097} {arXiv:hep-ph/0109097} \BibitemShut
  {NoStop}%
\bibitem [{\citenamefont {Bona}\ \emph {et~al.}(2006)\citenamefont {Bona} \emph
  {et~al.}}]{Bona:2006sa}%
  \BibitemOpen
  \bibfield  {author} {\bibinfo {author} {\bibfnamefont {M.}~\bibnamefont
  {Bona}} \emph {et~al.} (\bibinfo {collaboration} {UTfit}),\ }\bibfield
  {title} {\bibinfo {title} {{Constraints on new physics from the quark mixing
  unitarity triangle}},\ }\href {https://doi.org/10.1103/PhysRevLett.97.151803}
  {\bibfield  {journal} {\bibinfo  {journal} {Phys. Rev. Lett.}\ }\textbf
  {\bibinfo {volume} {97}},\ \bibinfo {pages} {151803} (\bibinfo {year}
  {2006})},\ \Eprint {https://arxiv.org/abs/hep-ph/0605213}
  {arXiv:hep-ph/0605213} \BibitemShut {NoStop}%
\bibitem [{\citenamefont {Altmannshofer}\ \emph
  {et~al.}(2020{\natexlab{b}})\citenamefont {Altmannshofer}, \citenamefont
  {Dev}, \citenamefont {Soni},\ and\ \citenamefont
  {Sui}}]{Altmannshofer:2020axr}%
  \BibitemOpen
  \bibfield  {author} {\bibinfo {author} {\bibfnamefont {W.}~\bibnamefont
  {Altmannshofer}}, \bibinfo {author} {\bibfnamefont {P.~S.~B.}\ \bibnamefont
  {Dev}}, \bibinfo {author} {\bibfnamefont {A.}~\bibnamefont {Soni}},\ and\
  \bibinfo {author} {\bibfnamefont {Y.}~\bibnamefont {Sui}},\ }\bibfield
  {title} {\bibinfo {title} {{Addressing R$_{D^{(*)}}$, R$_{K^{(*)}}$, muon
  $g-2$ and ANITA anomalies in a minimal $R$-parity violating supersymmetric
  framework}},\ }\href {https://doi.org/10.1103/PhysRevD.102.015031} {\bibfield
   {journal} {\bibinfo  {journal} {Phys. Rev. D}\ }\textbf {\bibinfo {volume}
  {102}},\ \bibinfo {pages} {015031} (\bibinfo {year} {2020}{\natexlab{b}})},\
  \Eprint {https://arxiv.org/abs/2002.12910} {arXiv:2002.12910 [hep-ph]}
  \BibitemShut {NoStop}%
\bibitem [{\citenamefont {Huang}\ \emph {et~al.}(2020)\citenamefont {Huang},
  \citenamefont {Morais},\ and\ \citenamefont {Santos}}]{Huang:2020ris}%
  \BibitemOpen
  \bibfield  {author} {\bibinfo {author} {\bibfnamefont {D.}~\bibnamefont
  {Huang}}, \bibinfo {author} {\bibfnamefont {A.~P.}\ \bibnamefont {Morais}},\
  and\ \bibinfo {author} {\bibfnamefont {R.}~\bibnamefont {Santos}},\
  }\bibfield  {title} {\bibinfo {title} {{Anomalies in $B$-meson decays and the
  muon $g-2$ from dark loops}},\ }\href
  {https://doi.org/10.1103/PhysRevD.102.075009} {\bibfield  {journal} {\bibinfo
   {journal} {Phys. Rev. D}\ }\textbf {\bibinfo {volume} {102}},\ \bibinfo
  {pages} {075009} (\bibinfo {year} {2020})},\ \Eprint
  {https://arxiv.org/abs/2007.05082} {arXiv:2007.05082 [hep-ph]} \BibitemShut
  {NoStop}%
\bibitem [{\citenamefont {Buras}\ \emph {et~al.}(2015)\citenamefont {Buras},
  \citenamefont {Girrbach-Noe}, \citenamefont {Niehoff},\ and\ \citenamefont
  {Straub}}]{Buras:2014fpa}%
  \BibitemOpen
  \bibfield  {author} {\bibinfo {author} {\bibfnamefont {A.~J.}\ \bibnamefont
  {Buras}}, \bibinfo {author} {\bibfnamefont {J.}~\bibnamefont {Girrbach-Noe}},
  \bibinfo {author} {\bibfnamefont {C.}~\bibnamefont {Niehoff}},\ and\ \bibinfo
  {author} {\bibfnamefont {D.~M.}\ \bibnamefont {Straub}},\ }\bibfield  {title}
  {\bibinfo {title} {{$ B\to {K}^{\left(\ast \right)}\nu \overline{\nu} $
  decays in the Standard Model and beyond}},\ }\href
  {https://doi.org/10.1007/JHEP02(2015)184} {\bibfield  {journal} {\bibinfo
  {journal} {JHEP}\ }\textbf {\bibinfo {volume} {02}},\ \bibinfo {pages}
  {184}},\ \Eprint {https://arxiv.org/abs/1409.4557} {arXiv:1409.4557 [hep-ph]}
  \BibitemShut {NoStop}%
\bibitem [{\citenamefont {Grygier}\ \emph {et~al.}(2017)\citenamefont {Grygier}
  \emph {et~al.}}]{Belle:2017oht}%
  \BibitemOpen
  \bibfield  {author} {\bibinfo {author} {\bibfnamefont {J.}~\bibnamefont
  {Grygier}} \emph {et~al.} (\bibinfo {collaboration} {Belle}),\ }\bibfield
  {title} {\bibinfo {title} {{Search for $\boldsymbol{B\to h\nu\bar{\nu}}$
  decays with semileptonic tagging at Belle}},\ }\href
  {https://doi.org/10.1103/PhysRevD.96.091101} {\bibfield  {journal} {\bibinfo
  {journal} {Phys. Rev. D}\ }\textbf {\bibinfo {volume} {96}},\ \bibinfo
  {pages} {091101} (\bibinfo {year} {2017})},\ \bibinfo {note} {[Addendum:
  Phys.Rev.D 97, 099902 (2018)]},\ \Eprint {https://arxiv.org/abs/1702.03224}
  {arXiv:1702.03224 [hep-ex]} \BibitemShut {NoStop}%
\bibitem [{\citenamefont {Aubert}\ \emph {et~al.}(2010)\citenamefont {Aubert}
  \emph {et~al.}}]{Aubert:2009ag}%
  \BibitemOpen
  \bibfield  {author} {\bibinfo {author} {\bibfnamefont {B.}~\bibnamefont
  {Aubert}} \emph {et~al.} (\bibinfo {collaboration} {BaBar}),\ }\bibfield
  {title} {\bibinfo {title} {{Searches for Lepton Flavor Violation in the
  Decays tau+- ---\ensuremath{>} e+- gamma and tau+- ---\ensuremath{>} mu+-
  gamma}},\ }\href {https://doi.org/10.1103/PhysRevLett.104.021802} {\bibfield
  {journal} {\bibinfo  {journal} {Phys. Rev. Lett.}\ }\textbf {\bibinfo
  {volume} {104}},\ \bibinfo {pages} {021802} (\bibinfo {year} {2010})},\
  \Eprint {https://arxiv.org/abs/0908.2381} {arXiv:0908.2381 [hep-ex]}
  \BibitemShut {NoStop}%
\bibitem [{\citenamefont {Amhis}\ \emph {et~al.}(2017)\citenamefont {Amhis}
  \emph {et~al.}}]{Amhis:2016xyh}%
  \BibitemOpen
  \bibfield  {author} {\bibinfo {author} {\bibfnamefont {Y.}~\bibnamefont
  {Amhis}} \emph {et~al.} (\bibinfo {collaboration} {HFLAV}),\ }\bibfield
  {title} {\bibinfo {title} {{Averages of $b$-hadron, $c$-hadron, and
  $\tau$-lepton properties as of summer 2016}},\ }\href
  {https://doi.org/10.1140/epjc/s10052-017-5058-4} {\bibfield  {journal}
  {\bibinfo  {journal} {Eur. Phys. J. C}\ }\textbf {\bibinfo {volume} {77}},\
  \bibinfo {pages} {895} (\bibinfo {year} {2017})},\ \Eprint
  {https://arxiv.org/abs/1612.07233} {arXiv:1612.07233 [hep-ex]} \BibitemShut
  {NoStop}%
\bibitem [{\citenamefont {Misiak}\ \emph {et~al.}(2015)\citenamefont {Misiak}
  \emph {et~al.}}]{Misiak:2015xwa}%
  \BibitemOpen
  \bibfield  {author} {\bibinfo {author} {\bibfnamefont {M.}~\bibnamefont
  {Misiak}} \emph {et~al.},\ }\bibfield  {title} {\bibinfo {title} {{Updated
  NNLO QCD predictions for the weak radiative B-meson decays}},\ }\href
  {https://doi.org/10.1103/PhysRevLett.114.221801} {\bibfield  {journal}
  {\bibinfo  {journal} {Phys. Rev. Lett.}\ }\textbf {\bibinfo {volume} {114}},\
  \bibinfo {pages} {221801} (\bibinfo {year} {2015})},\ \Eprint
  {https://arxiv.org/abs/1503.01789} {arXiv:1503.01789 [hep-ph]} \BibitemShut
  {NoStop}%
\bibitem [{\citenamefont {Misiak}\ \emph {et~al.}(2017)\citenamefont {Misiak},
  \citenamefont {Rehman},\ and\ \citenamefont {Steinhauser}}]{Misiak:2017woa}%
  \BibitemOpen
  \bibfield  {author} {\bibinfo {author} {\bibfnamefont {M.}~\bibnamefont
  {Misiak}}, \bibinfo {author} {\bibfnamefont {A.}~\bibnamefont {Rehman}},\
  and\ \bibinfo {author} {\bibfnamefont {M.}~\bibnamefont {Steinhauser}},\
  }\bibfield  {title} {\bibinfo {title} {{NNLO QCD counterterm contributions to
  $\bar B \to X_{s\gamma}$ for the physical value of $m_c$}},\ }\href
  {https://doi.org/10.1016/j.physletb.2017.05.008} {\bibfield  {journal}
  {\bibinfo  {journal} {Phys. Lett. B}\ }\textbf {\bibinfo {volume} {770}},\
  \bibinfo {pages} {431} (\bibinfo {year} {2017})},\ \Eprint
  {https://arxiv.org/abs/1702.07674} {arXiv:1702.07674 [hep-ph]} \BibitemShut
  {NoStop}%
\bibitem [{\citenamefont {Esteban}\ \emph {et~al.}(2020)\citenamefont
  {Esteban}, \citenamefont {Gonzalez-Garcia}, \citenamefont {Maltoni},
  \citenamefont {Schwetz},\ and\ \citenamefont {Zhou}}]{Esteban:2020cvm}%
  \BibitemOpen
  \bibfield  {author} {\bibinfo {author} {\bibfnamefont {I.}~\bibnamefont
  {Esteban}}, \bibinfo {author} {\bibfnamefont {M.~C.}\ \bibnamefont
  {Gonzalez-Garcia}}, \bibinfo {author} {\bibfnamefont {M.}~\bibnamefont
  {Maltoni}}, \bibinfo {author} {\bibfnamefont {T.}~\bibnamefont {Schwetz}},\
  and\ \bibinfo {author} {\bibfnamefont {A.}~\bibnamefont {Zhou}},\ }\bibfield
  {title} {\bibinfo {title} {{The fate of hints: updated global analysis of
  three-flavor neutrino oscillations}},\ }\href
  {https://doi.org/10.1007/JHEP09(2020)178} {\bibfield  {journal} {\bibinfo
  {journal} {JHEP}\ }\textbf {\bibinfo {volume} {09}},\ \bibinfo {pages}
  {178}},\ \Eprint {https://arxiv.org/abs/2007.14792} {arXiv:2007.14792
  [hep-ph]} \BibitemShut {NoStop}%
\bibitem [{\citenamefont {Dohmen}\ \emph {et~al.}(1993)\citenamefont {Dohmen}
  \emph {et~al.}}]{Dohmen:1993mp}%
  \BibitemOpen
  \bibfield  {author} {\bibinfo {author} {\bibfnamefont {C.}~\bibnamefont
  {Dohmen}} \emph {et~al.} (\bibinfo {collaboration} {SINDRUM II}),\ }\bibfield
   {title} {\bibinfo {title} {{Test of lepton flavor conservation in mu
  ---\ensuremath{>} e conversion on titanium}},\ }\href
  {https://doi.org/10.1016/0370-2693(93)91383-X} {\bibfield  {journal}
  {\bibinfo  {journal} {Phys. Lett. B}\ }\textbf {\bibinfo {volume} {317}},\
  \bibinfo {pages} {631} (\bibinfo {year} {1993})}\BibitemShut {NoStop}%
\bibitem [{\citenamefont {Tornow}(2014)}]{Tornow:2014vta}%
  \BibitemOpen
  \bibfield  {author} {\bibinfo {author} {\bibfnamefont {W.}~\bibnamefont
  {Tornow}} (\bibinfo {collaboration} {KamLAND-Zen}),\ }\bibfield  {title}
  {\bibinfo {title} {{Search for Neutrinoless Double-Beta Decay}},\ }in\
  \href@noop {} {\emph {\bibinfo {booktitle} {{34th International Symposium on
  Physics in Collision}}}}\ (\bibinfo {year} {2014})\ \Eprint
  {https://arxiv.org/abs/1412.0734} {arXiv:1412.0734 [nucl-ex]} \BibitemShut
  {NoStop}%
\bibitem [{\citenamefont {Dor\v{s}ner}\ \emph {et~al.}(2016)\citenamefont
  {Dor\v{s}ner}, \citenamefont {Fajfer}, \citenamefont {Greljo}, \citenamefont
  {Kamenik},\ and\ \citenamefont {Ko\v{s}nik}}]{Dorsner:2016wpm}%
  \BibitemOpen
  \bibfield  {author} {\bibinfo {author} {\bibfnamefont {I.}~\bibnamefont
  {Dor\v{s}ner}}, \bibinfo {author} {\bibfnamefont {S.}~\bibnamefont {Fajfer}},
  \bibinfo {author} {\bibfnamefont {A.}~\bibnamefont {Greljo}}, \bibinfo
  {author} {\bibfnamefont {J.~F.}\ \bibnamefont {Kamenik}},\ and\ \bibinfo
  {author} {\bibfnamefont {N.}~\bibnamefont {Ko\v{s}nik}},\ }\bibfield  {title}
  {\bibinfo {title} {{Physics of leptoquarks in precision experiments and at
  particle colliders}},\ }\href {https://doi.org/10.1016/j.physrep.2016.06.001}
  {\bibfield  {journal} {\bibinfo  {journal} {Phys. Rept.}\ }\textbf {\bibinfo
  {volume} {641}},\ \bibinfo {pages} {1} (\bibinfo {year} {2016})},\ \Eprint
  {https://arxiv.org/abs/1603.04993} {arXiv:1603.04993 [hep-ph]} \BibitemShut
  {NoStop}%
\bibitem [{\citenamefont {Kim}\ \emph {et~al.}(2019)\citenamefont {Kim},
  \citenamefont {Ko}, \citenamefont {Li}, \citenamefont {Park},\ and\
  \citenamefont {Wu}}]{Kim:2018oih}%
  \BibitemOpen
  \bibfield  {author} {\bibinfo {author} {\bibfnamefont {T.~J.}\ \bibnamefont
  {Kim}}, \bibinfo {author} {\bibfnamefont {P.}~\bibnamefont {Ko}}, \bibinfo
  {author} {\bibfnamefont {J.}~\bibnamefont {Li}}, \bibinfo {author}
  {\bibfnamefont {J.}~\bibnamefont {Park}},\ and\ \bibinfo {author}
  {\bibfnamefont {P.}~\bibnamefont {Wu}},\ }\bibfield  {title} {\bibinfo
  {title} {{Correlation between $ {R}_{D^{\left(\ast \right)}} $ and top quark
  FCNC decays in leptoquark models}},\ }\href
  {https://doi.org/10.1007/JHEP07(2019)025} {\bibfield  {journal} {\bibinfo
  {journal} {JHEP}\ }\textbf {\bibinfo {volume} {07}},\ \bibinfo {pages}
  {025}},\ \Eprint {https://arxiv.org/abs/1812.08484} {arXiv:1812.08484
  [hep-ph]} \BibitemShut {NoStop}%
\bibitem [{\citenamefont {Bobeth}\ \emph {et~al.}(2000)\citenamefont {Bobeth},
  \citenamefont {Misiak},\ and\ \citenamefont {Urban}}]{Bobeth:1999mk}%
  \BibitemOpen
  \bibfield  {author} {\bibinfo {author} {\bibfnamefont {C.}~\bibnamefont
  {Bobeth}}, \bibinfo {author} {\bibfnamefont {M.}~\bibnamefont {Misiak}},\
  and\ \bibinfo {author} {\bibfnamefont {J.}~\bibnamefont {Urban}},\ }\bibfield
   {title} {\bibinfo {title} {{Photonic penguins at two loops and $m_t$
  dependence of $BR[B \to X_s l^+ l^-]$}},\ }\href
  {https://doi.org/10.1016/S0550-3213(00)00007-9} {\bibfield  {journal}
  {\bibinfo  {journal} {Nucl. Phys. B}\ }\textbf {\bibinfo {volume} {574}},\
  \bibinfo {pages} {291} (\bibinfo {year} {2000})},\ \Eprint
  {https://arxiv.org/abs/hep-ph/9910220} {arXiv:hep-ph/9910220} \BibitemShut
  {NoStop}%
\bibitem [{\citenamefont {Huber}\ \emph {et~al.}(2006)\citenamefont {Huber},
  \citenamefont {Lunghi}, \citenamefont {Misiak},\ and\ \citenamefont
  {Wyler}}]{Huber:2005ig}%
  \BibitemOpen
  \bibfield  {author} {\bibinfo {author} {\bibfnamefont {T.}~\bibnamefont
  {Huber}}, \bibinfo {author} {\bibfnamefont {E.}~\bibnamefont {Lunghi}},
  \bibinfo {author} {\bibfnamefont {M.}~\bibnamefont {Misiak}},\ and\ \bibinfo
  {author} {\bibfnamefont {D.}~\bibnamefont {Wyler}},\ }\bibfield  {title}
  {\bibinfo {title} {{Electromagnetic logarithms in $\bar B \to X_s l^+
  l^-$}},\ }\href {https://doi.org/10.1016/j.nuclphysb.2006.01.037} {\bibfield
  {journal} {\bibinfo  {journal} {Nucl. Phys. B}\ }\textbf {\bibinfo {volume}
  {740}},\ \bibinfo {pages} {105} (\bibinfo {year} {2006})},\ \Eprint
  {https://arxiv.org/abs/hep-ph/0512066} {arXiv:hep-ph/0512066} \BibitemShut
  {NoStop}%
\bibitem [{\citenamefont {Descotes-Genon}\ \emph {et~al.}(2011)\citenamefont
  {Descotes-Genon}, \citenamefont {Ghosh}, \citenamefont {Matias},\ and\
  \citenamefont {Ramon}}]{DescotesGenon:2011yn}%
  \BibitemOpen
  \bibfield  {author} {\bibinfo {author} {\bibfnamefont {S.}~\bibnamefont
  {Descotes-Genon}}, \bibinfo {author} {\bibfnamefont {D.}~\bibnamefont
  {Ghosh}}, \bibinfo {author} {\bibfnamefont {J.}~\bibnamefont {Matias}},\ and\
  \bibinfo {author} {\bibfnamefont {M.}~\bibnamefont {Ramon}},\ }\bibfield
  {title} {\bibinfo {title} {{Exploring New Physics in the C7-C7' plane}},\
  }\href {https://doi.org/10.1007/JHEP06(2011)099} {\bibfield  {journal}
  {\bibinfo  {journal} {JHEP}\ }\textbf {\bibinfo {volume} {06}},\ \bibinfo
  {pages} {099}},\ \Eprint {https://arxiv.org/abs/1104.3342} {arXiv:1104.3342
  [hep-ph]} \BibitemShut {NoStop}%
\bibitem [{\citenamefont {Aaij}\ \emph
  {et~al.}(2017{\natexlab{b}})\citenamefont {Aaij} \emph
  {et~al.}}]{Aaij:2017xqt}%
  \BibitemOpen
  \bibfield  {author} {\bibinfo {author} {\bibfnamefont {R.}~\bibnamefont
  {Aaij}} \emph {et~al.} (\bibinfo {collaboration} {LHCb}),\ }\bibfield
  {title} {\bibinfo {title} {{Search for the decays $B_s^0\to\tau^+\tau^-$ and
  $B^0\to\tau^+\tau^-$}},\ }\href
  {https://doi.org/10.1103/PhysRevLett.118.251802} {\bibfield  {journal}
  {\bibinfo  {journal} {Phys. Rev. Lett.}\ }\textbf {\bibinfo {volume} {118}},\
  \bibinfo {pages} {251802} (\bibinfo {year} {2017}{\natexlab{b}})},\ \Eprint
  {https://arxiv.org/abs/1703.02508} {arXiv:1703.02508 [hep-ex]} \BibitemShut
  {NoStop}%
\bibitem [{\citenamefont {Lees}\ \emph {et~al.}(2017)\citenamefont {Lees} \emph
  {et~al.}}]{TheBaBar:2016xwe}%
  \BibitemOpen
  \bibfield  {author} {\bibinfo {author} {\bibfnamefont {J.~P.}\ \bibnamefont
  {Lees}} \emph {et~al.} (\bibinfo {collaboration} {BaBar}),\ }\bibfield
  {title} {\bibinfo {title} {{Search for $B^{+}\rightarrow K^{+}
  \tau^{+}\tau^{-}$ at the BaBar experiment}},\ }\href
  {https://doi.org/10.1103/PhysRevLett.118.031802} {\bibfield  {journal}
  {\bibinfo  {journal} {Phys. Rev. Lett.}\ }\textbf {\bibinfo {volume} {118}},\
  \bibinfo {pages} {031802} (\bibinfo {year} {2017})},\ \Eprint
  {https://arxiv.org/abs/1605.09637} {arXiv:1605.09637 [hep-ex]} \BibitemShut
  {NoStop}%
\bibitem [{\citenamefont {Capdevila}\ \emph
  {et~al.}(2018{\natexlab{b}})\citenamefont {Capdevila}, \citenamefont
  {Crivellin}, \citenamefont {Descotes-Genon}, \citenamefont {Hofer},\ and\
  \citenamefont {Matias}}]{Capdevila:2017iqn}%
  \BibitemOpen
  \bibfield  {author} {\bibinfo {author} {\bibfnamefont {B.}~\bibnamefont
  {Capdevila}}, \bibinfo {author} {\bibfnamefont {A.}~\bibnamefont
  {Crivellin}}, \bibinfo {author} {\bibfnamefont {S.}~\bibnamefont
  {Descotes-Genon}}, \bibinfo {author} {\bibfnamefont {L.}~\bibnamefont
  {Hofer}},\ and\ \bibinfo {author} {\bibfnamefont {J.}~\bibnamefont
  {Matias}},\ }\bibfield  {title} {\bibinfo {title} {{Searching for New Physics
  with $b\to s\tau^+\tau^-$ processes}},\ }\href
  {https://doi.org/10.1103/PhysRevLett.120.181802} {\bibfield  {journal}
  {\bibinfo  {journal} {Phys. Rev. Lett.}\ }\textbf {\bibinfo {volume} {120}},\
  \bibinfo {pages} {181802} (\bibinfo {year} {2018}{\natexlab{b}})},\ \Eprint
  {https://arxiv.org/abs/1712.01919} {arXiv:1712.01919 [hep-ph]} \BibitemShut
  {NoStop}%
\bibitem [{\citenamefont {Li}\ and\ \citenamefont {Liu}(2020)}]{Li:2020bvr}%
  \BibitemOpen
  \bibfield  {author} {\bibinfo {author} {\bibfnamefont {L.}~\bibnamefont
  {Li}}\ and\ \bibinfo {author} {\bibfnamefont {T.}~\bibnamefont {Liu}},\
  }\bibfield  {title} {\bibinfo {title} {{$b\to s\tau^+\tau^-$ Physics at
  Future $Z$ Factories}},\ }\href@noop {} {\  (\bibinfo {year} {2020})},\
  \Eprint {https://arxiv.org/abs/2012.00665} {arXiv:2012.00665 [hep-ph]}
  \BibitemShut {NoStop}%
\bibitem [{\citenamefont {Murgui}\ \emph {et~al.}(2019)\citenamefont {Murgui},
  \citenamefont {Pe\~nuelas}, \citenamefont {Jung},\ and\ \citenamefont
  {Pich}}]{Murgui:2019czp}%
  \BibitemOpen
  \bibfield  {author} {\bibinfo {author} {\bibfnamefont {C.}~\bibnamefont
  {Murgui}}, \bibinfo {author} {\bibfnamefont {A.}~\bibnamefont {Pe\~nuelas}},
  \bibinfo {author} {\bibfnamefont {M.}~\bibnamefont {Jung}},\ and\ \bibinfo
  {author} {\bibfnamefont {A.}~\bibnamefont {Pich}},\ }\bibfield  {title}
  {\bibinfo {title} {{Global fit to $b \to c \tau \nu$ transitions}},\ }\href
  {https://doi.org/10.1007/JHEP09(2019)103} {\bibfield  {journal} {\bibinfo
  {journal} {JHEP}\ }\textbf {\bibinfo {volume} {09}},\ \bibinfo {pages}
  {103}},\ \Eprint {https://arxiv.org/abs/1904.09311} {arXiv:1904.09311
  [hep-ph]} \BibitemShut {NoStop}%
\bibitem [{\citenamefont {Shi}\ \emph {et~al.}(2019)\citenamefont {Shi},
  \citenamefont {Geng}, \citenamefont {Grinstein}, \citenamefont {J\"ager},\
  and\ \citenamefont {Martin~Camalich}}]{Shi:2019gxi}%
  \BibitemOpen
  \bibfield  {author} {\bibinfo {author} {\bibfnamefont {R.-X.}\ \bibnamefont
  {Shi}}, \bibinfo {author} {\bibfnamefont {L.-S.}\ \bibnamefont {Geng}},
  \bibinfo {author} {\bibfnamefont {B.}~\bibnamefont {Grinstein}}, \bibinfo
  {author} {\bibfnamefont {S.}~\bibnamefont {J\"ager}},\ and\ \bibinfo {author}
  {\bibfnamefont {J.}~\bibnamefont {Martin~Camalich}},\ }\bibfield  {title}
  {\bibinfo {title} {{Revisiting the new-physics interpretation of the $b\to
  c\tau\nu$ data}},\ }\href {https://doi.org/10.1007/JHEP12(2019)065}
  {\bibfield  {journal} {\bibinfo  {journal} {JHEP}\ }\textbf {\bibinfo
  {volume} {12}},\ \bibinfo {pages} {065}},\ \Eprint
  {https://arxiv.org/abs/1905.08498} {arXiv:1905.08498 [hep-ph]} \BibitemShut
  {NoStop}%
\bibitem [{\citenamefont {Blanke}\ \emph {et~al.}(2019)\citenamefont {Blanke},
  \citenamefont {Crivellin}, \citenamefont {Kitahara}, \citenamefont {Moscati},
  \citenamefont {Nierste},\ and\ \citenamefont
  {Ni\v{s}and\v{z}i\'c}}]{Blanke:2019qrx}%
  \BibitemOpen
  \bibfield  {author} {\bibinfo {author} {\bibfnamefont {M.}~\bibnamefont
  {Blanke}}, \bibinfo {author} {\bibfnamefont {A.}~\bibnamefont {Crivellin}},
  \bibinfo {author} {\bibfnamefont {T.}~\bibnamefont {Kitahara}}, \bibinfo
  {author} {\bibfnamefont {M.}~\bibnamefont {Moscati}}, \bibinfo {author}
  {\bibfnamefont {U.}~\bibnamefont {Nierste}},\ and\ \bibinfo {author}
  {\bibfnamefont {I.}~\bibnamefont {Ni\v{s}and\v{z}i\'c}},\ }\bibfield  {title}
  {\bibinfo {title} {{Addendum to \textquotedblleft{}Impact of polarization
  observables and $B_c\to \tau \nu$ on new physics explanations of the $b\to c
  \tau \nu$ anomaly''}}\ }\href {https://doi.org/10.1103/PhysRevD.100.035035}
  {10.1103/PhysRevD.100.035035} (\bibinfo {year} {2019}),\ \bibinfo {note}
  {[Addendum: Phys.Rev.D 100, 035035 (2019)]},\ \Eprint
  {https://arxiv.org/abs/1905.08253} {arXiv:1905.08253 [hep-ph]} \BibitemShut
  {NoStop}%
\bibitem [{\citenamefont {Kumbhakar}\ \emph {et~al.}(2020)\citenamefont
  {Kumbhakar}, \citenamefont {Alok}, \citenamefont {Kumar},\ and\ \citenamefont
  {Sankar}}]{Kumbhakar:2019avh}%
  \BibitemOpen
  \bibfield  {author} {\bibinfo {author} {\bibfnamefont {S.}~\bibnamefont
  {Kumbhakar}}, \bibinfo {author} {\bibfnamefont {A.~K.}\ \bibnamefont {Alok}},
  \bibinfo {author} {\bibfnamefont {D.}~\bibnamefont {Kumar}},\ and\ \bibinfo
  {author} {\bibfnamefont {S.~U.}\ \bibnamefont {Sankar}},\ }\bibfield  {title}
  {\bibinfo {title} {{A global fit to $b\rightarrow c\tau\bar{\nu}$ anomalies
  after Moriond 2019}},\ }\href {https://doi.org/10.22323/1.364.0272}
  {\bibfield  {journal} {\bibinfo  {journal} {PoS}\ }\textbf {\bibinfo {volume}
  {EPS-HEP2019}},\ \bibinfo {pages} {272} (\bibinfo {year} {2020})},\ \Eprint
  {https://arxiv.org/abs/1909.02840} {arXiv:1909.02840 [hep-ph]} \BibitemShut
  {NoStop}%
\bibitem [{\citenamefont {Lees}\ \emph {et~al.}(2012)\citenamefont {Lees} \emph
  {et~al.}}]{Lees:2012xj}%
  \BibitemOpen
  \bibfield  {author} {\bibinfo {author} {\bibfnamefont {J.~P.}\ \bibnamefont
  {Lees}} \emph {et~al.} (\bibinfo {collaboration} {BaBar}),\ }\bibfield
  {title} {\bibinfo {title} {{Evidence for an excess of $\bar{B} \to D^{(*)}
  \tau^-\bar{\nu}_\tau$ decays}},\ }\href
  {https://doi.org/10.1103/PhysRevLett.109.101802} {\bibfield  {journal}
  {\bibinfo  {journal} {Phys. Rev. Lett.}\ }\textbf {\bibinfo {volume} {109}},\
  \bibinfo {pages} {101802} (\bibinfo {year} {2012})},\ \Eprint
  {https://arxiv.org/abs/1205.5442} {arXiv:1205.5442 [hep-ex]} \BibitemShut
  {NoStop}%
\bibitem [{\citenamefont {Lees}\ \emph {et~al.}(2013)\citenamefont {Lees} \emph
  {et~al.}}]{Lees:2013uzd}%
  \BibitemOpen
  \bibfield  {author} {\bibinfo {author} {\bibfnamefont {J.~P.}\ \bibnamefont
  {Lees}} \emph {et~al.} (\bibinfo {collaboration} {BaBar}),\ }\bibfield
  {title} {\bibinfo {title} {{Measurement of an Excess of $\bar{B} \to
  D^{(*)}\tau^- \bar{\nu}_\tau$ Decays and Implications for Charged Higgs
  Bosons}},\ }\href {https://doi.org/10.1103/PhysRevD.88.072012} {\bibfield
  {journal} {\bibinfo  {journal} {Phys. Rev. D}\ }\textbf {\bibinfo {volume}
  {88}},\ \bibinfo {pages} {072012} (\bibinfo {year} {2013})},\ \Eprint
  {https://arxiv.org/abs/1303.0571} {arXiv:1303.0571 [hep-ex]} \BibitemShut
  {NoStop}%
\bibitem [{\citenamefont {Aaij}\ \emph
  {et~al.}(2015{\natexlab{b}})\citenamefont {Aaij} \emph
  {et~al.}}]{Aaij:2015yra}%
  \BibitemOpen
  \bibfield  {author} {\bibinfo {author} {\bibfnamefont {R.}~\bibnamefont
  {Aaij}} \emph {et~al.} (\bibinfo {collaboration} {LHCb}),\ }\bibfield
  {title} {\bibinfo {title} {{Measurement of the ratio of branching fractions
  $\mathcal{B}(\bar{B}^0 \to
  D^{*+}\tau^{-}\bar{\nu}_{\tau})/\mathcal{B}(\bar{B}^0 \to
  D^{*+}\mu^{-}\bar{\nu}_{\mu})$}},\ }\href
  {https://doi.org/10.1103/PhysRevLett.115.111803} {\bibfield  {journal}
  {\bibinfo  {journal} {Phys. Rev. Lett.}\ }\textbf {\bibinfo {volume} {115}},\
  \bibinfo {pages} {111803} (\bibinfo {year} {2015}{\natexlab{b}})},\ \bibinfo
  {note} {[Erratum: Phys.Rev.Lett. 115, 159901 (2015)]},\ \Eprint
  {https://arxiv.org/abs/1506.08614} {arXiv:1506.08614 [hep-ex]} \BibitemShut
  {NoStop}%
\bibitem [{\citenamefont {Aaij}\ \emph
  {et~al.}(2018{\natexlab{a}})\citenamefont {Aaij} \emph
  {et~al.}}]{Aaij:2017deq}%
  \BibitemOpen
  \bibfield  {author} {\bibinfo {author} {\bibfnamefont {R.}~\bibnamefont
  {Aaij}} \emph {et~al.} (\bibinfo {collaboration} {LHCb}),\ }\bibfield
  {title} {\bibinfo {title} {{Test of Lepton Flavor Universality by the
  measurement of the $B^0 \to D^{*-} \tau^+ \nu_{\tau}$ branching fraction
  using three-prong $\tau$ decays}},\ }\href
  {https://doi.org/10.1103/PhysRevD.97.072013} {\bibfield  {journal} {\bibinfo
  {journal} {Phys. Rev. D}\ }\textbf {\bibinfo {volume} {97}},\ \bibinfo
  {pages} {072013} (\bibinfo {year} {2018}{\natexlab{a}})},\ \Eprint
  {https://arxiv.org/abs/1711.02505} {arXiv:1711.02505 [hep-ex]} \BibitemShut
  {NoStop}%
\bibitem [{\citenamefont {Aaij}\ \emph
  {et~al.}(2018{\natexlab{b}})\citenamefont {Aaij} \emph
  {et~al.}}]{Aaij:2017uff}%
  \BibitemOpen
  \bibfield  {author} {\bibinfo {author} {\bibfnamefont {R.}~\bibnamefont
  {Aaij}} \emph {et~al.} (\bibinfo {collaboration} {LHCb}),\ }\bibfield
  {title} {\bibinfo {title} {{Measurement of the ratio of the $B^0 \to D^{*-}
  \tau^+ \nu_{\tau}$ and $B^0 \to D^{*-} \mu^+ \nu_{\mu}$ branching fractions
  using three-prong $\tau$-lepton decays}},\ }\href
  {https://doi.org/10.1103/PhysRevLett.120.171802} {\bibfield  {journal}
  {\bibinfo  {journal} {Phys. Rev. Lett.}\ }\textbf {\bibinfo {volume} {120}},\
  \bibinfo {pages} {171802} (\bibinfo {year} {2018}{\natexlab{b}})},\ \Eprint
  {https://arxiv.org/abs/1708.08856} {arXiv:1708.08856 [hep-ex]} \BibitemShut
  {NoStop}%
\bibitem [{\citenamefont {Abdesselam}\ \emph
  {et~al.}(2019{\natexlab{c}})\citenamefont {Abdesselam} \emph
  {et~al.}}]{Abdesselam:2019dgh}%
  \BibitemOpen
  \bibfield  {author} {\bibinfo {author} {\bibfnamefont {A.}~\bibnamefont
  {Abdesselam}} \emph {et~al.} (\bibinfo {collaboration} {Belle}),\ }\bibfield
  {title} {\bibinfo {title} {{Measurement of $\mathcal{R}(D)$ and
  $\mathcal{R}(D^{\ast})$ with a semileptonic tagging method}},\ }\href@noop {}
  {\  (\bibinfo {year} {2019}{\natexlab{c}})},\ \Eprint
  {https://arxiv.org/abs/1904.08794} {arXiv:1904.08794 [hep-ex]} \BibitemShut
  {NoStop}%
\bibitem [{\citenamefont {Colangelo}\ \emph {et~al.}(2020)\citenamefont
  {Colangelo}, \citenamefont {De~Fazio},\ and\ \citenamefont
  {Loparco}}]{Colangelo:2020jmb}%
  \BibitemOpen
  \bibfield  {author} {\bibinfo {author} {\bibfnamefont {P.}~\bibnamefont
  {Colangelo}}, \bibinfo {author} {\bibfnamefont {F.}~\bibnamefont
  {De~Fazio}},\ and\ \bibinfo {author} {\bibfnamefont {F.}~\bibnamefont
  {Loparco}},\ }\bibfield  {title} {\bibinfo {title} {{Probes of Lepton Flavor
  Universality in $b \to u$ Transitions}},\ }\href
  {https://doi.org/10.3390/particles3010012} {\bibfield  {journal} {\bibinfo
  {journal} {Particles}\ }\textbf {\bibinfo {volume} {3}},\ \bibinfo {pages}
  {145} (\bibinfo {year} {2020})}\BibitemShut {NoStop}%
\bibitem [{\citenamefont {Arkani-Hamed}\ \emph {et~al.}(1998)\citenamefont
  {Arkani-Hamed}, \citenamefont {Dimopoulos},\ and\ \citenamefont
  {Dvali}}]{ArkaniHamed:1998rs}%
  \BibitemOpen
  \bibfield  {author} {\bibinfo {author} {\bibfnamefont {N.}~\bibnamefont
  {Arkani-Hamed}}, \bibinfo {author} {\bibfnamefont {S.}~\bibnamefont
  {Dimopoulos}},\ and\ \bibinfo {author} {\bibfnamefont {G.~R.}\ \bibnamefont
  {Dvali}},\ }\bibfield  {title} {\bibinfo {title} {{The Hierarchy problem and
  new dimensions at a millimeter}},\ }\href
  {https://doi.org/10.1016/S0370-2693(98)00466-3} {\bibfield  {journal}
  {\bibinfo  {journal} {Phys. Lett. B}\ }\textbf {\bibinfo {volume} {429}},\
  \bibinfo {pages} {263} (\bibinfo {year} {1998})},\ \Eprint
  {https://arxiv.org/abs/hep-ph/9803315} {arXiv:hep-ph/9803315} \BibitemShut
  {NoStop}%
\bibitem [{\citenamefont {Antoniadis}\ \emph {et~al.}(1998)\citenamefont
  {Antoniadis}, \citenamefont {Arkani-Hamed}, \citenamefont {Dimopoulos},\ and\
  \citenamefont {Dvali}}]{Antoniadis:1998ig}%
  \BibitemOpen
  \bibfield  {author} {\bibinfo {author} {\bibfnamefont {I.}~\bibnamefont
  {Antoniadis}}, \bibinfo {author} {\bibfnamefont {N.}~\bibnamefont
  {Arkani-Hamed}}, \bibinfo {author} {\bibfnamefont {S.}~\bibnamefont
  {Dimopoulos}},\ and\ \bibinfo {author} {\bibfnamefont {G.~R.}\ \bibnamefont
  {Dvali}},\ }\bibfield  {title} {\bibinfo {title} {{New dimensions at a
  millimeter to a Fermi and superstrings at a TeV}},\ }\href
  {https://doi.org/10.1016/S0370-2693(98)00860-0} {\bibfield  {journal}
  {\bibinfo  {journal} {Phys. Lett. B}\ }\textbf {\bibinfo {volume} {436}},\
  \bibinfo {pages} {257} (\bibinfo {year} {1998})},\ \Eprint
  {https://arxiv.org/abs/hep-ph/9804398} {arXiv:hep-ph/9804398} \BibitemShut
  {NoStop}%
\bibitem [{\citenamefont {Randall}\ and\ \citenamefont
  {Sundrum}(1999)}]{Randall:1999ee}%
  \BibitemOpen
  \bibfield  {author} {\bibinfo {author} {\bibfnamefont {L.}~\bibnamefont
  {Randall}}\ and\ \bibinfo {author} {\bibfnamefont {R.}~\bibnamefont
  {Sundrum}},\ }\bibfield  {title} {\bibinfo {title} {{A Large mass hierarchy
  from a small extra dimension}},\ }\href
  {https://doi.org/10.1103/PhysRevLett.83.3370} {\bibfield  {journal} {\bibinfo
   {journal} {Phys. Rev. Lett.}\ }\textbf {\bibinfo {volume} {83}},\ \bibinfo
  {pages} {3370} (\bibinfo {year} {1999})},\ \Eprint
  {https://arxiv.org/abs/hep-ph/9905221} {arXiv:hep-ph/9905221} \BibitemShut
  {NoStop}%
\bibitem [{\citenamefont {Arkani-Hamed}\ and\ \citenamefont
  {Schmaltz}(2000)}]{ArkaniHamed:1999dc}%
  \BibitemOpen
  \bibfield  {author} {\bibinfo {author} {\bibfnamefont {N.}~\bibnamefont
  {Arkani-Hamed}}\ and\ \bibinfo {author} {\bibfnamefont {M.}~\bibnamefont
  {Schmaltz}},\ }\bibfield  {title} {\bibinfo {title} {{Hierarchies without
  symmetries from extra dimensions}},\ }\href
  {https://doi.org/10.1103/PhysRevD.61.033005} {\bibfield  {journal} {\bibinfo
  {journal} {Phys. Rev. D}\ }\textbf {\bibinfo {volume} {61}},\ \bibinfo
  {pages} {033005} (\bibinfo {year} {2000})},\ \Eprint
  {https://arxiv.org/abs/hep-ph/9903417} {arXiv:hep-ph/9903417} \BibitemShut
  {NoStop}%
\bibitem [{\citenamefont {Chang}\ and\ \citenamefont
  {Ng}(2002)}]{Chang:2002ww}%
  \BibitemOpen
  \bibfield  {author} {\bibinfo {author} {\bibfnamefont {W.-F.}\ \bibnamefont
  {Chang}}\ and\ \bibinfo {author} {\bibfnamefont {J.~N.}\ \bibnamefont {Ng}},\
  }\bibfield  {title} {\bibinfo {title} {{CP violation in 5-D split fermions
  scenario}},\ }\href {https://doi.org/10.1088/1126-6708/2002/12/077}
  {\bibfield  {journal} {\bibinfo  {journal} {JHEP}\ }\textbf {\bibinfo
  {volume} {12}},\ \bibinfo {pages} {077}},\ \Eprint
  {https://arxiv.org/abs/hep-ph/0210414} {arXiv:hep-ph/0210414} \BibitemShut
  {NoStop}%
\bibitem [{\citenamefont {Chang}\ \emph {et~al.}(2009)\citenamefont {Chang},
  \citenamefont {Ng},\ and\ \citenamefont {Wu}}]{Chang:2008vx}%
  \BibitemOpen
  \bibfield  {author} {\bibinfo {author} {\bibfnamefont {W.-F.}\ \bibnamefont
  {Chang}}, \bibinfo {author} {\bibfnamefont {J.~N.}\ \bibnamefont {Ng}},\ and\
  \bibinfo {author} {\bibfnamefont {J.~M.~S.}\ \bibnamefont {Wu}},\ }\bibfield
  {title} {\bibinfo {title} {{Flavour Changing Neutral Current Constraints from
  Kaluza-Klein Gluons and Quark Mass Matrices in RS1}},\ }\href
  {https://doi.org/10.1103/PhysRevD.79.056007} {\bibfield  {journal} {\bibinfo
  {journal} {Phys. Rev. D}\ }\textbf {\bibinfo {volume} {79}},\ \bibinfo
  {pages} {056007} (\bibinfo {year} {2009})},\ \Eprint
  {https://arxiv.org/abs/0809.1390} {arXiv:0809.1390 [hep-ph]} \BibitemShut
  {NoStop}%
\bibitem [{\citenamefont {Chang}\ \emph {et~al.}(2008)\citenamefont {Chang},
  \citenamefont {Ng},\ and\ \citenamefont {Wu}}]{Chang:2008zx}%
  \BibitemOpen
  \bibfield  {author} {\bibinfo {author} {\bibfnamefont {W.-F.}\ \bibnamefont
  {Chang}}, \bibinfo {author} {\bibfnamefont {J.~N.}\ \bibnamefont {Ng}},\ and\
  \bibinfo {author} {\bibfnamefont {J.~M.~S.}\ \bibnamefont {Wu}},\ }\bibfield
  {title} {\bibinfo {title} {{Testing Realistic Quark Mass Matrices in the
  Custodial Randall-Sundrum Model with Flavor Changing Top Decays}},\ }\href
  {https://doi.org/10.1103/PhysRevD.78.096003} {\bibfield  {journal} {\bibinfo
  {journal} {Phys. Rev. D}\ }\textbf {\bibinfo {volume} {78}},\ \bibinfo
  {pages} {096003} (\bibinfo {year} {2008})},\ \Eprint
  {https://arxiv.org/abs/0806.0667} {arXiv:0806.0667 [hep-ph]} \BibitemShut
  {NoStop}%
\bibitem [{\citenamefont {Chang}\ \emph {et~al.}(2011)\citenamefont {Chang},
  \citenamefont {Chen},\ and\ \citenamefont {Liou}}]{Chang:2010ic}%
  \BibitemOpen
  \bibfield  {author} {\bibinfo {author} {\bibfnamefont {W.-F.}\ \bibnamefont
  {Chang}}, \bibinfo {author} {\bibfnamefont {I.-T.}\ \bibnamefont {Chen}},\
  and\ \bibinfo {author} {\bibfnamefont {S.-C.}\ \bibnamefont {Liou}},\
  }\bibfield  {title} {\bibinfo {title} {{Neutrino Masses via the Zee Mechanism
  in 5D split fermions model}},\ }\href
  {https://doi.org/10.1103/PhysRevD.83.025017} {\bibfield  {journal} {\bibinfo
  {journal} {Phys. Rev. D}\ }\textbf {\bibinfo {volume} {83}},\ \bibinfo
  {pages} {025017} (\bibinfo {year} {2011})},\ \Eprint
  {https://arxiv.org/abs/1008.5095} {arXiv:1008.5095 [hep-ph]} \BibitemShut
  {NoStop}%
\end{thebibliography}%
\end{document}